%% file: SC_13.tex
\documentclass[12pt]{article}
\usepackage{mathrsfs}
\usepackage{graphicx}
\usepackage{epstopdf}
\usepackage{amsthm}
\usepackage{amsmath}
\usepackage{amssymb}
\usepackage{amstext}
\usepackage{amsfonts}
\usepackage{enumerate}
\usepackage[authoryear,bibstyle]{natbib} 
\usepackage{footnote}
\makesavenoteenv{tabular}
\makesavenoteenv{table}
\usepackage[onehalfspacing]{setspace}
\usepackage[shortlabels, inline]{enumitem}
\usepackage{multirow}
\usepackage[title]{appendix}
\usepackage{xcolor}
\usepackage{cancel}  
\usepackage{tikz}
\usetikzlibrary{shapes,arrows}
\usepackage{tikz-cd}
\usepackage{sectsty}
\usepackage{caption, subcaption} 
\usepackage[colorinlistoftodos,textsize=tiny]{todonotes}
\usepackage{booktabs}
\usepackage{comment}
\usepackage[nottoc]{tocbibind} 
\usepackage[titles]{tocloft}
\makeatletter
\renewcommand{\@makefntext}[1]{%
  \noindent\@makefnmark\,%
  \begingroup
  \setlength{\parindent}{0pt}%
  #1%
  \endgroup
}
\makeatother

\AtBeginDocument{%
  \setlength{\abovedisplayskip}{6pt}%
  \setlength{\belowdisplayskip}{6pt}%
  \setlength{\abovedisplayshortskip}{3pt}%
  \setlength{\belowdisplayshortskip}{3pt}%
}

\begin{document}
\baselineskip=.22in\parindent=30pt

\newtheorem{tm}{Theorem}
\newtheorem{dfn}{Definition}
\newtheorem{lma}{Lemma}
\newtheorem{assu}{Assumption}
\newtheorem{prop}{Proposition}
\newtheorem{cro}{Corollary}
\newtheorem*{theorem*}{Theorem}
\newtheorem{example}{Example}
\newtheorem{observation}{Observation}
\newtheorem{axm}{Axiom}
\newcommand{\exm}{\begin{example}}
\newcommand{\exmm}{\end{example}}
\newcommand{\obs}{\begin{observation}}
\newcommand{\obss}{\end{observation}}
\newcommand{\cor}{\begin{cro}}
\newcommand{\corr}{\end{cro}}
\newtheorem{exa}{Example}
\newcommand{\ex}{\begin{exa}}
\newcommand{\exx}{\end{exa}}
\newtheorem{remak}{Remark}
\newcommand{\rmk}{\begin{remak}}
\newcommand{\rmkk}{\end{remak}}
\newcommand{\thm}{\begin{tm}}
\newcommand{\nt}{\noindent}
\newcommand{\thmm}{\end{tm}}
\newcommand{\lm}{\begin{lma}}
\newcommand{\lmm}{\end{lma}}
\newcommand{\ass}{\begin{assu}}
\newcommand{\asss}{\end{assu}}
\newcommand{\df}{\begin{dfn}  }
\newcommand{\dff}{\end{dfn}}
\newcommand{\prp}{\begin{prop}}
\newcommand{\prpp}{\end{prop}}
\newcommand{\axiom}{\begin{axm}}
\newcommand{\axiomm}{\end{axm}}

\newcommand{\bqu}{\sloppy \small \begin{quote}}
\newcommand{\equ}{\end{quote} \sloppy \large}
\newcommand\cites[1]{\citeauthor{#1}'s\ (\citeyear{#1})}

\renewcommand{\leq}{\leqslant}
\renewcommand{\geq}{\geqslant}

\newcommand{\eq}{\begin{equation}}
\newcommand{\eqq}{\end{equation}}
\newtheorem{claim}{\it Claim}
\newcommand{\cl}{\begin{claim}}
\newcommand{\cll}{\end{claim}}
\newcommand{\bit}{\begin{itemize}}
\newcommand{\eit}{\end{itemize}}
\newcommand{\ben}{\begin{enumerate}}
\newcommand{\een}{\end{enumerate}}
\newcommand{\bcen}{\begin{center}}
\newcommand{\ecen}{\end{center}}
\newcommand{\fn}{\footnote}
\newcommand{\ds}{\begin{description}}
\newcommand{\dss}{\end{description}}
\newcommand{\prf}{\begin{proof}}
\newcommand{\prff}{\end{proof}}
\newcommand{\cs}{\begin{cases}}
\newcommand{\css}{\end{cases}}
\newcommand{\ml}{\item}
\newcommand{\lb}{\label}
\newcommand{\ra}{\rightarrow}
\newcommand{\tra}{\twoheadrightarrow}
\newcommand*{\supp}{\operatornamewithlimits{sup}\limits}
\newcommand*{\inff}{\operatornamewithlimits{inf}\limits}
\newcommand{\nf}{\normalfont}
\renewcommand{\Re}{\mathbb{R}}
\newcommand*{\mmax}{\operatornamewithlimits{max}\limits}
\newcommand*{\mmin}{\operatornamewithlimits{min}\limits}
\newcommand*{\argmax}{\operatornamewithlimits{arg max}\limits}
\newcommand*{\argmin}{\operatornamewithlimits{arg min}\limits}
\newcommand{\uhr}{\!\! \upharpoonright  \!\! }

\newcommand{\CR}{\mathcal R}
\newcommand{\CC}{\mathcal C}
\newcommand{\CT}{\mathcal T}
\newcommand{\CS}{\mathcal S}
\newcommand{\CM}{\mathcal M}
\newcommand{\CL}{\mathcal L}
\newcommand{\CP}{\mathcal P}
\newcommand{\CN}{\mathcal N}

\newtheorem{innercustomthm}{Theorem}
\newenvironment{customthm}[1]
  {\renewcommand\theinnercustomthm{#1}\innercustomthm}
  {\endinnercustomthm}
\newtheorem{einnercustomthm}{Extended Theorem}
\newenvironment{ecustomthm}[1]
  {\renewcommand\theeinnercustomthm{#1}\einnercustomthm}
  {\endeinnercustomthm}
  
  \newtheorem{innercustomcor}{Corollary}
\newenvironment{customcor}[1]
  {\renewcommand\theinnercustomcor{#1}\innercustomcor}
  {\endinnercustomcor}
\newtheorem{einnercustomcor}{Extended Theorem}
\newenvironment{ecustomcor}[1]
  {\renewcommand\theeinnercustomcor{#1}\einnercustomcor}
  {\endeinnercustomcor}
    \newtheorem{innercustomlm}{Lemma}
\newenvironment{customlm}[1]
  {\renewcommand\theinnercustomlm{#1}\innercustomlm}
  {\endinnercustomlm}

\newcommand{\red}{\textcolor{red}}
\newcommand{\blue}{\textcolor{blue}}
\newcommand{\purple}{\textcolor{purple}}
\newcommand{\mred}[1]{\color{red}{#1}\color{black}}
\newcommand{\mblue}[1]{\color{blue}{#1}\color{black}}
\newcommand{\mpurple}[1]{\color{purple}{#1}\color{black}}

\newcommand{\norm}[1]{\left\lVert#1\right\rVert}

\newcommand{\citea}[1]{\citeauthor{#1}}
\renewcommand{\cite}[1]{\citeauthor{#1} (\citeyear{#1})}
\renewcommand{\citep}[2]{\citeauthor{#1} (\citeyear{#1},\space{#2})}
\newcommand{\citeaa}[1]{\citeauthor*{#1}}
\newcommand{\citeaay}[1]{\citeauthor*{#1} (\citeyear{#1})}
\newcommand*{\citeaayp}[2]{\citeauthor*{#1} (\citeyear{#1},\space{#2})}

\subsubsectionfont{\normalfont\itshape}

\makeatletter
\newcommand{\customlabel}[2]{%
\protected@write \@auxout {}{\string \newlabel {#1}{{#2}{}}}}
\makeatother

\renewcommand\bibname{References}

\renewcommand{\baselinestretch}{1.15} 


\def\qed{\hfill\vrule height4pt width4pt
depth0pt}
\def\reff #1\par{\noindent\hangindent =\parindent
\hangafter =1 #1\par}
\def\title #1{\begin{center}
{\Large {\bf #1}}
\end{center}}
\def\author #1{\begin{center} {\large #1}
\end{center}}
\def\date #1{\centerline {\large #1}}
\def\place #1{\begin{center}{\large #1}
\end{center}}

\def\date #1{\centerline {\large #1}}
\def\place #1{\begin{center}{\large #1}\end{center}}
\def\intr #1{\stackrel {\circ}{#1}}
\def\R{{\rm I\kern-1.7pt R}}
 \def\N{{\rm I}\hskip-.13em{\rm N}}
 \newcommand{\cprod}{\Pi_{i=1}^\ell}
\let\Large=\large
\let\large=\normalsize


\begin{titlepage}

\def\thefootnote{\fnsymbol{footnote}}

\title{On Continuity of Separately Convex  Preferences and Correspondences\fn{Some of the material of this paper was previously circulated under the title  \textit{Separately Convex and Separately Continuous Preferences:
On Results of Schmeidler, Shafer and Bergstrom-Parks-Rader}, and presented  at the {\it  Midwest Economic Theory and International Trade Meetings} held at the University of Rochester in October 2024,   and at the  \textit{Summer Workshop in Economic Theory (SWET)} at Universit\'e Paris-1 Panth\'eon-Sorbonne in December 2025. Versions of the work were also presented  at departmental seminars  at the University of California at Riverside (April  2025), at Rutgers University (February 2026) and at the University of Iowa (February 2026). The authors were especially heartened by the reaction of Jerry Green and the probing questions of Oriol Carbonell-Nicolau. They also acknowledge with pleasure the encouraging comments of  Paulo Barelli, Eric Fisher, Hari Govindan, Davide Carpentiere, Alfio Giarlotta, Jason Lepore, Hiroki Nishimura, Han Ozyolev, Angelo Petralia, Kevin Reffett, Nobusumi Sagara, Ed Scheinerman, Eddie Schlee, Siyang Xiong, Akira Yamazaki  and  Nicholas Yannelis.  }}


\author{Aniruddha Ghosh\fn{Orfalea College of Business, Cal Poly, San Luis Obispo, CA 93401. {\bf E-mail} {aghosh10@calpoly.edu}.} ~~ M. Ali Khan\fn{Department of Economics, Johns Hopkins University, Baltimore, MD 21218. {\bf E-mail}  {akhan@jhu.edu}} ~~ Metin Uyanik\fn{School of Economics, University of Queensland, Brisbane, QLD 4072.  {\bf E-mail}
{m.uyanik@uq.edu.au}.}  }

\vskip 0.10em

\date{\today}

\vskip 2.0em

\baselineskip=.18in

\noindent{\bf Abstract:} We study separate convexity for preferences and correspondences, and show that this weakening of the usual convexity postulate is strong enough to recover standard equivalences among continuity assumptions. For complete and transitive preferences, we establish equivalence theorems linking separate continuity, mixture continuity, Archimedean-type postulates, solvability and graph continuity, successively on product mixture sets and on Euclidean spaces. The results highlight the role of weaker axiomatic assumptions by  yielding representations for multilinear cardinal utility, continuous separately quasiconcave ordinal utility in $n$-person decision problems, and a scalar Anscombe--Aumann setting. For non-ordered preferences, formulated as correspondences, we characterize the open graph property under separate convexity and weak section-continuity, generalizing results of Schmeidler, Shafer, and Bergstrom-Parks-Rader. Examples identify the boundaries of our results.

 \vskip 0.6cm

\noindent {\it Key Words:} Separate convexity, separate continuity, open graph property, utility representation

\bigskip

\noindent {\it JEL} Classification: C60, D01
\smallskip

\bigskip 

\end{titlepage}









%

\setcounter{footnote}{0}


\setlength{\abovedisplayskip}{-10pt}
\setlength{\belowdisplayskip}{-10pt}





\tableofcontents 

\vskip\medskipamount 
\leaders\vrule width \textwidth\vskip0.6pt 
\vskip\medskipamount 
\nointerlineskip


\bqu {\it When one examines the main contents of received theory of resource allocation and competitive markets it is found that its
propositions depend essentially on convexity assumptions
with regard to both production possibilities
and preference structures.... The convexity
concept therefore enables us to state minimum assumptions
for the validity of an important part of
existing economic theory, thus helping to reduce this
part of our knowledge to its logical and mathematical
essentials.}\fn{See page 25 in \cite{ko57}, and also mutual  apologies by Koopmans and Farrell   to each other in the  elaboration of the context of the statement; see \cite{ko61} and \citet{fa59, fa61a}. 
}    \hfill{\cite{ko57}} \equ  
\vspace{-15pt}

\bqu {\it The assumption $P(x) \equiv \{y \in \mathbb R^n: y \succ x\}$ is convex demands more convexity than is needed for many purposes in general equilibrium analysis.} \hfill{\cite{bpr76}} \equ
\vspace{-15pt}

\section{Introduction}
\vspace{-5pt}

The convexity postulate on sets, functions, correspondences and binary relations plays a fundamental role in a limitation and  qualification of these objects in so far as they are used in the articulation of  modern economic theory. The fact that convexity can be substituted by a multiplicity of agents, and their consequent economic negligibility, is also by now well-understood as a consequence of work done in the seventies.\fn{There is by now a rich literature on this subject; the reader can begin with Chapters 7 and 8 in \cite{ah71}, \cite{hi74}, \cite{ks02} and their references.   }  
In this paper, rather than to a substitute for it, we turn to the 
  postulate itself, and  study a weaker algebraic notion, {\it separate convexity,} that lies on the spectrum between full convexity and non-convexity, and thereby offers  a less restrictive property for capturing what Koopmans refers to as the \lq\lq essentials."  We investigate how the notion of separate convexity not only captures more general preference and correspondence structures but also consolidates and generalizes existing results, which can now be regarded as pioneering. In particular, the main contribution of this paper is threefold: (i) we show that the widely used continuity postulates for ordered preferences that arise in economics, mathematics, and mathematical psychology are equivalent under the separate convexity assumption; (ii) we illustrate how separate convexity naturally arises in multi-person decision problems, and provide new axiomatizations for utility representation in $n$-person games; and (iii) motivated by the weak continuity concepts for functions, we provide new continuity notions for non-ordered preferences and establish equivalences of these concepts and the open graph property under separate convexity.

  In the context of individual choice, the convexity assumption implies that if an individual prefers one outcome to another, then every mixture of the two outcomes is weakly preferred to the less-preferred one.\fn{Under choices involving risk and uncertainty, convexity reflects a preference for diversification or risk aversion, whereby combining risks is generally favored over concentrating them. It is of interest that the interpretation of the convexity postulate, insofar as it interacts with other postulates such as continuity and monotonicity --- comonotonicity and {\it anticomonotonicity} --- continues in recent work; see \cite{pww24} and the references therein.} In contrast, non-convexity allows for preferences that deviate from this property, and such preferences often arise in settings involving risk, ambiguity, and other-regarding concerns. We say that a binary relation $\succsim$ on $\mathbb{R}^{n}$ is said to be \textit{separately convex} if, when restricted to any line parallel to a coordinate axis in $\mathbb{R}^{n}$, its weakly better-than and worse-than sections are convex. For example, consider the preference relation on $\mathbb{R}^{2}$ represented by the utility function, $u(x,y)=\max\{x,y\}$. The weakly better-than sets of this relation are not convex, and hence the preference relation itself is not convex. However, when either coordinate is fixed, the weakly better-than sections, and also the weakly worse-than sections, are convex. Thus the preference relation is separately convex. In what follows, we define separate convexity far more generally for binary relations and correspondences in Sections 2 and 4, respectively.

In terms of our \textit{first} specific contribution, we begin by  revisiting the foundational insight established by \cite{gp84} that \lq\lq separate continuity does not imply continuity"  and inspired by it,  deconstruct the \textit{convexity} concept as it is applied in economic theory.\footnote{The relationship between joint continuity and separate continuity, and the structure of discontinuity for separately continuous functions, is well studied, even during the time of Cauchy in the early 19th century. It remains a standard topic in multivariate calculus textbooks and leads to significant developments by mathematicians such as Heine, Baire, and Lebesgue. This work culminates in the benchmark results of \cite{yo10qjpam}, \cite{ro55} and \cite{kd69amm}; see, for example, \cite{cm16} and the references therein. For further discussion on its consolidation in the context of mathematical psychology and mathematical economics, see \cite{gku22games}.}  From an analytical point of view, therefore, this  contribution of the  paper then lies not so much in pushing  the internal investigation of the topological (continuity) register further, but in supplementing it with the sister-registers of linearity and order. 
It is in this spirit that in Section 2, we offer \textit{three} equivalence theorems among the widely used continuity postulates for ordered binary relations under the separate convexity assumption: one for mixture spaces, one for the relationship between separate and ordinary section-continuity on coordinate Euclidean domains, and one for the finite-dimensional graph/solvability equivalence. These theorems illustrate that as we impose more structure on the domain of the preference relation, we obtain additional  equivalences under the separate convexity assumption. Two examples, built on a separately continuous bilinear form on an infinite-dimensional Hilbert space endowed with its weak topology, show that the finite-dimensional structure in the second and third of these theorems is essential and cannot be removed (Example \ref{ex_bilinear_pref}). These results extend and consolidate previous findings in mathematical economics and mathematical psychology regarding complete and transitive binary relations; the resulting two-sided equivalence relationships are summarized in Figures \ref{fig_relation_partial} and \ref{fig_relation}.\fn{A detailed discussion of these figures highlights how our results generalize earlier findings by (inessentially) weakening continuity and (essentially) weakening the convexity assumption. For a further  exploration of the notions of {\it hiddenness} and {\it essentiality}, introduced in \cite{kr86}, see \cite{uk22jme}. \label{fn:hidden}}

We provide our \textit{second} specific contribution in Section 3 by illustrating how separate convexity naturally arises in multi-person decision problems. We apply our results sequentially to provide new axiomatizations for multilinear and quasiconcave utility in $n$-person games, covering both ordinal and cardinal utility representations. The ordinal representation is based on \cite{de52}, while the cardinal one is based on \cite{fi82}. For ordinal settings, our results  enable the construction of continuous and quasiconcave (in own action) utility functions under separate convexity and separate continuity;  while for cardinal settings, multilinear utility representations become possible through separate independence and separate continuity. This representation extends the framework of \cite{fr78} and addresses challenges in representing preferences that exhibit structure in some dimensions but lack global convexity (Proposition \ref{thm_cardinal_game}). This flexibility is particularly relevant in decision-making contexts involving multiple attributes. We also use the interplay between separate convexity and separate continuity to recover the completeness assumption implicit in subjective expected utility. In a scalar Anscombe--Aumann environment, we do not impose completeness as a primitive postulate. Rather, we recover it from transitivity, weak independence requirements, and continuity assumptions imposed only along state-wise coordinate sections. Once completeness is recovered, we show that the usual independence and state-independence assumptions deliver a subjective expected utility representation with a common affine utility index and state weights.

We shift our focus in Section 4 to non-ordered preferences, formulated more generally as correspondences, and revisit the relationship between section continuity and graph continuity. From the viewpoint of mainstream mathematical economics, we single out \citet{sc69}, \cite{sh74}, and \cite{bpr76} to explore the relationship between section and graph continuity assumptions for a preference relation, which are known to be equivalent under completeness and transitivity. \citet{sc69} drops completeness but retains transitivity and establishes equivalence under monotonicity. \citet{sh74} demonstrates that under a strong convexity assumption, section and graph continuity are equivalent for a complete but non-transitive preference relation. \citet{bpr76} generalize these results, providing a comprehensive treatment of the relationship between section and graph continuity.  There is also a substantive literature in economic theory on relaxing continuity assumptions on preferences and correspondences.\fn{See, for example, \cite{yp83}, \cite{dm86}, \cite{re99}, \cite{ca11}, \cite{hy16} and \cite{re20are}.} Our focus here, however, is not on the equilibrium-existence literature, but on the axiomatization literature -- we show that, under weak convexity conditions, continuity assumptions commonly employed in economics, mathematical psychology, and mathematics are equivalent to stronger continuity requirements.\fn{In this line of research, previous work of the authors examines the relationship among different continuity assumptions in mathematical economics and decision theory under monotonicity or full convexity assumptions on preferences; see, for example, \cite{uk22jme}, \citet{gku22games,gku23td}, and references therein. Also see \citet{ka07}, \citet{du11}, and \citet{gku19} for studies on the relationship among scalar continuity assumptions in decision theory.}

Building on this literature, we present our \textit{third} specific contribution by defining a notion of \textit{separate} convexity for correspondences. Under separate convexity, Theorems \ref{thm_separate_main} and \ref{thm_linear_main} characterize the open graph property and restore equivalences among several notions of correspondence continuity. Since a binary relation can be viewed as the graph of a correspondence, these results, together with their corollaries, substantially generalize \citet[5.1]{sc69}, \citet[main Lemma]{sh74}, and \citet[Theorem 3]{bpr76} by considerably weakening their continuity and convexity/monotonicity assumptions, and by allowing more general domains. These results and Example \ref{exm_yamazaki} also address a question left open by \citet{bpr76}: whether Shafer's results can be generalized to infinite-dimensional settings or arbitrary convex subsets of $\mathbb{R}^n$. Finally, we revisit and partially extend the infinite-dimensional results of \citet{ya83b}, thereby clarifying the relationship between separate convexity, section continuity, and the open graph property.  
In particular, Example \ref{ex_bilinear_pref} shows that the open-graph
characterizations do not extend, under their present hypotheses, to
infinite-dimensional coordinate factors: any extension would require
additional assumptions, in the spirit of Yamazaki's local finiteness
property, on which the example is silent.

In terms of   substantive work, there are several reasons why we think separate convexity is an appealing subject for investigation. First, through its focus on correspondences in general, it connects to work in operations research, providing a framework to analyze problems where traditional convexity assumptions may not hold. 
 Example \ref{ex_hpz} in particular is taken from \citet{hpz17} who also highlight the implications of non-convexity for techniques in revealed preferences. 
Second, our results   resonate with  \citet{rr19}  who  propose a generalized version of the convexity postulate, $\Psi$-convexity, for non-algebraic structures. In Example \ref{separate-rr}, we show that our notion of separate convexity is equivalent to a notion of separate
$\Psi$-convexity in their sense, where convexity is restricted to coordinate directions. In this connection, 
 one could ask what specifications on the choice set would yield a separate convexity-like assumption for their model and whether such conditions can even arise. Finally, going beyond the representation results in $n$-person games, separate convexity has a strong appeal for the general representation of binary relations under various continuity postulates, offering insights into the structural properties of both the functions and the relations at issue.\footnote{The practical relevance of separate convexity is evident in several economic and technical examples. In finance, risk preferences often exhibit separate convexity, reflecting well-behaved risk-averse behavior along individual asset classes. In production theory, the concept parallels pointwise convexity as developed by \cite{he73}, enabling a more nuanced analysis of marginal rates under weaker convexity assumptions.}  We hope that our paper contributes to this ongoing discussion on convexity, engaging with both the economically substantive and technical perspectives. 

\medskip
\noindent{\it Plan of the paper.} The rest of the paper is organized as follows. Section 2 introduces separate convexity for binary relations and presents three equivalence results for continuity postulates: product mixture sets, the relationship between separate and ordinary section-continuity on coordinate Euclidean domains, and the finite-dimensional graph/solvability equivalence. Section 3 applies these results to $n$-person decision problems and derives utility representations for both cardinal and ordinal formulations. Section 4 turns to correspondences, develops  the corresponding notion of separate convexity and characterizes the open graph property under weak section-continuity assumptions. Section 5 concludes with a discussion of the results and points to some directions for further work. Throughout these sections, we include a host of examples to illustrate the concepts and the subsequent results. The proofs are collected in the Section 6 which may alternatively be regarded as an appendix to the paper. For the reader's orientation: Theorems \ref{thm_equiv_pref_mixture}--\ref{thm_linear_main} and Propositions \ref{thm_cardinal_game}--\ref{thm_additional_directional_linear} are new; Corollaries \ref{thm_shafer} and \ref{thm_schmeidler} restate the classical results that they specialize to; Example \ref{exm_yamazaki} resolves the Bergstrom--Parks--Rader open problem in the negative; Example \ref{ex_bilinear_pref} locates the finite-dimensional boundary of the theory.

\section{Continuity of Separately Convex Preferences}\label{sec_ordered}

We begin with  the preliminaries.  Let $X$ be a set. A subset $\succsim$ of $X\times X$ denotes a {\it binary relation} on $X.$ We denote an element $(x,y)\in ~\!\!\! \succsim$ as $x\succsim y.$ The {\it  asymmetric part} $\succ$ of $\succsim$ is defined as $x\succ y$ if $x\succsim y$ and  $y\not\succsim x$, and its {\it symmetric part} $\sim$ is defined as $x\sim y$ if $x\succsim y$ and $y\succsim x.$     We call $x\bowtie y$ if $x\not\succsim y$ and $y\not\succsim x$. 
The inverse of $\succsim$ is defined as  $x\precsim y$ if $y\succsim x$. Its asymmetric part $\prec$ is defined analogously and its symmetric part is $\sim$. 
%
For any $x\in X$, let    
 $
A_\succsim(x)=\{y\in X| y\succsim x\}$ denote the {\it upper section} of $\succsim$ at $x$,  and  $A_\precsim(x)=\{y\in X| y\precsim x\}$ its {\it lower section} at $x$. 

\df
The binary relation $\succsim$ on $X$ is  {\it convex} if  for all $x\in X$, $A_{\succsim}(x)$ is convex.\fn{A set \(C\) is \emph{convex} if for every \(x,y \in C\) and every \(\lambda \in [0,1]\), we have $\lambda x + (1-\lambda)y \in C.$}  

\dff

\noindent A preference relation   $\succsim$  on a set $X$ is {\it reflexive} if $x \succsim x$ for all $x \in X$, {\it complete} if $x \succsim y$ or $y \succsim x$  for all $x, y \in X$, {\it transitive} if  $x \succsim y \succsim z \Rightarrow x \succsim z$ for all $x, y, z \in X$, {\it order-dense} if $x \succ y$ implies there exists $z\in X \ \mbox{such that} \ x \succ z\succ y \ \forall \ x, y \in X$.  
The results in this section are defined for binary relations that are complete and transitive.

In what follows, we shall need the following notation and concept, which is central to our paper.  Let $I=\{1, \ldots, n\}$, $X=\prod_{i\in I} X_i$ be a Cartesian product of sets, $Y\subseteq X$, and $\succsim$ be a binary relation on $Y$. For each $y\in Y$ and $i\in I$, the  $i$-th section of $Y$ at $y$ is defined as follows,\fn{This concept is related to the sections of a set; see for example \cite{fa66convex} for many applications of sets with convex sections.}   
\begin{align}
    Y_{i,y}=\{z\in Y|z_{-i}=y_{-i}\},
\end{align}  
where $z$ equals $y$ in all except the $i$-th coordinate ($z_{-i}=y_{-i}$).
Note that a \textit{straight line} in a set $X$ in a vector space is defined as the intersection of $X$ with a one-dimensional affine subset of the affine hull of $X$. It is easy to see that if each $X_i$ is a subset of $\Re$, then the $i$-th section of $Y$ at $y\in Y$ denotes the straight line parallel to coordinate $i$ passing through $y$.  We now present the definition of an upper and lower  separately convex binary relation.

\df
A binary relation $\succsim$ on $Y$ is upper (lower) separately convex if for all $x, y\in Y$ and all $i\in I$, $A_\succsim(x)\cap Y_{i, y}$ ($A_\precsim(x)\cap Y_{i,y}$) is convex. 
\dff

  Example \ref{ex_hpz} provides two instances of non-convex preferences that are \textit{separately} convex (both upper and lower separately convex), also illustrated in Figure~\ref{fig_hpz}.

\ex
\label{ex_hpz} {\nf   Consider the following two binary relations from \cite{hpz17} that are defined on  $\mathbb R^{2}_{+}$ and are represented by the following utility functions, 
\begin{equation}
\begin{array}{rl}
u_{1}(x, y)= \begin{cases}x^3 y & \text { if } x \geq y \\ x y^3 & \text { if } x\leq y\end{cases} 
&  \text{ and } ~~
u_{2}(x, y)= \sqrt{\max\{x,y\}}+\dfrac{1}{4}\sqrt{\min{\{x,y\}}}. 
\end{array}
\label{exm1}
\end{equation}

\begin{figure}[t]
\centering
\begin{subfigure}[t]{.45\textwidth}
\centering
\resizebox{0.8\linewidth}{!}{\input{figure_sep_ex.tex}}
\caption{$u_1(x,y)=x^3y$ if $x\ge y$, $\ xy^3$ if $x\le y$}
\label{fig:sub1}
\end{subfigure}\hfill
 \begin{subfigure}[t]{.45\textwidth}
 \centering
 \resizebox{0.8\linewidth}{!}{\input{figure4.tex}}
 \caption{$u_2(x,y)=\sqrt{\max\{x,y\}}+\tfrac14\sqrt{\min\{x,y\}}$}
 \label{fig:sub2}
 \end{subfigure}
\caption{The non-convex but separately convex preferences of Example~\ref{ex_hpz}. Panels (a) and (b) illustrate the indifference curves corresponding to the two utility functions, $u_{1}$ and $u_{2}$. Notice that while the preferences are non-convex, they are upper and lower separately convex: given two points on an indifference curve, a linear combination of these points need not lie within  the weakly preferred set of outcomes, thereby violating the full convexity requirement. However, restricted to line parallel to the horizontal or the vertical axis, the weakly better-than and worse than sets are convex.}
\label{fig_hpz}
\end{figure}}
\exx

In the remainder of this section, we present continuity postulates and three theorems that describe the relationships among these postulates for binary relations defined first on product mixture sets and then on convex subsets of $\mathbb R^n$ under the assumption of separate convexity. These theorems demonstrate that adding more structure to the space where the binary relation is defined leads to additional equivalences. In terms of structures, we proceed as follows:
\[\text{Mixture sets} \;\Longrightarrow\; \text{Coordinate section-continuity on } \mathbb{R}^n \;\Longrightarrow\; \text{Graph and solvability equivalences on } \mathbb{R}^n.\]

\subsection{Preferences on a Mixture Set}

A set $X$ is said to be a {\it mixture set} if for any $x, y \in X$ and any $\lambda\in [0,1]$, where the interval $[0,1]$ is endowed with the usual topology, we can associate an element of $X$, which we write as $x \lambda y$ (equal to $\lambda x+ (1-\lambda)y$ when $X$ is a convex subset of a vector space), such that for all $x, y \in X$ and all $\lambda, \mu\in [0,1]$, the following three conditions hold: (a) $x 1 y=x$, (b) $x \lambda y=y(1-\lambda) x$, and (c) $(x \mu y) \lambda y=x(\lambda \mu) y$. When $X=\prod_{i=1}^n X_i$ is a product of mixture sets, the mixture operation is understood coordinate-wise, so that $x\lambda y=(x_i\lambda y_i)_{i=1}^n$. A subset $C$ of a mixture set $X$ is {\it mixture-convex} if $x\lambda y\in C$ for all $x, y\in C$ and all $\lambda\in [0,1]$; when $X$ is a convex subset of a vector space with $x\lambda y=\lambda x+(1-\lambda)y$, mixture-convexity coincides with ordinary convexity. On a product of mixture sets, the separate convexity of a binary relation is understood in this mixture sense: $\succsim$ is upper (lower) separately convex if $A_\succsim(x)\cap X_{i,y}$ ($A_\precsim(x)\cap X_{i,y}$) is mixture-convex for all $x, y\in X$ and all $i\in I$.

We now define the continuity postulates on a mixture set. Let $\succsim$ be a binary relation defined on a finite product of mixture sets, $X=\prod_{i=1}^nX_i$. We begin with the classical mixture continuity and Archimedean properties. A binary relation  $\succsim$ is {\it upper (lower) mixture continuous} if, for any $x,y,z\in X$,  $\{\lambda\in [0,1]| x\lambda y\succsim z\}$ is closed ($ \{\lambda\in [0,1]| x\lambda y\precsim z\}  $ is closed) in the unit interval $[0,1]$,  
and    {\it mixture continuous} if $\succsim$ is  upper   and lower mixture  continuous;  and $\succsim$  is  
{\it upper $\left(\text{lower}\right)$ Archimedean} if, for any $x,y,z\in X, x\succ y$ implies that there exists $\lambda\in (0,1)$ $\left( \delta\in (0,1)\right)$  such that $x\lambda z\succ y$ $\left(x\succ y\delta z\right)$, and {\it  Archimedean} if it is upper and lower Archimedean.  

We now introduce the {\it separate} versions of mixture continuity and Archimedean properties. A binary relation $\succsim$  
is {\it upper (lower) separate mixture continuous} if, for any $x,y,z\in X$ and any $i=1,\ldots, n$, $x_{-i}=y_{-i},$  $\{\lambda\in [0,1]| x\lambda y\succsim z\}$ is closed ($ \{\lambda\in [0,1]| x\lambda y\precsim z\}  $ is closed) in the unit interval $[0,1]$,  
and    {\it separate mixture continuous} if $\succsim$ is  upper  and lower separate  mixture  continuous; and   $\succsim$  is {\it upper (lower) separate  Archimedean} if, for any $x,y,z\in X$ and any $i=1,\ldots, n$,  $x\succ y$ and $x_{-i}=z_{-i}$ ($y_{-i}=z_{-i}$) imply that there exists $\lambda\in (0,1)$ $\left( \delta\in (0,1)\right)$  such that $x\lambda z\succ y$ $\left(x\succ y\delta z\right)$, and {\it  separate Archimedean} if it is upper and lower separate Archimedean. 

The following solvability postulate is commonly used in decision theory -- a binary relation $\succsim$  defined on a mixture set $Y\subseteq X$ is  \textit{restricted solvable}  if for all $i\in \{1,\ldots,n\},$ all $x, y\in Y$ and all $(a_{i},  y_{-i} ), (b_{i},  y_{-i} )\in Y$ with  $(a_{i},  y_{-i} )\succsim x\succsim (b_{i}, y_{-i} )$, there exists $c_{i}$ with $(c_{i}, y_{-i} )\in Y$ such that $x\sim (c_{i}, y_{-i} ).$  Next, we present {\it solvability}-type continuity postulates that rely on mixture operation.  A binary relation $\succsim$  is {\it weakly Wold-continuous} if it is order-dense and $x\succ y\succ z$ implies that there exists $\lambda\in (0,1)$ such that  $x\lambda z\sim y$; $\succsim$ satisfies the
{\it intermediate value property (IVP)} if, for all $x,y,z\in X$ with $x\succsim y\succsim z$,  there exists $\lambda\in [0,1]$ such that $x\lambda z\sim y$; and it satisfies the {\it separate IVP}  if, for all $i\in \{1,\ldots,n\},$ all $x, y\in X$ and all $(a_{i},  y_{-i} ), (b_{i},  y_{-i} )\in X$ with  $(a_{i},  y_{-i} )\succsim x\succsim (b_{i}, y_{-i} )$, there exist $\lambda\in [0,1]$ and $c_{i}, d_i\in X_{i}$, with $a_{i} =c_i\delta d_i$ and  $b_{i} =c_i\delta' d_i$ for some $\delta, \delta'\in [0,1]$, 
such that $(c_{i}\lambda d_i, y_{-i} )\in X$ and $x\sim (c_{i}\lambda d_i, y_{-i} ).$ It is not difficult to verify that restricted solvability is stronger than separate IVP, but for subsets of $\Re^n$; see Claim \ref{thm_rs_sivp} in Section 6 for a proof.

\rmk {\nf Mixture continuity and Archimedean properties are standard in decision theory; see, for example, \cite{fi82} and \cite{wa89}. Weak-Wold continuity was introduced by \cite{wj53}, who extended the seminal work of \cite{wo43} on the numerical representation of a binary relation by a continuous function.\fn{This continuity postulate was independently introduced in decision theory by \cite{na50b} and \cite{ma50}; see \cite{bl16} for historical remarks.} The IVP is motivated by the corresponding property for functions and is analogous to weak-Wold continuity. Further, motivated by the separate continuity of functions, we introduce the weaker separate versions of these continuity postulates. 
There are nested relationships among these continuity postulates which are illustrated in Figures \ref{fig_relation_partial} and \ref{fig_relation}, and explicitly stated and proved in Section \ref{sec_proofs}.\fn{$A \Rightarrow B$ means $A$ is the stronger assumption and automatically ensures the weaker assumption $B$. Equivalently, $\{\,\text{theories satisfying }A\}\subseteq\{\,\text{theories satisfying }B\},$ and hence, they are nested.
}}\rmkk

\thm Let $\succsim$ be a complete and transitive binary relation on a non-empty
mixture set $X = \prod_{i \in I} X_i$, where $I = \{1, \ldots, n\}$ is a
finite index set.

\ben[{\nf (a)}, topsep=2pt]
\setlength{\itemsep}{-1pt} 
\item If $\succsim$ is upper separately convex, then $\succsim$ is lower mixture
continuous if and only if it is upper separate Archimedean. Symmetrically, if
$\succsim$ is lower separately convex, then $\succsim$ is upper mixture
continuous if and only if it is lower separate Archimedean.

\item If $\succsim$ is order dense, and both upper and lower separately convex,
then $\succsim$ is mixture continuous if and only if it satisfies the separate
IVP.
\een
\label{thm_equiv_pref_mixture}
\thmm

\rmk \nf{
In Theorem \ref{thm_equiv_pref_mixture}, we present equivalence between the strongest and weakest continuity postulates under separate convexity. We make three observations here. First, in part (a), the directional pairing is important. Under upper separate convexity, {\it lower mixture continuity $\Leftrightarrow$ lower separate mixture continuity $\Leftrightarrow$ upper separate Archimedeanity}; while under lower separate convexity, {\it upper mixture continuity $\Leftrightarrow$ upper separate mixture continuity $\Leftrightarrow$ lower separate Archimedeanity}. Thus upper separate convexity is paired with lower mixture-continuity, and lower separate convexity is paired with upper mixture-continuity. This is the same directional feature as in Shafer's open-graph argument: convexity of upper sections is used to obtain openness of strict upper sections. Second, in part (b), when both upper and lower separate convexity hold, the directional distinction disappears and the full chain of equivalences is: {\it mixture continuity $\Leftrightarrow$ separate mixture continuity $\Leftrightarrow$ Archimedeanity $\Leftrightarrow$ separate Archimedeanity $\Leftrightarrow$ weak-Wold continuity $\Leftrightarrow$ IVP $\Leftrightarrow$ separate IVP}. Third, in proving the equivalence between separate Archimedeanity and separate mixture continuity, separate convexity is required only in the backward direction. The equivalence between separate mixture continuity and mixture continuity also uses the same directional pairing: lower separate mixture-continuity is upgraded to lower mixture-continuity under upper separate convexity, while upper separate mixture-continuity is upgraded to upper mixture-continuity under lower separate convexity. Theorem \ref{thm_equiv_pref_mixture} contributes to this literature by showing that these relationships are preserved when the postulates are imposed only along coordinate sections.\fn{Further, note that when each factor $X_i$ is a subset of the real line, the convexity assumption in part (b) of the theorem is equivalent to the following monotonicity property:  $\succsim$ is {\it separately monotone}  if for all $i=1, \ldots, n$ and all $z \in X$, 
 $(x_i,z_{-i}) \succsim (y_i,z_{-i})$ for all $(x_i,z_{-i}),  (y_i,z_{-i}) \in X$ such that $x_i > y_i$,  or  $(x_i,z_{-i}) \precsim (y_i,z_{-i})$ for all 
$(x_i,z_{-i}),  (y_i,z_{-i}) \in X$ with $x_i > y_i$. Moreover, as noted in  \cite{gku23td}, a transitive binary relation $\succsim$ is separately monotone if and only if it is monotone. Some of the relationships among these scalar continuity concepts are partially studied in the literature; see for example \cite{du11}, \cite{ks15} and \cite{gku19}. Theorem \ref{thm_equiv_pref_mixture} contributes to this literature by  studying these relationships when the postulates are imposed only along coordinate sections.} 
}
\label{rmk_s_convexity}
\rmkk

\subsection{Preferences on Euclidean Spaces}
\begin{figure}[h!]
\begin{center}
\includegraphics[width=1\textwidth
]{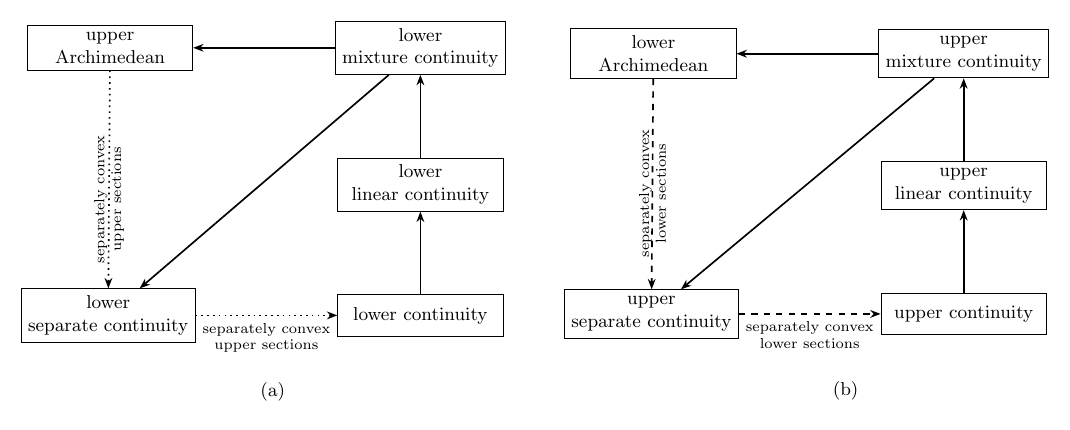}
\end{center} 
\vspace{-15pt} 
\caption{Equivalence among continuity postulates for complete and transitive preferences on convex subsets of $\mathbb R^n$. Panel (a) depicts the relationships obtained under separately convex upper sections, with the crossed implication to the lower-continuity family; panel (b) depicts the symmetric relationships obtained under separately convex lower sections, with the crossed implication to the upper-continuity family.}
\label{fig_relation_partial}
\end{figure}

\begin{figure}[h!]
\begin{center}
\includegraphics[width=0.95\textwidth
]{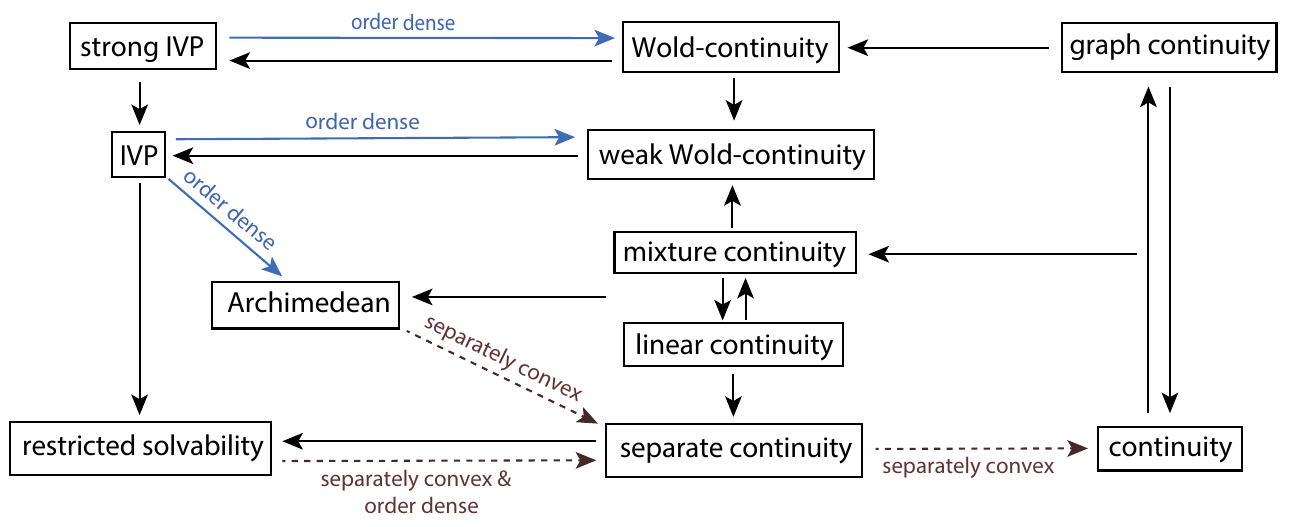}
\end{center} 
\vspace{-15pt} 
\caption{Equivalence relationships among continuity postulates for complete and transitive preferences on convex subsets of $\mathbb R^n$ under upper and lower separate convexity and order denseness.}
\label{fig_relation}
\end{figure}

The continuity postulates for mixture sets apply to convex subsets of topological vector spaces, as any convex set is a mixture set. In this subsection, we define additional commonly used continuity postulates for preferences on a convex subset of a Euclidean space, and present two further equivalence theorems: the first upgrades separate section-continuity to ordinary section-continuity, and the second links graph continuity to restricted solvability. Both are stated for coordinate Euclidean domains; Example \ref{ex_bilinear_pref} in Section \ref{sec_proofs} shows that this finite-dimensional restriction is essential and cannot be removed. Let $\succsim$ be a binary relation defined on a non-empty convex subset $X\subseteq \Re^n$, where $I=\{1,\ldots,n\}$ is the set of indices. The relation $\succsim$ is {\it graph continuous} if its graph is a closed subset of $X \times X$. It is {\it upper continuous} if it has closed upper sections, {\it lower continuous} if it has closed lower sections, and {\it continuous} if it has both closed upper and lower sections.\fn{The graph and section continuity postulates are commonly used in mathematical economics and theoretical economics. The linear and separate continuity postulates are analogous to their counterparts of functions; see for example \cite{cm16} and \cite{uk22jme} for details.  
There are nested relationships among these continuity postulates that are illustrated in Figures \ref{fig_relation_partial} and \ref{fig_relation}, and proved in Section \ref{sec_proofs}.   
}   

We define two continuity postulates that are motivated by linear and separate continuity of functions. The relation $\succsim$ is {\it upper (lower) linearly continuous} if the restriction of the upper (lower) sections of $\succsim$ to any straight line $L$ in $X$ is closed in $L$, and {\it linearly continuous} if $\succsim$ is upper and lower linearly continuous; and it is 
{\it upper (lower) separately continuous} if for any $i=1,\ldots, n$ and any $x\in X$, the restriction of the upper (lower) sections of $\succsim$ to $X_{i,x}$, the $i$-th section of $X$ at $x$, is closed in $X_{i,x}$, and {\it separately continuous} if $\succsim$ is upper and lower separately continuous. 

Finally, we define {\it solvability}-type continuity properties. The relation  $\succsim$ is  
{\it Wold-continuous} if it is order-dense and for all $x,y,z\in X$ with $x\succ y\succ z$ and all   curves $C_{xz}$ connecting $x$ and $z$,   there exists $c\in C_{xz}$ such that $c\sim y$, where a {\it curve} on $X$ is the image of a continuous injective function $m:[0,1]\ra X$; and it  (has) the {\it strong intermediate value property (strong IVP)} if for all $x,y,z\in X$ with $x\succsim y\succsim z$ and all   curves $C_{xz}$ connecting $x$ and $z$,   there exists $c\in C_{xz}$ such that $c\sim y$.

As a convex set is a mixture set, the equivalences shown in Theorem \ref{thm_equiv_pref_mixture} apply to convex Euclidean domains. Our second theorem establishes an additional equivalence result that complements Theorem \ref{thm_equiv_pref_mixture} and contributes to the literature\fn{We briefly discuss this literature in the first paragraph of Section 5.} by showing that, under separate convexity, separate section-continuity upgrades to ordinary section-continuity. The directional pairing is crossed: convexity of upper sections opens strict upper sections and therefore closes weak lower sections, while convexity of lower sections gives the symmetric conclusion.
Before presenting it, we mention a property and introduce a weak form of separate convexity that are required for its validity.

\medskip

\noindent {\bf Property A:}   
\textit{Let $Y\subseteq\mathbb R^n$ be a convex set, where $Y$ is either open or of the form $Y=\prod_{i=1}^nY_i$ with $Y_i\subseteq\mathbb R$ for all $i\in I=\{1,\ldots,n\}$.}

\customlabel{PrA}{{\bf A}}
\medskip

\nt The role of property {\bf \ref{PrA}} is geometric: the product (or open) structure guarantees that the coordinate-wise ``boxes'' assembled in the proofs remain inside the domain, and that the coordinate sections through nearby points are locally uniform. The two remarks following the proof of Theorem \ref{thm_separate_main} in the Section 6 make this precise, and Example \ref{exm_yamazaki} shows that the property cannot be dropped.
\medskip

\nt For \(J\subseteq I=\{1,\ldots,n\}\), we say that the upper sections of
\(\succsim\) are {\it separately convex in the coordinates \(J\)} if, for every
\(x,y\in X\) and every \(j\in J\), $A_{\succsim}(x)\cap X_{j,y}$ 
is convex. The lower sections are {\it separately convex in the coordinates \(J\)} if \(A_{\precsim}(x)\cap X_{j,y}\) is convex for every \(x,y\in X\) and every
\(j\in J\). We say that the upper, respectively lower, sections are {\it separately
convex in \(n-1\) indices} if this property holds for some \(J\subseteq I\)
with \(|J|=n-1\).

\thm  
 Let $\succsim$  be a complete and transitive binary relation on a non-empty convex  set  $X\subseteq \Re^n$  with property  {\bf \ref{PrA}}.   
If the upper sections of \(\succsim\) are separately convex in \(n-1\) indices, then \(\succsim\) is lower continuous if and only if it is lower separately continuous.  
Symmetrically, if the lower sections of \(\succsim\) are separately convex in \(n-1\) indices, then \(\succsim\) is upper continuous if and only if it is upper separately continuous.
\label{thm_equiv_pref}
\thmm

Our third result presents equivalences between the continuity concepts in Euclidean spaces under separate convexity. It not only combines the equivalences presented in the first two theorems above, but also links the restricted solvability concept to other continuity postulates that are used in mathematical economics and decision theory.

\thm  Let  $\succsim$  be a complete and transitive binary relation on a non-empty and convex  set  $X\subseteq \mathbb R^n$ with property  {\bf \ref{PrA}} where $\{1,\ldots,n\}$ is the set of indices. If $\succsim$ is order dense, and both upper and lower  separately convex, then $\succsim$ is graph continuous if and only if it is restricted solvable.
\label{thm_equiv_pref_finite}
\thmm

\rmk \nf{A direct comparison can be made between our Theorem \ref{thm_equiv_pref_finite} with Theorem 3 in \cite{uk22jme} and Theorems 15 and 16 in \cite{gku22games}, wherein the authors assume full convexity or monotonicity. In contrast, our theorem weakens the convexity assumption to its two-sided separate form and dispenses with the monotonicity assumption. Since global upper convexity neither implies nor is implied by two-sided separate convexity, the two sets of hypotheses are non-nested; the theorem thereby covers preferences, such as those in Example \ref{ex_hpz}, that lie outside the reach of the earlier results. For functions, the corresponding relationship between separate and joint continuity is established by \cite{yo10qjpam} and \cite{kd69amm} under a weak monotonicity assumption; Theorem \ref{thm_equiv_pref} contributes to this literature by working with preferences on the domains delineated by property {\bf \ref{PrA}}, and under an assumption that is weaker than monotonicity. 
}
\rmkk

\rmk
\nf
Theorems \ref{thm_equiv_pref} and \ref{thm_equiv_pref_finite} are  stated for coordinate Euclidean domains, and the restriction is essential rather than expositional. \citet{nt95} show that mixture continuity does not imply the usual continuity in infinite-dimensional spaces under a strong convexity assumption; Example \ref{ex_bilinear_pref} in Section \ref{sec_proofs} sharpens this boundary for the product structure studied here, since even the conjunction of two-sided separate convexity, two-sided separate continuity and full mixture continuity does not deliver continuity when the coordinate factors are infinite-dimensional. On mixture sets, by contrast, Theorem \ref{thm_equiv_pref_mixture} is free of any dimensionality restriction, that is, the finite-dimensional structure is needed exactly where separate postulates are upgraded to joint ones.
\rmkk

Figures \ref{fig_relation_partial} and \ref{fig_relation} summarize the Euclidean relationships among the continuity postulates. Figure \ref{fig_relation_partial} depicts the equivalences obtained when separate convexity is imposed only on upper or lower sections. Figure \ref{fig_relation} depicts the full equivalence class obtained in Theorem \ref{thm_equiv_pref_finite} when both upper and lower separate convexity, together with order denseness, are imposed.

Example \ref{ex_separate_function} below shows that just upper separate convexity is not enough to obtain the equivalences as seen in Theorem 3. 

\ex{\nf 
This example illustrates that just upper separate convexity is not enough to obtain equivalence between Archimedeanity and mixture continuity and between separate continuity and continuity as well as  restricted solvability and continuity. It is motivated by the classic counterexample of  \cite{gp84}. 
Let $X=\Re^2$ and  $f:X\ra \Re$ defined as follows:
\[
f(x_1,x_2)=
\begin{cases}
\displaystyle
\frac{2x_1x_2}{x_1^2+x_2^2}+\min\{x_1,x_2\},
& \text{if } (x_1,x_2)\in\mathbb{R}_+^2\setminus\{(0,0)\},\\[1.2ex]
0,
& \text{otherwise.}
\end{cases}
\] 
Let $\succsim$ be the binary relation on $X$ induced by $f$, that is $x\succsim y$ if and only if $f(x)\geq f(y)$. In this case, it is easy to show that $\succsim$ is upper separately convex and satisfies upper Archimedeanity, separate Archimedeanity, separate continuity, and restricted solvability. However, $\succsim$ fails lower Archimedeanity and upper mixture continuity along the $45^\circ$ line, hence also fails continuity. The proofs of these properties are presented   in Section \ref{sec_proofs} for completeness. 
}\label{ex_separate_function}
\exx

We end this section with an example  showing that separate convexity is equivalent to separate
$\Psi$-convexity in the sense of \cite{rr19}, where $\Psi$ is restricted to coordinate
directions. This motivates our notion within the \cite{rr19} framework and shows its adaptability to an environment of substantive economic interest.

\ex 
{\nf
Let $X\subseteq \mathbb{R}^n$ be a closed and convex set. Each non-zero vector $v\neq 0$
defines an algebraic linear ordering by $x \geq_v y$ if $v \cdot x \geq v \cdot y$, and
$\Psi := \{v\in \mathbb{R}^{n}\mid v\neq 0\}$ denotes the set of all algebraic linear
orderings. Following \cite{rr19}, a complete and transitive preference relation $\succsim$
on $X$ is $\Psi$-convex if for every $a, b \in X,$ the following condition holds: if for
every $\geq_k \in \Psi$ there is a $y_k$ such that $b \geq_k y_k$ and $y_k \succsim a$,
then $b \succsim a$.
 
For separate $\Psi_i$-convexity, let $\Psi_i := \{\lambda e_i : \lambda \neq 0\}$ denote the
set of algebraic orderings defined by the $i$-th coordinate direction, where $e_i$ is the
$i$-th standard basis vector. A preference relation $\succsim$ is $\Psi_i$-convex if for
every $a, b \in X$ the following condition holds: if for every $\geq_k \in \Psi_i$ there
is a $y_k = (t, b_{-i})$ for some $t$, such that $(b_i, b_{-i}) \geq_k y_k$ and
$y_k \succsim (a_i, b_{-i})$, then $(b_i, b_{-i}) \succsim (a_i, b_{-i})$. The relation
$\succsim$ is \textit{separately $\Psi$-convex} if and only if it is $\Psi_i$-convex for
all $i = 1, \ldots, n$.
 
Analogous to \cite{rr19}, we prove that for a continuous preference relation on $X$,
separate convexity is equivalent to separate $\Psi$-convexity. The proof is presented in Section \ref{sec_proofs} and follows along
the lines of \cite{rr19}, with the separating hyperplane theorem now applied to the $i$-th coordinate section. 
} \label{separate-rr}
 \exx

\section{Applications: Axiomatization of Utility in $n$-person Games} \label{sec_ordinal_G}
We now proceed to apply the above theorems to provide new formulations for representation theorems in $n$-person games, both for ordinal and cardinal utility representations. The three applications of this section share a single logic: in each, a joint postulate of the classical axiomatization - Archimedeanity in the cardinal setting of Subsection 3.1, joint continuity in the ordinal setting of Subsection 3.2, and completeness together with full mixture-continuity in the subjective expected utility setting of Subsection 3.3 - is replaced by its separate, coordinate-wise counterpart, and the equivalence theorems of Section \ref{sec_ordered} recover the joint postulate.

\subsection{Multilinear Utility Representation}

We present a result on the axiomatization of multilinear utility in strategic form games that contributes to the result presented in  \cite{fr78} and \cite{fi82}. A utility function \( u \) is cardinal, meaning that it is unique up to a positive affine transformation.\fn{Our sense of cardinality follows that of \citet{wz99}.} In other words, if \( u \) represents preferences in a cardinal sense, then any transformation of the form \( u' = au + b \), where \( a > 0 \) and \( b \) is a constant, will preserve the same ordinal ranking of choices while maintaining the meaningfulness of differences in utility levels. This notion of cardinality is crucial wherein the magnitude of utility differences carries interpretative significance, such as in expected utility theory. Let $n\geq 1$ be an integer, $I= \{1, \ldots, n\}$, $X_i$ be a mixture set for all $i\in I$ and $X=\prod_{i\in I}X_i$. Following \cite{fi82}, we call a real-valued function  $u$ on $X$  \textit{multilinear} if for all $x,y\in X$ and all $i\in I$ with $x_{-i}=y_{-i}$, 
\[
u(x\lambda y)=\lambda u(x)+ (1-\lambda)u(y).
\]
A preference relation $\succsim$  on $X$ is {\it separately independent} if for all $x,y,z,w\in X$, all $i,j\in I$ and all $\lambda\in (0,1)$ with $x_{-i}=z_{-i}$ and  $y_{-j}=w_{-j}$, $x\succ y \text{ and } z\sim w \text{ implies } x\lambda z \succ y\lambda w$. Moreover,   $\succsim$  is {\it separately Archimedean$^*$} if for all $x,y,z\in X$ and all $i\in I$  with $x_{-i}=z_{-i}$,  $x\succ y \succ z$  implies there exist $\lambda,\delta\in (0,1)$ such that $x\lambda z \succ y\succ x\delta z$. 

\cite{fr78} and \citet[Theorem 1, p.88]{fi82} show that completeness, transitivity, \textit{separate independence} and \textit{separate Archimedean$^*$} are necessary and sufficient for multilinear utility representation. The following result provides an alternative characterization of multilinear utility. 

\prp 
Let  $\succsim$ be a complete, transitive, and order-dense preference relation on a non-empty mixture set  $X=\prod_{i}X_i,$ $i=\{1,\ldots, n\}.$  Then  there is a multilinear function $u$ on $X$ such that, for all $x, y \in X, x \succsim y \iff u(x)\geq u(y)$  if and only if $\succsim$ is separately independent and satisfies separate IVP. In addition, such a multilinear function $u$ is unique up to a positive affine transformation. 
\label{thm_cardinal_game}
\prpp

\rmk{\nf 
In the proof of the proposition, we show that separate independence implies separate convexity, and that, under separate independence, the separate Archimedean$^*$ property is equivalent to separate Archimedeanity. Then, by Theorem \ref{thm_equiv_pref_mixture}, separate IVP in Proposition \ref{thm_cardinal_game} can be replaced with any of the following continuity postulates: mixture continuity, separate mixture continuity, Archimedean, separate Archimedean, weak Wold-continuity and IVP.  
\label{rmk_fishburn}
}\rmkk

\subsection{Separately Quasi-concave Utility Representation}

We first define a separately quasi-concave utility representation. A real-valued function $u(x_1, \ldots, x_n)$ defined on a convex subset of \(\mathbb{R}^n\) is said to be quasiconcave (quasiconvex) in \(x_i\) if, for every fixed $x_{-i} = \left(x_1, \ldots, x_{i-1}, x_{i+1}, \ldots, x_n\right),
$ the function \(u(\cdot, x_{-i})\) is quasiconcave (quasiconvex) in the variable \(x_i\).\footnote{The general reader should note an abuse of notation whereby an element \(x \in \mathbb{R}^n\) is written as \((x_i, x_{-i})\); alternatively, \((a, x_{-i})\) denotes an element of \(\mathbb{R}^n\) in which \(a \in \mathbb{R}\) is substituted in the \(i\)th position.} A utility function \(u\) is  \textit{separately} quasiconcave (\textit{separately} quasiconvex) if it is quasiconcave (quasiconvex) in each variable. Here, we \textit{axiomatize} quasiconcave and continuous payoffs in $n$-person games with Euclidean action sets.\footnote{Notice that the action sets need not be open in this case; Cartesian products suffice.} The exact equivalence uses continuity together with separate upper convexity -- under the crossed implication in Theorem \ref{thm_equiv_pref}, upper continuity may be weakened to upper separate continuity only when lower sections satisfy the corresponding separate-convexity hypothesis.

We note here that the quasi-concavity of the utility function on own actions in a game is related to separate upper convexity of the underlying preference relation. Our next result is presented in two parts. In part (a), we work with separate quasiconcavity in every coordinate and in part (b), we impose quasiconcavity only in the agent's own action. The final sentence in each part \textit{emphasises} the extra lower-section convexity needed if one wants to replace upper continuity by upper separate continuity.\fn{We can also axiomatize upper semi-continuous games (or games where preferences satisfy open lower sections property.) Moreover, we state the proposition for Euclidean action sets as in \cite{de52}: in the light of Example \ref{ex_bilinear_pref}, the separate-to-joint continuity upgrades of Theorem \ref{thm_equiv_pref} are unavailable beyond finite-dimensional domains, and on general topological vector spaces the only equivalences at our disposal are the mixture-set results of Theorem \ref{thm_equiv_pref_mixture}.} 

\prp Let \( I=\{1,\ldots, n\} \) be a finite set, and let \( X_{i} \) be a non-empty and convex subset of a Euclidean space \(\Re^{k_i}\) for each \( i\in I \). Define the Cartesian product as \( X = \prod_{i \in I} X_i \subseteq \Re^{N}\), where \(N=\sum_{i\in I}k_i\), which consists of all tuples \(x= (x_1, \dots, x_n) \) where each \( x_i \in X_i \).
\bit
\item[(a)] Let \( \succsim \) be a complete and transitive preference relation on \( X \). Then there exists a continuous and separately quasiconcave utility representation \( u: X \to \mathbb{R} \) of $\succsim$ if and only if \( \succsim \) satisfies upper separate convexity and continuity. If, in addition, the lower sections of \(\succsim\) are separately convex in \(N-1\) of the \(N\) scalar coordinates of \(\Re^{N}\) and \(X\subseteq\Re^{N}\) satisfies property {\bf \ref{PrA}}, then upper continuity in this condition can be replaced by upper separate continuity.
\item[(b)] Let \( \succsim_i \) be a complete and transitive preference relation of agent \( i \) on \( X \). Then there exists a continuous utility representation \( u_i: X \to \mathbb{R} \) of $\succsim_i$ that is quasiconcave in \( X_i \) if and only if \( \succsim_i \) is upper convex in coordinate \(i\) and continuous. If, in addition, the lower sections of \(\succsim_i\) are separately convex in \(N-1\) of the \(N\) scalar coordinates of \(\Re^{N}\) and \(X\subseteq\Re^{N}\) satisfies property {\bf \ref{PrA}}, then upper continuity in this condition can be replaced by upper separate continuity.
\eit
\label{thm_ordinal_game}
\prpp
\rmk
\nf In his paper on existence of an equilibrium in a \textit{social} system, \cite{de52} requires the payoff function to be quasiconcave.\fn{Debreu covers a more general setup of ``generalized games" which adds a constrained correspondence. See also \cite{ro65ecma}, \cite{bd04wp} and \cite{to22geb} for a variant of this generalized games setup which assumes a convex set of jointly feasible alternatives. Debreu's setup allows a separately convex jointly feasible set.} In part (b) of Proposition \ref{thm_ordinal_game},  we axiomatize the quasiconcavity of the utility function with separately upper-convex sections.
\rmkk

\subsection{Subjective Expected Utility Representation}
\label{subsec_seu}

We next present a subjective expected utility implication of the {\it hiddenness}
results developed in this branch of the literature.\fn{See \cite{ku19a} and
also the references in Footnote \ref{fn:hidden} above for this branch of the
literature.} We focus on the scalar Anscombe--Aumann environment. Equivalently,
this is the binary-consequence Anscombe--Aumann case, where each objective
lottery can be identified with a number in an interval. This scalar formulation
makes clear how the separate continuity assumptions recover completeness and
full mixture-continuity.

Let \(S=\{1,\ldots,n\}\) be a finite set of states, and let
\(Y\subset\Re\) be a compact non-degenerate interval. Let \(X=Y^S\) be the set
of acts. Thus, for \(f\in X\), \(f_s\in Y\) is the scalar outcome assigned to
state \(s\). Mixtures are defined coordinate-wise. If \(p\in Y\), then
\((p_s,f_{-s})\) denotes the act obtained from \(f\) by replacing only its
\(s\)th coordinate with \(p\), and
\[
        X_s(f_{-s})=\{(p_s,f_{-s})\mid p\in Y\}
\]
denotes the {\it coordinate section} at \(f_{-s}\).

A state \(s\) is {\it null} if \((p_s,f_{-s})\sim(q_s,f_{-s})\) for all
\(p,q\in Y\) and all \(f\in X\). We now define some properties for a binary
relation \(\succsim\) on \(X=Y^S\). A binary relation \(\succsim\) is
\emph{state-independent} if, for any two non-null states \(s,t\in S\), all
\(p,q\in Y\), and all \(f_{-s}\in Y^{S\setminus\{s\}}\) and
\(g_{-t}\in Y^{S\setminus\{t\}}\),
\[
        (p_s,f_{-s})\succsim(q_s,f_{-s})
        \quad\Longleftrightarrow\quad
        (p_t,g_{-t})\succsim(q_t,g_{-t});
\]
it is {\it independent} if, for all \(f,g,h\in X\) and all
\(\lambda\in(0,1)\), \(f\succsim g\) implies
\(f\lambda h\succsim g\lambda h\); it is \emph{weakly separately independent}
if, for all \(f,g,h,k\in X\), all \(s,t\in S\), and all
\(\lambda\in(0,1)\), whenever \(f_{-s}=h_{-s}\), \(g_{-t}=k_{-t}\),
\(f\sim g\), and \(h\sim k\), we have $f\lambda h\sim g\lambda k$;
it is {\it non-trivial} if there exist \(f,g\in X\) such that \(f\succ g\);
and finally, it satisfies \emph{coordinate-wise comparability} if, for every
\(s\in S\) and every \(f_{-s}\in Y^{S\setminus\{s\}}\), there exist distinct
\(p,q\in Y\) such that
$(p_s,f_{-s})\succsim(q_s,f_{-s})\quad\text{or}\quad (q_s,f_{-s})\succsim(p_s,f_{-s}).$

\noindent Weak separate independence is used first to propagate indifference on
coordinate sections. Once completeness has been recovered, it matches
Fishburn's multilinearity independence axiom for products of mixture sets; see
\citet[Chapter 7, Theorem 1]{fi82}. Coordinate-wise comparability requires that,
on every coordinate section, some two distinct acts be comparable, either by
strict preference or by indifference.

The environment is that of \cite{aa63}, in its scalar form. The following result is in the flavor of \cite{sc71} wherein completeness, a
multi-linear representation, and full mixture-continuity are hidden under
separate scalar continuity, weak separate independence, and coordinate-wise
comparability.

\prp
Let \(X=Y^S\), where \(S\) is finite and \(Y\subset\Re\) is a compact
non-degenerate interval. Let \(\succsim\) be a reflexive and transitive binary
relation on \(X\) that satisfies coordinate-wise comparability. Suppose that
\(\succsim\) is weakly separately independent, upper and lower separately
mixture-continuous, and upper and lower separately Archimedean. Then
\(\succsim\) is complete, mixture-continuous and admits a continuous representation \(V:X\to\Re\) that is affine in each coordinate separately.\fn{We note that the multilinear conclusion should be distinguished from additive separability.
Weak separate independence permits interaction terms across states. For example,
on \([0,1]^2\), a representation of the form 
$V(f_1,f_2)=f_1+f_2+\gamma f_1f_2$ is affine in each coordinate separately, but it is not additively separable
when \(\gamma\neq 0\). Additive representations require the usual additive
cancellation or no-interaction axioms; see \citet[Chapter 6]{fi82}. Under such
an additional condition, the multi-affine representation reduces to an additive
one. This additive route is not needed below. \label{fn_additive_separability}}   
\label{prop_hidden_complete_scalar}
\prpp

The subjective expected utility implication is now immediate. The independence
axiom implies weak separate independence. Hence the preceding proposition
recovers completeness and full mixture-continuity. Full independence then gives
the affine representation on the whole mixture space, and state independence
converts this affine representation into a subjective expected utility
representation with a common utility index.

\cor
Let \(S=\{1,\ldots,n\}\) be finite, let \(Y\subset\Re\) be a compact
non-degenerate interval, and let \(X=Y^S\). Let \(\succsim\) be a reflexive and
transitive binary relation on \(X\). Suppose that \(\succsim\) is independent,
upper and lower separately mixture-continuous, and upper and lower separately
Archimedean. Suppose also that \(\succsim\) satisfies coordinate-wise
comparability and is state-independent. Then \(\succsim\) is complete and
mixture-continuous, and there exist a probability vector \(\pi\in\Delta(S)\)
and an affine function \(u:Y\to\Re\) such that, for all \(f,g\in X\),  
$       f\succsim g$ 
        if and only if  
        $\sum_{s\in S}\pi_s u(f_s)
        \geq
        \sum_{s\in S}\pi_s u(g_s).
$ 
If, in addition, \(\succsim\) is non-trivial, then \(u\) can be chosen
non-constant and \(\pi_s=0\) whenever \(s\) is null.
\label{cor_behavioral_seu}
\corr

\rmk {\nf Two comments are in order. First, coordinate-wise comparability cannot be dropped from Proposition \ref{prop_hidden_complete_scalar}: the equality relation on $X$, under which $f\succsim g$ if and only if $f=g$, is reflexive, transitive, weakly separately independent, upper and lower separately mixture-continuous, and vacuously separately Archimedean, yet it is incomplete. Second, the displayed assumptions are also implied by the conclusions: a continuous representation that is affine in each coordinate separately delivers weak separate independence, the separate continuity and Archimedean postulates, and coordinate-wise comparability; and the subjective expected utility form delivers state independence. Proposition \ref{prop_hidden_complete_scalar} and Corollary \ref{cor_behavioral_seu} can therefore be read as characterizations, and we state them in the sufficiency direction for economy of exposition.
}\rmkk

The scalar formulation corresponds to the binary-consequence Anscombe--Aumann
environment. We expect the hiddenness argument to extend to coordinates given by a full simplex \(\Delta(C)\) with \(|C|>2\), with coordinate-wise comparability strengthened to the requirement that each coordinate section be either non-trivial or fully indifferent, and with Fishburn's multilinear representation theorem applied to the product of the coordinate mixture sets. Since we do not supply a proof, we emphasise this extension as a conjecture and leave its formal treatment to future work.

We end this section by noting that we keep transitivity as a primitive
assumption in this subsection. A natural step forward is to ask whether transitivity
can {\it also} be weakened in this product Anscombe--Aumann setting under
separate continuity and independence assumptions. This would connect the
present SEU implication more directly to the
Eilenberg--Sonnenschein--Schmeidler program extended in the authors' recent
work: Schmeidler's hiddenness argument recovers completeness, while
Sonnenschein's analysis points to weaker consistency requirements than full
transitivity. Such an extension would clarify whether transitivity, like
completeness and full mixture-continuity, is also hidden in suitable separate
continuity and independence postulates.

\section{Continuity of Separately Convex Correspondences}\label{sec_correspondence}

We finally turn to the continuity of possibly non-ordered preferences in the general context of correspondences and develop a notion of \textit{separate} convexity for correspondences and apply it to present two theorems on the characterization of the open graph property that has been commonly used to define the (strong) continuity of a possibly non-ordered preference relation in mathematical economics. Our results consolidate and generalize the existing results on the relationship between section and graph continuity of correspondences presented in \cite{sc69}, \cite{sh74} and \cite{bpr76}.

A {\it correspondence} from a non-empty set $X$ into a set $Y$ is a mapping $F:X\tra Y$ that assigns every $x\in X$ to a subset of $Y$. Define the {\it graph of $F$} as $\operatorname{gr}F=\{(x,y)\in X\times Y~|~y\in F(x)\}$. For every $x\in X$, $F(x)$ denotes the {\it upper section} of $F$ at $x$, and for every $y\in Y$, $F^{-1}(y)=\{x\in X~|~y\in F(x)\}$ denotes the {\it lower section} of $F$ at $y.$

In the results below, $X$ lies in a topological space and $Y\subseteq\Re^n$ is convex. Let $I=\{1, \ldots, n\}.$ For each $y\in Y$ and $i\in I$, define the $i$-th coordinate section of $Y$ at $y$ by
\[
        L_{i,y}=\{z\in Y\mid z_{-i}=y_{-i}\}.
\]
A correspondence $F:X\tra Y$ is  \textit{separately convex} if for all $x\in X$, all $y\in Y$ and all $i\in I$, $F(x)\cap L_{i,y}$ is convex.\fn{A straight line in a set $Y$ in a vector space is defined as the intersection of $Y$ with a one-dimensional affine subset of the affine hull of $Y$. A subset $X$ of a real vector space is called {\it affine} if for all $x,y\in X$ and $\lambda\in \Re,$ $\lambda x+(1-\lambda)y\in X.$} A correspondence $F$ has \textit{open} sections if it has both open upper and lower sections. It has {\it separately open upper sections} if for every $x\in X$, every $i\in I$ and every $y\in Y$, $L_{i,y}\cap F(x)$ is open in $L_{i,y}$; and it has {\it linearly open upper sections} if for every $x\in X$ and every straight line $L$ in $Y$, $L\cap F(x)$ is open in $L$. Given the above definitions, the following nested relationships hold among the continuity postulates for a correspondence, and each implication is strict in general -- see Examples \ref{exm_yamazaki}, \ref{ex_l_open_not_open} and \ref{ex_s_open_not_open}, respectively, below:
\begin{align}
\text{open graph} \Longrightarrow \text{open  sections} \Longrightarrow \text{linearly open sections} \Longrightarrow \text{separately open  sections.} 
\label{eq_nested}
\end{align}

\nt \cite{sh74} and \cite{bpr76} show that for a correspondence $F:\Re_+^n\tra \Re_+^n$ with convex upper sections, the converse of the first relationship also holds, that is, $F$ has an open graph if and only if it has open upper and lower sections.\fn{See Corollary \ref{thm_shafer} below for the statement of this result.} \cite{sc69} proves this relationship under the monotonicity assumption. We extend their results by allowing more general domains and ranges, by adding two other relationships concerning the linear and separate continuity concepts, which are weaker than graph and section continuity, and  by showing that the converse relationships hold under separate convexity, which is weaker than their monotonicity and convexity assumptions.     

\subsection{Counterexample to an Open Problem}

Before we present our results, we note an open problem stated by \cite{bpr76}: ``the generalization of this result to an arbitrary convex set in $\mathbb{R}^n$ remains an open problem." Example \ref{exm_yamazaki} demonstrates that the result of \cite{sh74} and \cite{bpr76} cannot be extended to a setting where the range (and the domain in their setting) $Y$ of the correspondence is an arbitrary convex set in $\mathbb{R}^n$. Thus, Example \ref{exm_yamazaki} resolves this open problem in the negative. It also provides a counterexample to Corollary 3 of \cite{ya83b}.\fn{\citet[Proposition 2]{ya83b} identifies the following property for a subspace of a topological vector space: a subset $X$ of a topological vector space is {\it locally finite} if for each $z\in X$ there exists a finite collection of points ${x^1, \ldots, x^k}$ in $X$ such that the convex hull of ${x^1, \ldots, x^k}$ is a neighborhood of $z$. Yamazaki's Proposition 2 provides a generalization of Shafer's result to sets satisfying this property. (He also works with a weakening of convex domains by restricting the preferences to the restricted domain.) The local finiteness property has its roots in the proof of \cites{sh74} and plays a crucial role. 
\label{fn_yamazaki}}

\begin{figure}[htb]
\begin{center}
\begin{tikzpicture}[scale=1.12,>=stealth,line cap=round,line join=round,
  every node/.style={font=\small}]


\coordinate (O) at (0,0);
\coordinate (Yone) at (4.15,0);
\coordinate (Ytwo) at (0,3.25);
\coordinate (Xone) at (-3.10,-1.78);
\coordinate (XoneYone) at (1.05,-1.78);
\coordinate (XoneYtwo) at (-3.10,1.47);
\coordinate (Mfront) at (0,0.82);

\path[fill=gray!18]
  (Yone)
  plot[domain=0:90,samples=100] ({4.15*cos(\x)},{3.25*sin(\x)})
  -- (XoneYtwo)
  plot[domain=90:0,samples=100] ({-3.10+4.15*cos(\x)},{-1.78+3.25*sin(\x)})
  -- cycle;
\path[fill=gray!18] (O) -- (Yone) -- (XoneYone) -- (Xone) -- cycle;
\path[fill=gray!18] (O) -- (Ytwo) -- (XoneYtwo) -- (Xone) -- cycle;
\path[fill=gray!18]
  (O) -- (Yone)
  plot[domain=0:90,samples=100] ({4.15*cos(\x)},{3.25*sin(\x)})
  -- cycle;
\path[fill=gray!18,draw=gray!18,line width=0.4pt]
  (O) -- (Yone) -- (Mfront) -- cycle;
  \draw[gray!18,line width=1.4pt] (Yone) -- (Mfront);

\draw[->,very thick] (O) -- ++(-3.85,-2.20) node[below left] {$x_1$};
\draw[->,very thick] (O) -- ++(4.95,0) node[right] {$y_1$};
\draw[->,very thick] (O) -- ++(0,3.95) node[above] {$y_2$};

\draw[semithick] (O) -- (Yone);
\draw[semithick] (O) -- (Ytwo);
\draw[semithick] (O) -- (Xone);
\draw[semithick] (Yone) -- (XoneYone);
\draw[semithick] (Ytwo) -- (XoneYtwo);
\draw[semithick] (Xone) -- (XoneYone);
\draw[semithick] (Xone) -- (XoneYtwo);
\draw[semithick] plot[domain=0:90,samples=100] ({4.15*cos(\x)},{3.25*sin(\x)});
\draw[semithick] plot[domain=0:90,samples=100] ({-3.10+4.15*cos(\x)},{-1.78+3.25*sin(\x)});


\foreach \t in {0,0.02,0.04,0.06,0.08,0.10,0.12,0.14,0.16,0.18,
                0.20,0.22,0.24,0.26,0.28,0.30,0.32,0.34,0.36,0.38,
                0.40,0.42,0.44,0.46,0.48,0.50,0.52,0.54,0.56,0.58,
                0.60,0.62,0.64,0.66,0.68,0.70,0.72,0.74,0.76,0.78,
                0.80,0.82,0.84,0.86,0.88,0.90,0.92,0.94,0.96,0.98}
{
  \draw[fill=white,draw=black!80,line width=0.45pt]
    ({-3.10*\t+4.15*sin(90*\t)},{-1.78*\t+3.25*cos(90*\t)})
    circle (1.4pt);
}

\fill[black] (XoneYone) circle (2.55pt);

\node at (-1.05,.45) {$\operatorname{gr}P$};
\node at (0.85,1.55) {$\operatorname{gr}f$};

\node[above right] at (Ytwo) {$(0,0,1)$};
\node[below right] at (Yone) {$(0,1,0)$};
\node[left] at (Xone) {$(1,0,0)$};
\node[below right] at (O) {$(0,0,0)$};
\node[below right] at (XoneYone) {$(1,1,0)$};

\draw[->,thin] (3.15,2.42) -- (1.40,1.60);
\node[right] at (3.15,2.42) {deleted curve $(x_1,f(x_1))$, $x_1<1$};
\draw[->,thin] (2.0,-1.50) -- (XoneYone);
\node[right] at (2.0,-1.50) {included limit point};

\end{tikzpicture}

\end{center}  
\vspace{-15pt}  
\caption{A convex binary relation with open sections whose graph is not open}
\vspace{-5pt}   
\label{fig_Yamazaki}
\end{figure}

\ex{\nf 
Let $X=\{x\in \Re_+^2| x_1^2+x_2^2\leq 1\}$.  Clearly, $X$ is convex. Consider a homeomorphism $f: [0,1]\ra \{x\in \Re_+^2| x_1^2+x_2^2 = 1\}$ as illustrated in Figure \ref{fig_Yamazaki} where $f(0)=(0,1)$ and $f(1)=(1,0)$.  

For expositional purposes, we first define a correspondence $P:[0,1]\tra X$ as $P(x)=X\backslash \{f(x)\}$ for all $x<1$ and $P(1)=X$. Figure \ref{fig_Yamazaki} illustrates the graph of $P$. For all $x\in [0,1]$, since $P(x)$ excludes at most one point in $X$ from the upper boundary $\{x\in \Re_+^2| x_1^2+x_2^2 = 1\}$ of $X$, it is convex and open. Since for all $y\in X$, $P^{-1}(y)$ satisfies one of $[0,1]$, $(0,1]$ and $[0,a)\cup(a,1]$, where $a\in (0,1)$, therefore it is open in $[0,1]$.  However, $grP$ is not open in $[0,1]\times X$ as for any open neighborhood $V$ of $(1,(1,0))$, there exists $a$ close to $1$ such that $(a,f(a))\in V$ but $(a,f(a))\notin grP$.

We next extend this setting to a preference relation. Define a correspondence $F: X\tra X$   as follows: $F(1, x_2)=X$ for all $x_2\in X$ and $F(x_1, x_2) = X\backslash \{f(x_1)\}$ for all $x\in X$ with $x_1\neq 1$.  Note that $F(x)$ is convex for all $x\in X$ and is constant in the second variable. Also, $F(x)$ is open in $X$ for all $x\in X$ since its complement is closed (either an empty set or a singleton).  Furthermore,
\[
F^{-1}(y)= \begin{cases} X & \text { if $y \notin  f([0,1))$ } \\ X\backslash( \{a\}\times \{z_2\in \Re_+~|~(a,z_2)\in X\}) & \text { if } y= f(a) \text{ for some } a \in [0,1).
\end{cases}
\] 
Note that for each $y\in \{x\in \Re_+^2| x_1^2+x_2^2 = 1\}$, $y\neq (1,0)$, there exists a unique such $a\in [0,1)$ such that $y=f(a)$. Hence, $F$ has open lower sections. 
However, \(F\) does not have an open graph. Indeed,
\(((1,0),(1,0))\in \operatorname{gr} F\), since \(F(1,0)=X\).
For \(a<1\), we have \(f(a)\notin F(a,0)\), so $((a,0),f(a))\notin \operatorname{gr} F.$
As \(a\to 1\), \(((a,0),f(a))\to ((1,0),(1,0))\). Hence every
neighborhood of \(((1,0),(1,0))\) meets the complement of
\(\operatorname{gr} F\), and \(\operatorname{gr} F\) is not open. 
%
%
In this example,  setting $P=\text{\nf gr}F$ implies that $P\subseteq X\times X$ is a binary relation on the convex set $X$ with open sections and convex values. However, $P$ is not open in $X\times X$. 
This example also shows why the domain restrictions in Theorems \ref{thm_separate_main} and \ref{thm_linear_main} cannot simply be dropped: the quarter disk is neither open nor a product set, so it fails property {\bf \ref{PrA}}, and it is not a polyhedron, so it fails property {\bf \ref{PrB}}.
}
\label{exm_yamazaki}
\exx  
 

 In the following subsection, we provide a two-fold generalization of \citet[Lemma, p. 914]{sh74} and   \citet[Theorem 3]{bpr76}: (i) extend the domain and the range of the correspondence and (ii) weaken the continuity and convexity assumptions on the correspondence. Our results highlight the trade-off between these two directions, in particular the trade-off between the convexity assumption and the structure of the range $Y$ of the correspondence.

\subsection{Open Graph Property under Separate Convexity}

Our first theorem in this section is on the graph continuity of a correspondence that provides a partial converse relationship among the continuity postulates listed in (\ref{eq_nested}).  

\thm
Let $X$ be a topological space, $Y\subseteq\mathbb R^n$ with property {\bf \ref{PrA}}, and $F:X\tra Y$ have separately convex upper sections.  
Then $F$ has open graph if and only if it has open lower sections and separately open upper sections.  
\label{thm_separate_main}
\thmm 

\nt  Notice that the separate convexity assumption in Theorem \ref{thm_separate_main} is weaker than assuming that $F$ has convex values. Further, notice that if we define a correspondence $F:Y\tra X$ from $Y$ into $X$ and define separately open lower sections analogous to separately open upper sections, then we obtain the following result that is symmetric to Theorem \ref{thm_separate_main}:  {\it if the lower sections of $F$ are separately convex, then   $F$ has open graph if and only if it has open upper sections and separately open lower sections.} In the next proposition, we establish an equivalence between the continuity postulates of the correspondence, given separately convex upper (or lower) sections. In fact, separate convexity is needed in only $n-1$ of the coordinates: for $i\in I$, say that $F$ has \textit{separately convex upper sections in the coordinates other than $i$} if $F(x)\cap L_{j,y}$ is convex for all $x\in X$, all $y\in Y$ and all $j\in I\setminus\{i\}$. We state the proposition in this weaker form because it is exactly the form invoked in the proof of Theorem \ref{thm_equiv_pref}, which assumes separate convexity in $n-1$ indices only.
\prp
Let $X$ be a topological space and  $Y$ a  convex subset of $\Re^n$ that satisfies property {\bf \ref{PrA}},  and $F: X \tra Y$ be a correspondence. 
If $F$ has separately convex upper sections in the coordinates other than $i$, for some $i\in I$, then the following are equivalent:  $F$ has (i) open upper sections, (ii)  linearly open upper  sections, (iii) separately open upper  sections. In particular, the equivalence holds whenever $F$ has separately convex upper sections.
\label{thm_additional_directional_separate}
\prpp

\nt Just as previously, the additional relationships among the {\it lower} continuity postulates are obtained by suitably adjusting the domains and the ranges of the correspondences and replacing ``upper" in Proposition \ref{thm_additional_directional_separate} by ``lower." Combining Proposition \ref{thm_additional_directional_separate} and Theorem \ref{thm_separate_main}, the  two results imply the following converse relationship among the continuity postulates listed in (\ref{eq_nested}).  
 
\cor
Let $X\subseteq \Re^n$ satisfy property  {\bf  \ref{PrA}},  and $F: X\tra X$ has separately convex upper sections and separately convex lower sections.  Then $F$ has an open graph if and only if it has separately open lower sections and separately open upper sections.  
\label{thm_convex_concave}
\corr

Examples \ref{ex_separate_S1} and \ref{ex_s_open_not_open}, together with Example \ref{ex_bilinear_pref} in Section \ref{sec_proofs}, illustrate the failure of the hypotheses in Theorem \ref{thm_separate_main}. In Example \ref{ex_separate_S1}, property {\bf \ref{PrA}} fails; in Example \ref{ex_s_open_not_open}, the separate convexity assumption fails; and in Example \ref{ex_bilinear_pref}, the range is a product of infinite-dimensional factors, therefore the Euclidean structure required by property {\bf \ref{PrA}} fails.
 
\ex
{\nf 
Let $X=\{x\in [0,1]^2|~x_2\geq x_1\}$,  $A=\{x\in [0,1]^2|~x_2> x_1\}\cup\{(1,1)\}\subset X$ and $F: X\tra X$ be a correspondence  defined as $F(x)=A$ for all $x\in X$. Note that $X$ is a bounded polyhedron (polytope), but it is not a product set, hence it fails property {\bf \ref{PrA}}.  $F$ has open, hence linearly open, lower sections, and separately open upper sections since for any straight line $L$ in $X$ that is parallel to a coordinate axis, $L\cap A$ is open in $L$. Moreover, $F$ has convex upper sections. However, the graph of $F$ is not open since every neighborhood of $((1,1), (1,1))$ contains a point outside of the graph of $F$. 
}
\label{ex_separate_S1}
\exx

\ex{\nf 
Let $X=\Re^2$,  $A=\Re^2\backslash \{x\in \Re^2|x_1=x_2, x\neq 0\}$ and $F: X\tra X$ be a correspondence  defined as $F(x)=A$ for all $x\in X$.   Clearly, $X$ satisfies property {\bf \ref{PrA}}. Notice that $F$ is not  separately convex. The intersection of  $A$ and any line parallel to a coordinate axis is either the real line or a union of two open intervals. Therefore,  $F$ has separately open upper sections. However, it does not have linearly open upper sections, since the intersection of $A$ with the diagonal line is $\{0\}$, which is not open in that line. Moreover, it does not have an open graph since every open ball containing the origin in $\Re^4$ contains a point in the complement of the graph of $F$.  
}\label{ex_s_open_not_open}
\exx

\subsection{Open Graph Property under Convexity}

\noindent Our final theorem is on the continuity of correspondences defined on subspaces that satisfy the following property.

\medskip

\noindent {\bf Property B:} {\it 
For a convex set $Y \subseteq \mathbb{R}^n$, either $Y$ is open or it is a polyhedron, where a polyhedron is defined as the intersection of a finite number of closed half-spaces.
}
\customlabel{PrB}{B}
\medskip

\thm
Let $X$ be a topological space,  $Y$ a  convex subset of $\Re^n$ that satisfies property {\bf \ref{PrB}}, and $F: X \tra Y$ a correspondence such that $F(x)$ is convex for all $x\in X$.  Then, $F$ has an open graph if and only if $F$ has open lower sections and linearly open upper sections. 
\label{thm_linear_main}
\thmm

\nt Analogous to Theorem \ref{thm_separate_main},  if we define a correspondence $F:Y\tra X$ from $Y$ into $X$, then we obtain the following result that is symmetric to Theorem \ref{thm_linear_main}:  {\it if the lower sections of $F$ are convex, then   $F$ has open graph if and only if it has open upper sections and linearly open lower sections.} 
The next proposition provides additional relationships among the continuity postulates. 

\prp
Let $X$ be a topological space and $Z$ a convex subset of $\Re^n$ that satisfies property {\bf \ref{PrB}},  and $G: X \tra Z$ be a correspondence.  If $G$  has convex upper  sections, then $G$ has open upper  sections if and only if it has linearly open upper   sections. 
\label{thm_additional_directional_linear}
\prpp

\nt Note that the additional relationships among the {\it lower} continuity postulates are obtained by suitably adjusting the domains and the ranges of the correspondences and replacing ``upper" in Proposition \ref{thm_additional_directional_linear} by ``lower." These two results imply the following converse relationship among the first three continuity postulates listed in Equation \ref{eq_nested}.  
 
\cor
Let $X\subseteq \Re^n$ satisfy property  {\bf  \ref{PrB}},  and $F: X\tra X$ has convex upper sections and convex lower sections.  Then $F$ has open graph if and only if it has linearly open sections.  
\label{thm_convex_concave_linear}
\corr

\nt Note that unlike Theorem \ref{thm_separate_main},  open graph property  in Theorem  \ref{thm_linear_main} is not equivalent to open lower sections and separately open upper sections. For instance, $X$ is a  polytope in Example \ref{ex_separate_S1} and hence satisfies property {\bf B} but the correspondence $F$ fails to have linearly open upper sections, hence does not have an open graph. Also notice that properties {\bf \ref{PrA}} and {\bf \ref{PrB}} are non-nested: the half-open product $[0,1)\times [0,1]$ satisfies property {\bf \ref{PrA}} but is neither open nor a polyhedron, while a non-degenerate triangle in $\Re^2$ satisfies property {\bf \ref{PrB}} but is neither open nor a product set. Since, further, the convexity and continuity assumptions in Theorem \ref{thm_linear_main} are stronger than those in Theorem \ref{thm_separate_main}, the assumptions in these two theorems are non-nested.\fn{A direct comparison can also be made between these two theorems and the literature on the characterization of open and closed sets in $\Re^n$: \cite{ha66amm} provides a necessary and sufficient condition for a convex set to be closed, and \cite{uk23bams} for the openness of a set whose sections are convex, and of a convex set; see also \cite{fa66convex} for sets with convex sections, and \cite{gkr68pams} and \cite{er13pams} for the role that polytopes play in the continuity theory of convex functions. Theorems \ref{thm_separate_main} and \ref{thm_linear_main} contribute to this literature by providing the corresponding characterizations for a correspondence defined on an arbitrary topological space, with values in a subset of $\Re^n$ satisfying property {\bf \ref{PrA}} or property {\bf \ref{PrB}}.}

Example \ref{ex_separate_S1} above illustrates a correspondence that is separately convex and has separately open upper sections but fails to have linearly open upper sections as well as an open graph. Example \ref{ex_l_open_not_open} illustrates that if the separate convexity assumption fails, then a correspondence may have both separately open and linearly open upper sections but still fail to exhibit graph continuity.

\ex{\nf 
Let $X=\Re^2$,  $A=\Re^2\backslash \{x\in \Re^2|x_2=x_1^2, x\neq 0\}$ and $F: X\tra X$ be a correspondence  defined as $F(x)=A$ for all $x\in X$.   Clearly, $X$ satisfies properties {\bf \ref{PrA}} and {\bf \ref{PrB}} but $F$ is neither  separately convex nor has  convex upper sections. 
For any straight line $L$,  $L\cap A$  excludes at most two points of $L$. Hence, $F$ has both separately open and linearly open upper sections. However, $A$ is not open since every open ball containing $0$ contains a point in the complement of $A$, hence $F$ fails to have open upper sections and an open graph.    }\label{ex_l_open_not_open}
\exx

\noindent  The following example illustrates that the convexity assumption 
 in Theorem  \ref{thm_linear_main} is essential even for a correspondence defined on an interval in $\Re$. 

\ex{\nf 
Let $X=[0,1]$ and $F: X\tra X$ such that $F(0)=(0, 1]$ and $F(x)=\{y\in X~|~y>x, \text{ and }  y\neq (1-x)\}$ for all $x>0$. It is easy to see that $F$ has open sections but $(0,1)$ has no open neighborhood contained in the graph of $F$, hence $F$ does not have an open graph. It is clear that $F(x)$ and $F^{-1}(y)$ are not convex for all $x\in (0,0.5)$ and for all $y\in (0.5, 1)$.    
\label{exm_onedim}
}\exx

As in Theorem  \ref{thm_separate_main}, considering a binary relation as the graph of a correspondence, Theorem  \ref{thm_linear_main} provides   necessary and sufficient conditions for a binary relation on a convex subspace of $\Re^n$ to be continuous. It provides a characterization of the continuity of a binary relation, or a correspondence, by using a topological property similar to the \textit{linear continuity} postulate.  As noted in Equation \ref{eq_nested}, open upper (lower) sections property is stronger than linearly open upper (lower) sections property that is stronger than separately open upper (lower) sections property.  
Therefore, Theorems \ref{thm_separate_main} and \ref{thm_linear_main}  generalize the following result of \citet[Lemma, pg. 914]{sh74} and   \citet[Theorem 3]{bpr76} on continuity of a binary relation by considerably weakening their continuity and convexity assumptions, and allowing a more general domain on which the binary relation is defined. 

We now introduce notation for our final two corollaries. Define a {\it binary relation} \( P \) on \( X \), where \( P \subseteq X \times X \). Note that for any binary relation \( P \) on \( X \), there exists a unique correspondence \( F: X \to X \) such that \( P = \text{\nf gr} F \). The {\it upper} and {\it lower sections} of a binary relation \( P \) at \( x \in X \) are defined as \( P(x) = F(x) \) and \( P^{-1}(x) = F^{-1}(x) \), respectively.

\cor[Shafer, 1974; Bergstrom-Parks-Rader, 1976]
Let $P: \mathbb{R}^n_+\tra \mathbb{R}^n_+$ be a correspondence such that $P$ has convex upper sections (or has convex lower sections). Then, $P$ has an open graph if and only if $P$ has open sections. 
\label{thm_shafer}
\corr

Along with Theorems \ref{thm_separate_main} and \ref{thm_linear_main}, Corollary \ref{thm_convex_concave} generalizes the following result of \cite{sc69} by dropping the transitivity assumption, substantially weakening the continuity and monotonicity assumptions\fn{It is easy to observe that if a strict binary relation is strongly monotone, then it has separately convex upper and lower sections, but the converse relationship does not hold as for example the relation need not be complete. A similar relationship holds for weak monotonicity whose proof is provided in Section \ref{sec_proofs}.} and allowing a more general domain on which the binary relation is defined. 

\cor[Schmeidler, 1969]
Let  $P$ be a transitive, irreflexive and strongly monotone binary relation on $\mathbb{R}^n_+$. Then $P$ has an open graph if and only if $P$ has open sections.  
\label{thm_schmeidler}
\corr

\citet[Theorem 1]{bpr76} provides another theorem that generalizes Schmeidler's result for a transitive and order-dense binary relation on general topological spaces. Their result and our results above are non-nested -- while we  impose weaker continuity assumptions and do not assume  transitivity or order-denseness -- they do not impose any convexity assumption and allow for a more general domain. Further, note that \cite{ge15} provides a result on the relationship between sections and graph continuity of a reflexive and transitive binary relation under the additivity assumption. The results in this paper focus on separate convexity and do not impose additivity. It is easy to show that the additivity and separate convexity assumptions are non-nested for preferences defined in both of these papers. Therefore, our results and those presented in \cite{ge15} are non-nested.\fn{ Moreover, \citet{ge15} works with a transitive and reflexive preference relation and shows that section and graph continuity are equivalent under additivity. Although additivity is not a standard assumption in mathematical economics, it and related forms have been used in decision theory; see \citet{ge10,ge13}. In a parallel inquiry, \citet[Proposition 1]{dmo04} show that under independence, graph mixture continuity is equivalent to graph continuity for an incomplete preference relation, while \citet{gmms10} prove that under independence, a reflexive and transitive weak preference relation has a closed graph if and only if it is mixture continuous.}

 The results that we have presented have focused on correspondences with open graph, or on open binary relations. While there are results in the literature that show equivalence between having closed sections and being closed for a binary relation -- see for example, \cite{wa54a} and    \cite{sh74}  under completeness or transitivity assumptions -- Example \ref{exm_closed_section} illustrates that a binary relation need not be closed even if it has closed sections and convex upper sections.\fn{In a related setting, \citet[Proposition 2]{zh95jmaa} replaces open lower sections assumption in Shafer's result with lower semicontinuity of a correspondence whose values are in $Y=\Re^n$ to provide a characterization of the open graph property; see also \cite{ir12} for a recent treatment. By using the arguments in this paper, it is possible to show that the open and convex upper sections in Zhou's result can be replaced with separately open and separately continuous upper sections. }

 \ex{\nf 
Let $X=[0,1]$ and $P\subseteq X\times X$ such that $P(x)=\{x\}$ for all $x<1$ and $P(1)=\{0\}$. It is clear that $P$ has both upper and lower closed sections. Also, $P(x)$ is convex-valued since it is a singleton for all $x\in X$. However, $P$ is not closed in $X\times X$ since $(1,1)\in P^c$ has no   open neighborhood contained in $P^c$. 
\label{exm_closed_section}
}\exx

\section{Concluding Remarks} \label{sec_remarks}

To conclude, we offer three remarks that connect our results to the antecedent literature and to some possible directions for further research.

The first concerns the {\it \cite{ba58adm} maximum principle} which states that  {\it a  continuous and convex function defined on a non-empty, compact and convex subset of a locally convex Hausdorff topological vector space  reaches its maximum on some extreme points.}    
Maximizing convex functions is an important problem in optimization theory, and their properties are  interesting especially for finding maxima.\fn{See for example \cite{ch69bk}, \cite{ho75bk},   \cite{ba23wp}, and also \citet[Theorem 7.69]{ab06} for the statement of and discussion on the Bauer maximum principle.} The Bauer maximum principle has been an important result in many fields, and it has been increasingly applied to problems in economic theory recently.\fn{See also \cite{ketal21ecma},  \cite{aetal23te}, \cite{rsy25wp} and \cite{yy25wp}.} Moreover, separate convexity of functions has been used in optimization problems in many fields and it would be interesting to study whether the Bauer maximum principle generalizes to separately quasi-convex functions and preferences with separately convex lower sections.\fn{On a related but different direction, \cite{ki25wp} shows that under monotonicity and supermodularity, separate concavity implies quasi-concavity. See also \cite{daz81} for different forms of concavity and quasi-concavity that are used in economics. Also, see \cite{af83mor}, \cite{ah86ijm}, 
 \cite{getal07mmor}, \cite{fetal11jet} and  \cite{setal21siam}. }   

The second remark concerns the continuity of \textit{separately continuous} functions. 
Let $X\subseteq E:=\prod_{i=1}^n E_i$, where $E_i$ is a non-empty set for all $i=1,\ldots,n$. A function $f:X\rightarrow \Re$ is  {\it separately continuous} if for all   $i=1,\ldots,n$ and all $x\in X$, the mapping $t \mapsto f(t, x_{-i})$ is continuous,  
that is, $f$ is continuous with respect to each variable separately. Moreover, $f$  is {\it jointly continuous} if it is continuous in the usual (topological) sense.
From the early work of \cite{ca21} and  \cite{gp84}, separate continuity of functions and its relationship to the joint continuity remain a topic of active study; see \cite{cm16} for a recent survey. \cite{gp84} provide an example showing that separate continuity is strictly weaker than joint continuity;  \cite{yo10qjpam} and \cite{kd69amm} show that for separately monotone functions defined on sets in $\Re^n$, continuity is equivalent to separate continuity.\fn{See also \cite{gku22games} and \cite{uk23bams} for results relating separate continuity of functions to different continuity postulates on binary relations.}  
It is easy to show that a function is monotone in index $i$ if and only if it is quasi-convex  and quasi-concave in index $i$. Since these convexity properties of functions are analogous to separately convex lower and upper sections of preference relations,  
Theorem \ref{thm_equiv_pref} and its proof provide a preference-theoretic analogue of the separate-to-joint continuity mechanism on coordinate Euclidean domains. Our method in this paper also provides an alternative proof of the classical result of \cite{yo10qjpam} and \cite{kd69amm} in that setting. The bilinear Example \ref{ex_bilinear_pref} shows that this separate-to-joint mechanism is intrinsically finite-dimensional.

The third remark concerns the behavioral implications of separate continuity under separate convexity along the directions initiated by \cite{ei41}, \cite{so65} and \cite{sc71}, and recently extended and applied to various settings by the authors of this paper. In particular, the results above can be used to obtain subjective expected utility representations under weaker behavioral assumptions and correspondingly weaker continuity requirements.  Finally, non-convexity in preferences continues to be an important avenue for exploration.\fn{\cite{tr84} writes on the role of non-convexity in the context of market demand and large economies. See also \cite{kmu25et} on using the excess-demand approach to show the equilibrium in a Walrasian economy without the convexity assumption.} Recent work by \cite{hpz17} suggests useful implications of non-convexity for techniques in revealed preferences. We hope that by bringing separate convexity to the picture, we can stimulate applications in this area.

\section{Proofs of the Results} \label{sec_proofs}

The proofs are presented in the order in which the results are stated in the text: the theorems of Section \ref{sec_ordered} first, then the propositions of Section \ref{sec_ordinal_G}, and finally the results of Section \ref{sec_correspondence}. One dependence should be flagged at the outset: the proofs of Theorems \ref{thm_equiv_pref} and \ref{thm_equiv_pref_finite} invoke Proposition \ref{thm_additional_directional_separate}, whose proof, given below among the results of Section \ref{sec_correspondence}, rests on Theorem \ref{thm_separate_main} alone; since neither of these two results draws on any result of Section \ref{sec_ordered}, no circularity is involved. The proofs of Theorem \ref{thm_linear_main} and Proposition \ref{thm_additional_directional_linear} make essential use of two results of \citet{ro70}: the finite generation of polyhedral convex sets (Theorem 19.1) and the lemma on relative interiors stated as Lemma \ref{thm_rockafellar} below.

\cl \nf{The observations of Remark \ref{rmk_s_convexity} are established in the course of the proof below. Two further facts, not stated there, also follow. First, in part (a), full upper Archimedeanity joins the chain: under upper separate convexity, {\it lower mixture continuity $\Leftrightarrow$ lower separate mixture continuity $\Leftrightarrow$ upper Archimedeanity $\Leftrightarrow$ upper separate Archimedeanity}, with the symmetric statement under lower separate convexity. Second, for the upgrade from separate (upper/lower) mixture continuity to (upper/lower) mixture continuity, separate (lower/upper) convexity in $n-1$ indices suffices: in the coordinate-release argument below, the first release uses only separate mixture continuity, and convexity is used only to release the remaining coordinates. There are nested relationships among the continuity postulates listed above even with no separate convexity assumption.
}\label{cl_thm1_remarks}

\cll

\prf[Proof of Theorem \ref{thm_equiv_pref_mixture}]
Throughout, recall that for a complete and transitive relation, upper separate convexity of the weak upper sections is equivalent to separate convexity of the strict upper sections, and likewise for the lower sections; see \citet[Lemma 6]{uk22jme}. We use the strict form freely below. The mixture operation on the product $X=\prod_i X_i$ is coordinate-wise. We also use the derived mixture-set identities (i) $x\lambda x=x$, (ii) $(x\lambda y)\delta x=x(\delta\lambda+1-\delta)y$, (iii) $(x\lambda y)\gamma y=x(\gamma\lambda)y,$ and (iv) $(x\alpha y)\beta(x\gamma y)=x(\beta\alpha+(1-\beta)\gamma)y;$ see \citet[p.~88]{fi82}. 

\noindent The first two identities follow directly from axioms (b) and (c) of a mixture set, and the last is the standard affine-combination identity in mixture-set notation.
 
\noindent{\bf (a)} We prove the assertion under upper separate convexity -- the assertion under lower separate convexity follows verbatim by passing to the inverse relation $\precsim$. Thus we show that, under upper separate convexity, $\succsim$ is lower mixture continuous if and only if it is upper separate Archimedean, through the cycle: {\it lower mixture $\Rightarrow$ upper separate Archimedean $\Rightarrow$ lower separate mixture $\Rightarrow$ lower mixture}.
\smallskip

 \noindent\textit{Step 1 (lower mixture continuity $\Rightarrow$ upper separate Archimedeanity -- no convexity is used).}
Fix $x,y,z\in X$ and $i\in I$ with $x\succ y$ and $x_{-i}=z_{-i}$. By completeness, $\{\lambda\in[0,1]\mid x\lambda z\succ y\}=\{\lambda\in[0,1]\mid x\lambda z\precsim y\}^{c}$, and the set on the right is closed by lower mixture continuity; hence $\{\lambda\mid x\lambda z\succ y\}$ is open in $[0,1]$. It contains $\lambda=1$, since $x1z=x\succ y$. Therefore it contains a half-open interval $(\bar\lambda,1]$, and any $\lambda\in(\bar\lambda,1)$ satisfies $x\lambda z\succ y$. The argument does not use the restriction $x_{-i}=z_{-i}$; hence lower mixture continuity implies full upper Archimedeanity, and therefore upper separate Archimedeanity.
 
\smallskip
\noindent\textit{Step 2 (upper separate Archimedeanity $\Rightarrow$ lower separate mixture continuity -- uses upper separate convexity).}
Fix $x,y,z\in X$ and $i\in I$ with $x_{-i}=y_{-i}$, and let $\lambda^{m}\to\lambda$ with $z\succsim x\lambda^{m}y$ for all $m$. We show $z\succsim x\lambda y$. Suppose not, so $x\lambda y\succ z$. Since $(x\lambda y)_{-i}=x_{-i}=y_{-i}$, upper separate Archimedeanity applied to $x\lambda y\succ z$ with the coordinate-$i$ alternatives $x$ and $y$ yields $\delta,\gamma\in(0,1)$ with $(x\lambda y)\delta x\succ z$ and $(x\lambda y)\gamma y\succ z$. By the mixture identities, these are $x(\delta\lambda+1-\delta)y\succ z$ and $x(\gamma\lambda)y\succ z$, and $\gamma\lambda\leq\lambda\leq\delta\lambda+1-\delta$, where at least one inequality is strict (both are strict when $\lambda\in(0,1)$; the left one is strict when $\lambda=1$, the right one when $\lambda=0$). The strict upper section $\{\mu\in[0,1]\mid x\mu y\succ z\}$ is convex and contains $\gamma\lambda$ and $\delta\lambda+1-\delta$, hence contains the whole interval $[\gamma\lambda,\,\delta\lambda+1-\delta]$. As $\lambda$ lies in this interval and $\lambda^{m}\to\lambda$, we have $\lambda^{m}\in[\gamma\lambda,\,\delta\lambda+1-\delta]$ for all large $m$, so $x\lambda^{m}y\succ z$, contradicting $z\succsim x\lambda^{m}y$. Hence $z\succsim x\lambda y$, and $\succsim$ is lower separate mixture continuous.
 
\smallskip
\noindent\textit{Step 3 (lower separate mixture continuity $\Rightarrow$ lower mixture continuity -- uses upper separate convexity).}
Fix $x,y,z\in X$. For $\alpha=(\alpha_i)_{i\in I}\in[0,1]^{I}$ write $y\alpha z:=(y_i\alpha_iz_i)_{i\in I}$. By completeness $\{\lambda\mid y\lambda z\precsim x\}=\{\lambda\mid y\lambda z\succ x\}^{c}$, so it suffices to prove that $B:=\{\alpha\in[0,1]^{I}\mid y\alpha z\succ x\}$
is open in $[0,1]^{I}$; restricting to the diagonal $\lambda\mapsto(\lambda,\dots,\lambda)$ then shows $\{\lambda\mid y\lambda z\succ x\}$ is open, i.e. $\{\lambda\mid y\lambda z\precsim x\}$ is closed, which is lower mixture continuity.
We prove openness of $B$ by freeing one coordinate at a time. For $J\subseteq I$ and $\bar\alpha_{J}\in[0,1]^{J}$, set $B(\bar\alpha_{J}):=\{\alpha_{I\setminus J}\in[0,1]^{I\setminus J}\mid y(\alpha_{I\setminus J},\bar\alpha_{J})z\succ x\}$, and claim that $B(\bar\alpha_{J})$ is open in $[0,1]^{I\setminus J}$ for every $J$ and every $\bar\alpha_{J}$; the case $J=\emptyset$ is the assertion. When exactly one coordinate is free, openness of $B(\bar\alpha_{J})$ is precisely lower separate mixture continuity in that coordinate, so this first coordinate release uses no convexity. Convexity is used only in the subsequent releases, and therefore separate convexity in $n-1$ indices is enough for this upgrade from lower separate mixture continuity to lower mixture continuity.
 
Assume the claim holds for the fixed set $J$, pick $i\in J$, and consider the fixed set $J\setminus\{i\}$ (one more free coordinate). Fix $\bar\alpha_{J\setminus\{i\}}$ and a point $(\alpha_{I\setminus J},\alpha_i)\in B(\bar\alpha_{J\setminus\{i\}})$. By lower separate mixture continuity in coordinate $i$, there is a relative interval $V_i\ni\alpha_i$ with $y(\alpha_{I\setminus J},t,\bar\alpha_{J\setminus\{i\}})z\succ x$ for all $t\in V_i$; choose $r^{-},r^{+}\in V_i$ with $r^{-}<\alpha_i<r^{+}$ when $\alpha_i\in(0,1)$, and with the corresponding one-sided inequalities when $\alpha_i$ is an endpoint of $[0,1]$. Applying the induction hypothesis with coordinate $i$ fixed at $r^{-}$ and at $r^{+}$ gives open neighbourhoods $U^{-},U^{+}$ of $\alpha_{I\setminus J}$ such that $y(\beta_{I\setminus J},r^{\pm},\bar\alpha_{J\setminus\{i\}})z\succ x$ for $\beta_{I\setminus J}\in U^{\pm}$. Let $U=U^{-}\cap U^{+}$. For $\beta_{I\setminus J}\in U$ and $t\in[r^{-},r^{+}]$, the point $y(\beta_{I\setminus J},t,\bar\alpha_{J\setminus\{i\}})z$ lies on the coordinate-$i$ segment between $y(\beta_{I\setminus J},r^{-},\bar\alpha_{J\setminus\{i\}})z$ and $y(\beta_{I\setminus J},r^{+},\bar\alpha_{J\setminus\{i\}})z$, both strictly preferred to $x$; since the strict upper section of $x$ is separately convex, $y(\beta_{I\setminus J},t,\bar\alpha_{J\setminus\{i\}})z\succ x$. Thus $U\times[r^{-},r^{+}]$ is a neighbourhood of $(\alpha_{I\setminus J},\alpha_i)$ contained in $B(\bar\alpha_{J\setminus\{i\}})$, proving openness. Releasing the coordinates of $J$ one by one yields openness of $B=B(\emptyset)$, so $\succsim$ is lower mixture continuous.
 
\noindent Steps 1--3 close the cycle, establishing the equivalence under upper separate convexity. Passing to the inverse relation $\precsim$ gives the symmetric statement under lower separate convexity, this proves (a).
 
\medskip
\noindent{\bf (b)} Assume $\succsim$ is order dense and both upper and lower separately convex.  
Applying part (a) in both directions, upper separate convexity yields the equivalence of lower mixture continuity, lower separate mixture continuity, and upper separate Archimedeanity, while lower separate convexity yields the equivalence of upper mixture continuity, upper separate mixture continuity, and lower separate Archimedeanity. Conjoining the two, mixture continuity, separate mixture continuity, and separate Archimedeanity are equivalent.  

Moreover, lower mixture continuity implies upper Archimedeanity and upper mixture continuity implies lower Archimedeanity (completeness and \citet[Proposition 1]{gku19}), while Archimedeanity trivially implies separate Archimedeanity; hence full Archimedeanity coincides with separate Archimedeanity here as well. We now bring in the solvability postulates. We show
\textit{mixture continuity} $\Rightarrow$ \textit{weak Wold continuity} $\Rightarrow$ \textit{IVP} $\Rightarrow$  \textit{separate IVP} $\Rightarrow$ \textit{separate mixture continuity},  
which closes the loop, since separate mixture continuity is already equivalent to mixture continuity.
 \medskip 
 
\noindent\textit{(i) Mixture continuity $\Rightarrow$ weak Wold continuity.} Let $x\succ y\succ z$. By mixture continuity, the sets $\{\lambda\mid x\lambda z\succsim y\}$ and $\{\lambda\mid x\lambda z\precsim y\}$ are closed, cover $[0,1]$ by completeness, and contain $1$ and $0$ respectively; connectedness of $[0,1]$ gives a common point $\lambda$ with $x\lambda z\sim y$. Here $\lambda\ne1$ (else $x\sim y$) and $\lambda\ne0$ (else $y\sim z$), so $\lambda\in(0,1)$. As $\succsim$ is order dense, this is weak Wold continuity.

 \medskip 
 
\noindent\textit{(ii) Weak Wold continuity $\Rightarrow$ IVP $\Rightarrow$ separate IVP.} Let $x\succsim y\succsim z$. If $x\sim y$ take $\lambda=1$; if $y\sim z$ take $\lambda=0$; otherwise $x\succ y\succ z$ and weak Wold continuity gives $\lambda\in(0,1)$ with $x\lambda z\sim y$. Hence IVP holds, and IVP $\Rightarrow$ separate IVP is immediate from the definitions.

 \medskip 
 
\noindent\textit{(iii) Separate IVP $\Rightarrow$ separate mixture continuity.} We prove upper separate mixture continuity; the lower case is symmetric, with the roles of upper and lower separate convexity interchanged. Suppose upper separate mixture continuity fails: there are $x,y,z\in X$ and $i\in I$ with $y_{-i}=z_{-i}$, and $\lambda^{m}\to\lambda$ with $y\lambda^{m}z\succsim x$ for all $m$ but $x\succ y\lambda z$. If $\lambda^{m}=\lambda$ for infinitely many $m$, then $y\lambda z\succsim x$, contradicting $x\succ y\lambda z$; hence, after passing to a subsequence, we may assume $\lambda^{m}\neq\lambda$ for all $m$. Passing to a further subsequence, assume $\lambda^{m}>\lambda$ for all $m$ (the case $\lambda^{m}<\lambda$ is analogous). Relabel the subsequence so that $\lambda^{m}\in(\lambda,\lambda^{1}]$ for every $m$. By order denseness choose $x'$ with $x\succ x'\succ y\lambda z$.

Since $y\lambda^{1}z\succsim x\succ x'\succ y\lambda z$, separate IVP applied to $a:=y\lambda^{1}z$ and $b:=y\lambda z$
yields $c_i,d_i\in X_i$ and $\delta,\delta',\tilde\lambda\in[0,1]$ with
$        a_i=c_i\delta d_i,
        \qquad
        b_i=c_i\delta' d_i,
        \qquad
        w:=(c_i\tilde\lambda d_i,y_{-i})\sim x'.$
Writing $c:=(c_i,y_{-i})$ and $d:=(d_i,y_{-i})$, all the acts in play lie on the coordinate-$i$ mixture line through $c$ and $d$. Indeed, for each $m$, define $s^{m}:=(\lambda^{m}-\lambda)/(\lambda^{1}-\lambda)\in[0,1]$ and $\nu^{m}:=s^{m}\delta+(1-s^{m})\delta'$. Using $(x\alpha y)\beta(x\gamma y)=x(\beta\alpha+(1-\beta)\gamma)y$, we have
\[        y\lambda^{m}z=a\,s^{m}\,b=c\nu^{m}d,
        \qquad\text{and}\qquad
        \nu^{m}\to\delta'. \]
The parameter set $U:=\{\nu\in[0,1]\mid c\nu d\succsim x'\}$ is convex by upper separate convexity. It contains $\delta$, $\tilde\lambda$ and every $\nu^{m}$, since $a\succsim x'$, $w\sim x'$ and $y\lambda^{m}z\succsim x\succ x'$. It does not contain $\delta'$, since $b=y\lambda z\prec x'$. Since $\nu^{m}\to\delta'$ while $\delta'\notin U$, convexity forces the whole set $U$, and in particular $\tilde\lambda$, to lie strictly on the $\nu^{m}$-side of $\delta'$ for all sufficiently large $m$.

Now consider the strict lower parameter set $L:=\{\nu\in[0,1]\mid x\succ c\nu d\}$. This set is convex by lower separate convexity. It contains $\delta'$ and $\tilde\lambda$, since $x\succ b=c\delta' d$ and $x\succ x'\sim w=c\tilde\lambda d$. Hence it contains the whole parameter interval between $\delta'$ and $\tilde\lambda$. Since $\nu^{m}\to\delta'$ from the $\tilde\lambda$-side, we have $\nu^{m}\in L$ for all large $m$. Therefore, $ x\succ c\nu^{m}d=y\lambda^{m}z$
for all large $m$, contradicting $y\lambda^{m}z\succsim x$. Hence upper separate mixture continuity holds.
 
\noindent Combining (i)--(iii) with the equivalences of part (a) shows that all the postulates in (b) are equivalent.
\prff

\prf[Proof of Theorem \ref{thm_equiv_pref}] 
We prove the assertion under upper separate convexity -- the assertion under lower separate convexity follows by passing to the inverse relation $\precsim$.

Assume first that $\succsim$ is lower continuous. Then each lower section is closed, and hence its restriction to any coordinate section is closed in that section. Therefore, $\succsim$ is lower separately continuous. Conversely, assume that $\succsim$ is lower separately continuous. By completeness, for every $x\in X$ and every coordinate section $X_{i,y}$, $A_\succ(x)\cap X_{i,y}=X_{i,y}\setminus\big(A_\precsim(x)\cap X_{i,y}\big)$. Thus lower separate continuity is equivalent to separate openness of the strict upper sections $A_\succ(x)$. Moreover, for a complete and transitive relation, upper separate convexity of the weak upper sections is equivalent to separate convexity of the strict upper sections; see \citet[Lemma 6]{uk22jme}. Define the correspondence $F:X\tra X$ by $F(x)=A_\succ(x)$. Then $F$ has separately open upper sections and separately convex upper sections in the coordinates other than the exempted index. Proposition \ref{thm_additional_directional_separate}, stated precisely for this $n-1$-coordinate form and proved below independently of the present theorem, implies that $A_\succ(x)$ is open in $X$ for every $x\in X$. By completeness again, $A_\precsim(x)=X\setminus A_\succ(x)$ is closed for every $x\in X$. Hence $\succsim$ is lower continuous. This proves the asserted equivalence under upper separate convexity. Passing to the inverse relation gives the symmetric statement: under lower separate convexity in $n-1$ indices, upper continuity is equivalent to upper separate continuity.

We also note two standard topological implications used in the Euclidean relationship figures.

\smallskip
\noindent {\it Continuity $\Longrightarrow$ Wold-continuity, under order denseness:}
Let $\succsim$ be a complete, transitive, order-dense and continuous relation on $X$. Fix $x,y,z\in X$ with $x\succ y\succ z$, and let $C_{xz}$ be any curve connecting $x$ and $z$. By continuity, $A_\succsim(y)\cap C_{xz}$ and $A_\precsim(y)\cap C_{xz}$ are closed in the subspace $C_{xz}$. They cover $C_{xz}$ by completeness, and they contain $x$ and $z$, respectively. Since $C_{xz}$ is connected as the continuous image of $[0,1]$, these two closed sets must intersect. Hence there exists $c\in C_{xz}$ such that $c\sim y$. Together with order denseness, this is Wold-continuity.

\smallskip
\noindent {\it Linear continuity $\Longleftrightarrow$ mixture-continuity:}
Assume $\succsim$ is linearly continuous. Pick $x,y,z\in X$. If $x=y$, the relevant mixture-continuity set is either $[0,1]$ or $\emptyset$, so there is nothing to prove. If $x\neq y$, let $L_{xy}$ be the line segment joining $x$ and $y$, and let $l_{xy}:[0,1]\to L_{xy}$ be the usual mixture-linear parametrization. Since $X\subseteq\mathbb R^n$ is Hausdorff, $l_{xy}$ is a homeomorphism from $[0,1]$ onto $L_{xy}$. Upper linear continuity implies that $I_\succsim(l_{xy},z):=\{\lambda\in[0,1]\mid l_{xy}(\lambda)\succsim z\}$ is closed in $[0,1]$. Hence $\succsim$ is upper mixture-continuous. The lower case is identical, so linear continuity implies mixture-continuity.

Conversely, assume $\succsim$ is mixture-continuous. Suppose for some straight line $L$ in $X$ and some $x\in X$, the set $A_\succsim(x)\cap L$ is not closed in $L$. Since $L$ is a one-dimensional Euclidean subspace, there are a sequence $y^m\in A_\succsim(x)\cap L$ and a point $y\in L$ such that $y^m\to y$ and $y\notin A_\succsim(x)$. Discarding any terms equal to $y$, infinitely many $y^m$ lie on one side of $y$ in the line order; choose $y^k$ on that side. Then the segment joining $y^k$ and $y$ contains infinitely many members of the sequence. Under the homeomorphism $l_{y^ky}:[0,1]\to L_{y^ky}$, the inverse image of $A_\succsim(x)\cap L_{y^ky}$ is not closed, contradicting upper mixture-continuity. Hence every upper section is closed on every straight line. The proof for lower sections is symmetric, and mixture-continuity implies linear continuity.
\prff

We include the following result for completeness, establishing the equivalence between restricted solvability and separate IVP for sets in $\Re^n$ where the mixture operation is the usual convex combination. 

\cl 
Let $X\subseteq \mathbb{R}^n$ be convex. A binary relation $\succsim$ on $X$ satisfies restricted solvability if and only if it satisfies separate IVP. 
\label{thm_rs_sivp} 
\cll

\prf[Proof of Claim \ref{thm_rs_sivp} ] 
For each $i\in\{1,\dots,n\}$ and each $y_{-i}\in\mathbb{R}^{n-1}$, let $X_i(y_{-i}) := \{r\in\mathbb{R} : (r,y_{-i})\in X\}$. Since $X$ is convex, each section $X_i(y_{-i})$ is a convex subset of $\mathbb{R}$, hence an interval. Suppose first that $\succsim$ satisfies separate IVP. Fix $i\in\{1,\dots,n\}$, $x\in X$, and $(a_i,y_{-i}), (b_i,y_{-i})\in X$ such that $(a_i,y_{-i})\succsim x \succsim (b_i,y_{-i})$. By separate IVP, there exist $c_i,d_i\in\mathbb{R}$ and $\lambda,\delta,\delta'\in[0,1]$ such that $a_i=\delta c_i+(1-\delta)d_i$, $b_i=\delta'c_i+(1-\delta')d_i$, and $x\sim (\lambda c_i+(1-\lambda)d_i,y_{-i})$. Setting $t_i:=\lambda c_i+(1-\lambda)d_i$, the convexity of the coordinate section gives $(t_i,y_{-i})\in X$, and $x\sim (t_i,y_{-i})$. Thus restricted solvability holds.
Conversely, suppose that $\succsim$ satisfies restricted solvability. Fix $i\in\{1,\dots,n\}$, $x\in X$, and $(a_i,y_{-i}), (b_i,y_{-i})\in X$ such that $(a_i,y_{-i})\succsim x \succsim (b_i,y_{-i})$. By restricted solvability, there exists $t_i\in\mathbb{R}$ such that $(t_i,y_{-i})\in X$ and $x\sim (t_i,y_{-i})$. Hence $a_i,b_i,t_i\in X_i(y_{-i})$. Since $X_i(y_{-i})$ is an interval, the three points $a_i$, $b_i$, and $t_i$ lie in the common interval with endpoints $c_i:=\min\{a_i,b_i,t_i\}$ and $d_i:=\max\{a_i,b_i,t_i\}$. In particular, $c_i,d_i\in X_i(y_{-i})$, so $(c_i,y_{-i}), (d_i,y_{-i})\in X$. Moreover, because $a_i,b_i,t_i\in[c_i,d_i]$, there exist $\delta,\delta',\lambda\in[0,1]$ such that $a_i=\delta c_i+(1-\delta)d_i$, $b_i=\delta'c_i+(1-\delta')d_i$, and $t_i=\lambda c_i+(1-\lambda)d_i$. Since $x\sim (t_i,y_{-i})$, it follows that $x\sim (\lambda c_i+(1-\lambda)d_i,y_{-i})$. Hence separate IVP holds. Therefore, restricted solvability and separate IVP are equivalent on convex subsets of $\mathbb{R}^n$.
\prff

\cl
Let  $\succsim$  be a complete and transitive binary relation on a non-empty and convex  set  $X\subseteq \mathbb R^n$ with property  {\bf \ref{PrA}} where $\{1,\ldots,n\}$ is the set of indices. 

\ben[{\nf (a)}, topsep=2pt]
\setlength{\itemsep}{-1pt}

\item If the upper (lower) sections of $\succsim$ are separately convex, then the following are equivalent for $\succsim$: lower (upper) continuity, lower (upper) mixture continuity, lower (upper) linear continuity, lower (upper) separate continuity, lower (upper) separate mixture continuity, upper (lower) Archimedeanity, and upper (lower) separate Archimedeanity.

\item If the sections of $\succsim$ are separately convex and $\succsim$ is order dense, then the following are equivalent for $\succsim$: graph continuity, continuity, linear continuity, mixture continuity, Archimedeanity, strong IVP, IVP, restricted (separate) IVP, Wold-continuity, weak Wold-continuity, and separate continuity.
\een
\label{thm_equiv_pref_finite_general}
\cll

\prf[Proof of Claim \ref{thm_equiv_pref_finite_general}]
We prove part (a) under upper separate convexity while the assertion under lower separate convexity follows by passing to the inverse relation. Lower continuity implies lower mixture continuity and lower linear continuity, and lower linear continuity implies lower separate continuity, by their definitions. Lower separate continuity coincides with lower separate mixture continuity, since closedness on a coordinate section of $X$ is a local property and every point of the section lies in a mixture segment of the section. Lower mixture continuity implies upper Archimedeanity by \citet[Proposition 1]{gku19}, and upper Archimedeanity implies upper separate Archimedeanity by definition. Upper separate Archimedeanity implies lower separate mixture continuity, and lower separate mixture continuity implies lower mixture continuity, under upper separate convexity in $n-1$ indices, by the coordinate-release argument in the proof of Theorem \ref{thm_equiv_pref_mixture}. Finally, lower separate continuity implies lower continuity by Theorem \ref{thm_equiv_pref}. This closes the cycle and proves part (a).

For part (b), upper and lower separate convexity allow us to apply part (a) in both directions. Hence continuity, mixture continuity, linear continuity, separate continuity, Archimedeanity and separate Archimedeanity are equivalent. Graph continuity implies continuity for any binary relation, and the converse holds here since $X$ is a connected and separable subset of $\mathbb R^n$ and a complete, transitive and continuous relation on $X$ admits a continuous representation; see \citet[Theorem 3.2.9]{bm95}. Continuity and order-denseness imply Wold-continuity, as shown in the proof of Theorem \ref{thm_equiv_pref}; Wold-continuity implies strong IVP and weak Wold-continuity, and these imply IVP, hence separate IVP, by their definitions. Separate IVP and restricted solvability are equivalent by Claim \ref{thm_rs_sivp}. Finally, under upper and lower separate convexity and order-denseness, the argument establishing separate mixture continuity from separate IVP in the proof of Theorem \ref{thm_equiv_pref_mixture} applies on coordinate sections and returns us to the above cycle. Thus all the postulates in part (b) are equivalent.
\prff

\prf[Proof of Theorem \ref{thm_equiv_pref_finite}]
The one-directional implications among the continuity postulates do not require convexity and are standard; see \citet[Proposition 9]{gku22games} and the references therein.

Suppose first that $\succsim$ is graph continuous. Then it is continuous, hence separately continuous. Fix $i$, $x\in X$ and $(a_i,y_{-i}),(b_i,y_{-i})\in X$ with $(a_i,y_{-i})\succsim x\succsim(b_i,y_{-i})$. On the segment joining $a_i$ and $b_i$ in $X_i(y_{-i})$, the sets $\{t\mid (t,y_{-i})\succsim x\}$ and $\{t\mid x\succsim(t,y_{-i})\}$ are closed by separate continuity, cover the segment by completeness, and contain $a_i$ and $b_i$, respectively. Since the segment is connected, they intersect, and any $t_i$ in the intersection satisfies $x\sim(t_i,y_{-i})$. Hence $\succsim$ is restricted solvable.

Conversely, suppose that $\succsim$ is restricted solvable. By Claim \ref{thm_rs_sivp}, restricted solvability and separate IVP are equivalent on convex subsets of $\mathbb R^n$. Under upper and lower separate convexity and order-denseness, the argument establishing separate mixture continuity from separate IVP in the proof of Theorem \ref{thm_equiv_pref_mixture} takes place entirely on a single coordinate section and therefore applies verbatim to any convex $X\subseteq\mathbb R^n$ with property {\bf \ref{PrA}}. Hence $\succsim$ is upper and lower separately mixture continuous. Along any coordinate section of $X$, closedness is a local property and every point of the section lies in a mixture segment of the section; hence separate mixture continuity is exactly the closedness of the upper and lower sections restricted to that section. Thus $\succsim$ is separately continuous.

We now convert separate continuity into continuity with the correct crossed pairing. By completeness, for every $x\in X$ and every coordinate section $X_{i,y}$, $A_\succ(x)\cap X_{i,y}=X_{i,y}\setminus\big(A_\precsim(x)\cap X_{i,y}\big)$ and $A_\prec(x)\cap X_{i,y}=X_{i,y}\setminus\big(A_\succsim(x)\cap X_{i,y}\big)$. Thus lower (upper) separate continuity is equivalent to separate openness of the strict upper (lower) sections. Moreover, for a complete and transitive relation, upper (lower) separate convexity of weak upper (lower) sections is equivalent to separate convexity of the corresponding strict sections; see \citet[Lemma 6]{uk22jme}. Applying Proposition \ref{thm_additional_directional_separate} to the correspondence $x\mapsto A_\succ(x)$ gives that $A_\succ(x)$ is open for every $x$, so $A_\precsim(x)$ is closed and $\succsim$ is lower continuous. Applying the same proposition to $x\mapsto A_\prec(x)$, using lower separate convexity, gives upper continuity. Therefore $\succsim$ is continuous. Finally, since $X$ is a connected and separable subset of $\mathbb R^n$, a complete, transitive and continuous relation on $X$ admits a continuous representation and therefore has a closed graph; see \citet[Theorem 3.2.9]{bm95}. Since graph continuity implies continuity, the proof is complete.
\prff

The following example, referenced in Sections \ref{sec_ordered} and \ref{sec_correspondence}, shows that Theorems \ref{thm_equiv_pref}, \ref{thm_equiv_pref_finite} and \ref{thm_separate_main} \textit{cannot} be extended beyond Euclidean domains: on products of infinite-dimensional factors, the separate postulates lose their power entirely, even in their strongest two-sided form.

\ex{\nf 
Let \(H=\ell^2\) with its weak topology \(H_w\) and standard orthonormal basis
\(\{e_m\}_{m\geq1}\), let \(X=H_w\times H_w\), a convex and open product
domain with the two copies of \(H_w\) as the coordinate factors, and let
\(\succsim\) be the complete and transitive relation induced by the inner
product \(u(y,z)=\langle y,z\rangle\), the classical example of a separately
continuous but not jointly continuous bilinear functional; see, for example,
\citet[Chapter 6]{ab06}. Freezing either coordinate, the restriction of any
weak upper or lower section of \(\succsim\) to a coordinate section is a set
of the form
\(\{y\in H\mid \langle y,z_0\rangle\geq\alpha\}\) or
\(\{y\in H\mid \langle y,z_0\rangle\leq\alpha\}\). Since
\(\langle\cdot,z_0\rangle\) is a weakly continuous linear functional, these
sets are weakly closed and convex. Thus \(\succsim\) is upper and lower
separately convex and upper and lower separately continuous. Moreover, along
any straight line, \(u\) restricts to a polynomial of degree at most two in
the line parameter, so \(\succsim\) is mixture continuous and linearly
continuous.

Yet continuity fails in both directions. Since \(e_m\to0\) in \(H_w\), both
\((e_m,e_m)\to(0,0)\) and \((e_m,-e_m)\to(0,0)\) in \(X\). For every \(m\),
\(u(e_m,e_m)=1=u(e_1,e_1)\), so the upper section
\(A_{\succsim}(e_1,e_1)\) contains \((e_m,e_m)\) for every \(m\), but not its
limit \((0,0)\). Hence \(A_{\succsim}(e_1,e_1)\) is not weakly closed, and
upper continuity fails. Symmetrically,
\(u(e_m,-e_m)=-1=u(e_1,-e_1)\), so the lower section
\(A_{\precsim}(e_1,-e_1)\) contains \((e_m,-e_m)\) for every \(m\), but not
\((0,0)\). Hence \(A_{\precsim}(e_1,-e_1)\) is not weakly closed, and lower
continuity fails. Therefore, the conclusions of Theorems
\ref{thm_equiv_pref} and \ref{thm_equiv_pref_finite} fail on products of
infinite-dimensional coordinate factors, even under two-sided separate
convexity, two-sided separate continuity, mixture continuity and linear
continuity.

The same computation also bounds Theorem \ref{thm_separate_main}. Regard the
strict lower section
\(A_{\prec}(e_1,e_1)
=\{(y,z)\in H\times H\mid \langle y,z\rangle<1\}\)
as the value of the constant correspondence
\(F:\{\ast\}\tra H_w\times H_w\) on a one-point domain. The lower sections of
\(F\) are either \(\emptyset\) or \(\{\ast\}\), and hence are open. Freezing
either coordinate factor, the corresponding section of \(F(\ast)\) is an
open convex half-space determined by a weakly continuous linear functional.
Thus \(F\) has separately convex and separately open upper sections, and its
codomain is both open and a product space. However,
\((0,0)\in F(\ast)\), while \((e_m,e_m)\notin F(\ast)\) for every \(m\), and
\((e_m,e_m)\to(0,0)\) in \(X\). Hence \(F(\ast)\), and therefore
\(\operatorname{gr}F\), is not weakly open. Consequently, the conclusion of
Theorem \ref{thm_separate_main} also fails when the coordinate factors are
infinite-dimensional, even for a constant correspondence and even though the
remaining hypotheses hold in their strongest form.

The mechanism throughout is the one isolated in Footnote
\ref{fn_yamazaki}: the convex hull of finitely many points of a section has
empty interior in an infinite-dimensional coordinate factor. Thus the local
finiteness that drives the coordinate-release argument in the proof of
Theorem \ref{thm_separate_main} is unavailable, and the same obstruction
simultaneously bounds Theorems \ref{thm_equiv_pref},
\ref{thm_equiv_pref_finite} and \ref{thm_separate_main}.
}
\label{ex_bilinear_pref}
\exx

Next, we prove that for a continuous preference relation presented in Example \ref{separate-rr}, 
separate convexity is equivalent to separate $\Psi$-convexity.

\prf[Proof of Convexity in Example \ref{separate-rr}] We first show the if-part.
 Assume that $\succsim$ is separately convex. Fix a coordinate
$i$ and take two points $a, b \in X$ such that for every $\geq_k \in \Psi_i$ there is a
$y_k = (t_k, b_{-i})$ with $(b_i, b_{-i}) \geq_k y_k$ and $y_k \succsim (a_i, b_{-i})$.
We show $(b_i, b_{-i}) \succsim (a_i, b_{-i})$ by contradiction. Suppose
$(a_i, b_{-i}) \succ (b_i, b_{-i})$. Since $\succsim$ is continuous and separately
convex, the $i$-th coordinate section of the upper contour set, $S := \{t \in \mathbb{R} : (t, b_{-i}) \succsim (a_i, b_{-i})\},$
is a closed and convex subset of $\mathbb{R}$, hence a closed interval. Since $(a_i, b_{-i}) \succ (b_i, b_{-i})$,
we have $b_i \notin S$, so either $b_i < \inf S$ or $b_i > \sup S$. In either case there
exists $\lambda \neq 0$ with $\lambda b_i < \inf_{t \in S} \lambda t$: take $\lambda > 0$
when $b_i < \inf S$ and $\lambda < 0$ when $b_i > \sup S$. The vector $v = \lambda e_i$
defines an ordering $\geq_l \in \Psi_i$ that strictly separates $b_i$ from $S$. By hypothesis
there is $y_l = ((y_l)_i, b_{-i})$ with $(b_i, b_{-i}) \geq_l y_l$, that is
$\lambda b_i \geq \lambda (y_l)_i$, and $y_l \succsim (a_i, b_{-i})$, that is $(y_l)_i \in S$.
But $(y_l)_i \in S$ gives $\lambda (y_l)_i \geq \inf_{t \in S} \lambda t > \lambda b_i$,
contradicting $\lambda b_i \geq \lambda (y_l)_i$. Since $i$ was
arbitrary, $\succsim$ is separately $\Psi$-convex.

Now we show the only-if part. Assume that $\succsim$ is separately $\Psi$-convex. Fix a
coordinate $i$ and take two points $a, b \in X$ with
$(b_i, b_{-i}) \succsim (a_i, b_{-i})$. Let
$c = (\lambda a_i + (1-\lambda) b_i,\, b_{-i})$ for some $\lambda \in [0,1]$. The
cases $\lambda = 0$ and $\lambda = 1$ are trivial: $\lambda = 1$ gives
$c = (a_i, b_{-i})$ and $\lambda = 0$ gives $c = (b_i, b_{-i})$, both of which
satisfy $c \succsim (a_i, b_{-i})$ directly. For $\lambda \in (0,1)$, we show
$c \succsim (a_i, b_{-i})$ using $\Psi_i$-convexity. Each $\geq_k \in \Psi_i$ is one of
the two coordinate orientations, and $c_i = \lambda a_i + (1-\lambda) b_i$ lies in the
interval $[\min(a_i,b_i), \max(a_i,b_i)]$. We exhibit a witness $y_k = (t, b_{-i})$ for each
orientation:
\bit
    \item If $\geq_k$ is the orientation $x \geq_k y \iff x_i \geq y_i$, set
    $y_k = (\min(a_i,b_i), b_{-i})$. Then $c_i \geq \min(a_i,b_i)$ gives $c \geq_k y_k$, and
    $y_k$ equals $(a_i, b_{-i})$ or $(b_i, b_{-i})$, so $y_k \succsim (a_i, b_{-i})$ by
    reflexivity or by hypothesis.
    \item If $\geq_k$ is the orientation $x \geq_k y \iff x_i \leq y_i$, set
    $y_k = (\max(a_i,b_i), b_{-i})$. Then $c_i \leq \max(a_i,b_i)$ gives $c \geq_k y_k$, and
    again $y_k$ equals $(a_i, b_{-i})$ or $(b_i, b_{-i})$, so $y_k \succsim (a_i, b_{-i})$.
\eit
Since for every $\geq_k \in \Psi_i$ such a $y_k$ exists,
$\Psi_i$-convexity yields $c \succsim (a_i, b_{-i})$. Since $i$ was arbitrary,
$\succsim$ is separately convex.
\prff

The next proof verifies the properties of the preference relation in Example \ref{ex_separate_function}.

\prf[Verification of Properties in Example \ref{ex_separate_function}] We verify the properties asserted in Example \ref{ex_separate_function}. Fix a horizontal line $x_2=c$; the vertical case is symmetric by the symmetry of $f$. If $c\leq 0$, then $f(t,c)=0$ for all $t$, and the section is fully indifferent. If $c>0$, write $g(t):=f(t,c)$. For $t<0$, $g(t)=0$; for $t\geq 0$, $g(t)=2tc/(t^2+c^2)+\min\{t,c\}$. On $[0,c]$ both summands are increasing, so $g$ increases from $0$ to $c+1$; on $[c,\infty)$ the first summand is decreasing and the second is constant at $c$, so $g$ decreases from $c+1$ and remains above $c$. Hence $g$ is quasiconcave on the line: every upper level set $\{t\mid g(t)\geq\alpha\}$ is an interval (empty for $\alpha>c+1$; a compact interval around $c$ for $\alpha\in(c,c+1]$; a half-line for $\alpha\in(0,c]$; the whole line for $\alpha\leq 0$). This gives upper separate convexity. Since $g$ is continuous on each coordinate line (the Genocchi--Peano term $2tc/(t^2+c^2)$ vanishes along both axes, including at the origin), the relation is separately continuous, and restricted solvability on coordinate sections follows from the intermediate value theorem applied to $g$; for upper Archimedeanity, note that $f$ is continuous on $\Re^2\setminus\{0\}$ and attains its minimum value $0$ at the origin, so every strict upper section $\{z\mid f(z)>\alpha\}$ with $\alpha\geq 0$ is open in $\Re^2$; given $x\succ y$, the segment from any $z$ to $x$ therefore enters the open set $\{f>f(y)\}$ for $\lambda$ close to $1$, and separate Archimedeanity follows a fortiori. For the failure of the lower postulates: take $x\in\Re^2_+$ with $f(x)=1/2$ (such $x$ exists on any horizontal line $x_2=c$ with $0<c<1/2$, by the intermediate value theorem), $y=(0,0)$ and $z=(1,1)$. Then $x\succ y$, and for every $\delta\in(0,1)$, $y\delta z=(1-\delta,1-\delta)$ lies on the strictly positive diagonal, where $f(1-\delta,1-\delta)=1+(1-\delta)>1>f(x)$; hence $x\nsucc y\delta z$ for every $\delta$, and lower Archimedeanity fails. Similarly, with $x'=(1,1)$ and $y'=(-1,-1)$, the mixture $x'\lambda y'=(2\lambda-1,2\lambda-1)$ satisfies $f(x'\lambda y')=2\lambda$ for $\lambda>1/2$ and $f(x'\lambda y')=0$ for $\lambda\leq 1/2$, so $\{\lambda\in[0,1]\mid x'\lambda y'\succsim x\}=(1/2,1]$ is not closed: upper mixture continuity fails along the $45^{\circ}$ line, and hence continuity fails.
\prff

\prf[Proof of Proposition \ref{thm_cardinal_game}]
The forward direction is straightforward. For the backward direction, we show that separate independence implies separate convexity, and that, under separate independence, separate Archimedean$^*$ and separate Archimedean properties are equivalent. Therefore, by Theorem \ref{thm_equiv_pref_mixture}, separate IVP can be used instead of separate Archimedean$^*$ property in Fishburn's theorem. Throughout, for $x,z\in X$ and $i\in I$, we write $xT_iz$ for Fishburn's relation of differing in at most coordinate $i$, that is, $x_{-i}=z_{-i}$.  

\medskip

\nt\textit{Separate Independence $\Longrightarrow$ Separate Convexity.} Assume that separate independence holds.  Pick $x,y,z\in X$ and $i\in I$ such that $x\succsim y, z\succsim y$ and $xT_i z$. Next, we show that for all $\lambda\in (0,1)$, $x\lambda z\succsim y$.  
By completeness, assume without loss of generality that $x\succsim z$. Hence, either $x\sim z$ or $x\succ z$.  
By K1 and  K3 in \citet[p. 90]{fi82},  we have for all $\lambda \in (0,1)$, $x\lambda z\succsim z$.\footnote{Note that K1 and K3 in Fishburn's argument do not rely on separate Archimedean$^*$ property. } Hence, it follows from transitivity and $z \succsim y$ that for all $\lambda \in (0,1)$, $x\lambda z\succsim y$. Hence $\succsim$ has separately convex upper sections. A symmetric argument shows that $\succsim$ has separately convex lower sections. 

\medskip

\nt \textit{Separate Archimedean$^*$ $\Longleftrightarrow$ Separate Archimedean.} 
It is easy to check from the definitions of these two axioms that separate Archimedeanity implies separate Archimedean$^*$. For the converse, assume separate Archimedean$^*$ holds. Pick $x, y, z\in X$ and $i\in I=\{1,\ldots, n\}$ such that $xT_iz$ and $x\succ y$.  We show that there exists $\lambda\in (0,1)$ such that $x\lambda z\succ y$. Assume towards a contradiction that for all $\lambda\in (0,1)$,  $x\lambda z\nsucc y$. By completeness,  for all $\lambda\in (0,1)$,  $y\succsim x\lambda z$.  Suppose that $x\lambda z \sim y$. Applying separate independence to the quadruple $(x, y, x\lambda z, y)$ -- note that $x\succ y$, $x\lambda z\sim y$, and $x_{-i}=(x\lambda z)_{-i}$ -- yields, for all $\delta\in (0,1)$, $x\delta(x\lambda z)\succ y\delta y=y$.\fn{The identity $w\delta w=w$ holds in every mixture set: conditions (a) and (b) give $x0w=w1x=w$ for any $x$, and condition (c) with $\mu=0$ then gives $w\delta w=(x0w)\delta w=x0w=w$.} Since $x\delta(x\lambda z)=x\gamma z$ for some $\gamma\in (0,1)$,\fn{In particular, $x\delta(x\lambda z)=(z(1-\lambda)x)(1-\delta)x=z(1-\lambda)(1-\delta)x=x(1-(1-\lambda)(1-\delta))z$, hence $\gamma=1-(1-\lambda)(1-\delta)$.} $x\gamma z\succ y$ yields a contradiction. Suppose that $y\succ x\lambda z$. Then $x\succ y\succ x\lambda z$ and by separate Archimedean$^*$ property,  there exists $\delta\in (0,1)$ such that $x\delta(x\lambda z)\succ y$. Since $x\delta(x\lambda z)=x\gamma z$ for some $\gamma\in (0,1)$,  $x\gamma z\succ y$ yields a contradiction. Therefore, there exists $\lambda\in (0,1)$ such that $x\lambda z\succ y$. Hence, $\succsim$ is upper separate Archimedean.

Next, we show that lower separate Archimedean holds.  
Pick $x, y, z\in X$ and $i\in I=\{1,\ldots, n\}$ such that $xT_iz$ and $y\succ z$.  We show that there exists $\lambda\in (0,1)$ such that $y\succ x\lambda z$. Assume towards a contradiction that for all $\lambda\in (0,1)$,  $y\nsucc x\lambda z$. By completeness,  for all $\lambda\in (0,1)$,  $x\lambda z\succsim y$.  Suppose that $x\lambda z \sim y$. Applying separate independence to the quadruple $(y, z, y, x\lambda z)$ -- note that $y\succ z$, $y\sim x\lambda z$, and $z_{-i}=(x\lambda z)_{-i}$ -- yields, for all $\delta\in (0,1)$, $y=y\delta y\succ z\delta(x\lambda z)$. Since $z\delta(x\lambda z)=x\gamma z$ for some $\gamma\in (0,1)$,  $y\succ x\gamma z$ yields a contradiction. Suppose that $x\lambda z\succ y$.  
Then $x\lambda z\succ y\succ z$ and by separate Archimedean$^*$ property, there exists $\delta\in (0,1)$ such that $y\succ  (x\lambda z)\delta z$. Since $(x\lambda z)\delta z=x\gamma z$ for some $\gamma\in (0,1)$,  $y\succ x\gamma z$ yields a contradiction. Therefore, there exists $\lambda\in (0,1)$ such that $y\succ x\lambda z$. Hence, $\succsim$ is lower separate Archimedean.

Hence, we have shown that under separate independence, the separate Archimedean$^*$ property holds if and only if  the binary relation satisfies the separate Archimedean property.  
\prff

\prf[Proof of Proposition \ref{thm_ordinal_game}]
Let $u:X\to\Re$ represent $\succsim$.  We first note the equivalence between separate quasiconcavity of $u$ and upper separate convexity of $\succsim$.  Fix $i\in I$, fix $a_{-i}$, and take two points $(y_i,a_{-i})$ and $(z_i,a_{-i})$ in the same coordinate section.  If $u$ is quasiconcave in coordinate $i$, then for every $x\in X$ and every $\lambda\in[0,1]$,
$(y_i,a_{-i})\succsim x \hbox{ and } (z_i,a_{-i})\succsim x$
imply $u(\lambda y_i+(1-\lambda)z_i,a_{-i})\geq \min\{u(y_i,a_{-i}),u(z_i,a_{-i})\}\geq u(x)$, and hence $(\lambda y_i+(1-\lambda)z_i,a_{-i})\succsim x$.  Thus the upper sections of $\succsim$ are separately convex.  Conversely, if the upper sections of $\succsim$ are separately convex, apply the definition to upper contour sets of $u$ to obtain the usual coordinate-wise quasiconcavity inequality.  Hence upper separate convexity of the represented relation is equivalent to separate quasiconcavity of $u$.

If $u$ is continuous, the represented relation has closed upper and lower sections; if $u$ is separately quasiconcave, its weak upper sections are separately convex. Conversely, upper separate convexity gives separate quasiconcavity of any representing utility, while continuity gives closed upper and lower sections. The standard representation theorem for continuous complete and transitive relations on second-countable spaces then yields a continuous utility representation; see \citet[Theorem 3.2.9]{bm95}. If one assumes lower separate convexity in $N-1$ of the $N$ scalar coordinates of $\Re^{N}$ and property {\bf \ref{PrA}} for $X\subseteq\Re^{N}$, Theorem \ref{thm_equiv_pref}, applied with the $N$ scalar coordinates as the relevant indices, allows upper continuity in this statement to be replaced by upper separate continuity. The coordinate-$i$ version in part (b) is identical, with upper convexity required only in the own-action coordinate. This proves the claim.
\prff
Next, we turn to the proofs of Proposition
\ref{prop_hidden_complete_scalar} and Corollary \ref{cor_behavioral_seu}.

\prf[Proof of Proposition \ref{prop_hidden_complete_scalar}]
We first prove completeness, beginning with the coordinate sections. Fix
\(s\in S\) and \(f_{-s}\in Y^{S\setminus\{s\}}\), and restrict \(\succsim\) to
the coordinate section \(X_s(f_{-s})\), a one-dimensional mixture set isomorphic
to \(Y\). On this section, every mixture line is a coordinate mixture line of
\(X\), so upper and lower separate mixture-continuity become ordinary
mixture-continuity and upper and lower separate Archimedeanity become ordinary
Archimedeanity; reflexivity and transitivity are inherited. We write
\(p\succsim_s q\) when \((p_s,f_{-s})\succsim(q_s,f_{-s})\), and write
\(\sim_s\) and \(\succ_s\) for its symmetric and asymmetric parts.

By coordinate-wise comparability there are distinct \(p,q\in Y\) comparable on
this section. If one is strictly preferred to the other, the restricted relation
is non-trivial, reflexive, mixture-continuous and Archimedean on a mixture set,
with a transitive symmetric part and, being transitive, semi-transitive;
\citet[Theorem 1]{gku19} then implies that it is complete. If instead
\(p\sim_s q\) with \(p\neq q\), then the whole section is indifferent. Indeed,
weak separate independence, with \(h=k=(r_s,f_{-s})\), gives $\lambda p+(1-\lambda)r\sim_s\lambda q+(1-\lambda)r$ for all $r\in Y$ and $\lambda\in(0,1)$. If \(p<q\), then for any \(a\in\operatorname{int}Y\) and sufficiently small
\(t>0\), choosing \(\lambda=t/(q-p)\) and
\(r=(a-\lambda p)/(1-\lambda)\) (so that \(r\in Y\), uniformly on compact
subintervals of \(\operatorname{int}Y\)) gives \(a\sim_s a+t\);
interchanging \(p\) and \(q\) gives \(a\sim_s a-t\). Thus indifference is
local on \(\operatorname{int}Y\). Transitivity chains these local
indifferences over compact subintervals of \(\operatorname{int}Y\), and
mixture-continuity adds the endpoints. Hence \(a\sim_s b\) for all
\(a,b\in Y\). In either case, the coordinate section is complete.

We now prove completeness on \(X\) by induction on \(|S|\). The one-coordinate
case has just been proved. Suppose the completeness conclusion has been proved
for products with at most \(m-1\) coordinates under the hypotheses of the
proposition, and let \(X=Y^m\). Take arbitrary \(x,y\in X\), write
\(x=(x_1,x_{-1})\), \(y=(y_1,y_{-1})\), and set
$q=(x_1,y_{-1}).$
Then \(q\) and \(y\) differ only in the first coordinate, so they lie in a
coordinate section, which has already been shown complete. Hence \(q\) and
\(y\) are comparable.

On the other hand, \(q\) and \(x\) both lie in the \((m-1)\)-dimensional face $F=\{x_1\}\times Y^{m-1}.$
We identify \(F\) with \(Y^{m-1}\) through the remaining coordinates. The
restriction of \(\succsim\) to this face inherits the hypotheses of the
proposition in the separate sense. Indeed, each coordinate section of \(F\) is
a coordinate section of the original space with the first coordinate fixed at
\(x_1\). Hence upper and lower separate mixture-continuity and upper and lower
separate Archimedeanity on \(F\) follow directly from the corresponding
separate postulates on \(X\). Reflexivity, transitivity and weak separate
independence are inherited as well. We also inherit coordinate-wise comparability on \(F\): for a remaining coordinate \(i\in\{2,\ldots,m\}\), fixing all
other coordinates in \(F\) is exactly the same as fixing all coordinates other
than \(i\) in \(X\), with coordinate \(1\) fixed at \(x_1\). Hence the
induction hypothesis applies on \(F\), and \(q\) and \(x\) are comparable.

Thus the induction hypothesis is used only to make the corner point \(q\)
comparable with \(x\). This comparison does not use any mixture-continuity along
the segment joining \(q\) and \(x\). In the outer corner argument below, the
only mixtures formed are \(q\lambda y\) and \(r\mu y\), and these move only in
the first coordinate. Since \(q\) is also comparable with \(y\), we can now use
the same corner argument as in the two-dimensional case.

Suppose, to the contrary, that \(x\bowtie y\). Since \(q\) is comparable with
both \(x\) and \(y\), transitivity of \(\succsim\), which is maintained as a
primitive assumption here, rules out $x\succsim q\succsim y$ and $y\succsim q\succsim x$. Hence either \(q\succsim x\) and \(q\succsim y\), or
\(x\succsim q\) and \(y\succsim q\).

First suppose \(q\succsim x\) and \(q\succsim y\). Let
$B=\{\lambda\in[0,1]\mid q\lambda y\succsim x\}.$
Since \(q\) and \(y\) differ only in the first coordinate, the path
\(q\lambda y\) lies in a single coordinate section. Hence \(B\) is closed by
separate mixture-continuity. Moreover, \(1\in B\) and \(0\notin B\). Let
\(\alpha=\min B\) and set \(r=q\alpha y\). Then \(r\succsim x\). If
\(r\succ x\), upper separate Archimedeanity gives \(\mu\in(0,1)\) such that
\(r\mu y\succ x\). But $r\mu y=(q\alpha y)\mu y=q(\alpha\mu)y$, with \(\alpha\mu<\alpha\), contradicting the definition of \(\alpha\). Hence
\(r\sim x\). Here \(r=q\alpha y\) has \(r_{-1}=y_{-1}\), so \(r\) and \(y\)
lie on the first-coordinate section through \(y_{-1}\). Since coordinate
sections have already been shown complete, \(r\) and \(y\) are comparable. If
\(y\succsim r\), then transitivity gives \(y\succsim x\), a contradiction.
Otherwise \(r\succ y\), and since \(r\sim x\), transitivity gives
\(x\succsim y\), again a contradiction.

The case \(x\succsim q\) and \(y\succsim q\) is symmetric. Let
$C=\{\lambda\in[0,1]\mid x\succsim q\lambda y\}.$
Again \(q\lambda y\) lies in a single coordinate section, so \(C\) is closed by
separate mixture-continuity. Also \(1\in C\) and \(0\notin C\). Let
\(\alpha=\min C\) and set \(r=q\alpha y\). Then \(x\succsim r\). If
\(x\succ r\), lower separate Archimedeanity gives \(\mu\in(0,1)\) such that
\(x\succ r\mu y\). But $r\mu y=(q\alpha y)\mu y=q(\alpha\mu)y$, with \(\alpha\mu<\alpha\), contradicting the definition of \(\alpha\). Hence
\(x\sim r\). Again \(r_{-1}=y_{-1}\), so \(r\) and \(y\) lie on the same
first-coordinate section and are comparable. If \(r\succsim y\), then
transitivity gives \(x\succsim y\), a contradiction. Otherwise \(y\succ r\),
and since \(x\sim r\), transitivity gives \(y\succsim x\), again a
contradiction. 

Therefore \(x\) and \(y\) are comparable. Since \(x,y\in Y^m\) were arbitrary,
\(\succsim\) is complete on \(Y^m\). By induction, \(\succsim\) is complete on
\(X=Y^S\).

It remains to obtain the representation. If \(|S|=1\), this is the
one-dimensional mixture-space case covered by the affine utility theorem for
mixture spaces; see \citet[Chapter 2, Theorem 1]{fi82}. Suppose now that
\(|S|\geq2\). Let \(\succ\) be the asymmetric part of \(\succsim\). Since
\(\succsim\) is complete and transitive, \(\succ\) is an asymmetric weak order.
For each coordinate \(i\), let \(T_i\) denote Fishburn's relation of differing
in at most coordinate \(i\).

We verify the hypotheses of Fishburn's multilinear utility theorem,
\citet[Chapter 7, Theorem 1]{fi82}. His axiom \(E1\) is the asymmetric weak
order just noted. His axiom \(E2\) is precisely the multilinearity independence
axiom for products of mixture sets: if \(xT_i z\), \(yT_j w\), \(x\sim y\),
and \(z\sim w\), then $x\lambda z\sim y\lambda w$. This is exactly weak separate independence, written with \(f=x\), \(h=z\),
\(g=y\), \(k=w\), \(s=i\), and \(t=j\). His axiom \(E3\) is the coordinate
Archimedean axiom, and it is implied by the upper and lower separate
Archimedean assumptions. Therefore \citet[Chapter 7, Theorem 1]{fi82} gives a
real-valued function \(V:X\to\Re\) that represents \(\succ\) and is affine in
each coordinate separately.

Since \(\succsim\) is complete, the same \(V\) represents \(\succsim\): $f\succsim g$ if and only if $V(f)\geq V(g)$. Indeed, if \(f\not\succsim g\), then completeness gives \(g\succ f\), hence
\(V(g)>V(f)\).

Because \(Y\) is an interval and \(S\) is finite, a function that is affine in
each coordinate separately is continuous and has the multi-affine form
$V(f)=\sum_{A\subseteq S} c_A\prod_{s\in A}\tau(f_s),$
where \(\tau:Y\to[0,1]\) is any affine normalization and the empty product is
\(1\). Since \(V\) is continuous, for all \(f,g,h\in X\), the sets $\{\lambda\in[0,1]\mid f\lambda g\succsim h\}=\{\lambda\in[0,1]\mid V(f\lambda g)\geq V(h)\}$ and $\{\lambda\in[0,1]\mid h\succsim f\lambda g\}=\{\lambda\in[0,1]\mid V(h)\geq V(f\lambda g)\}$ are closed. Hence \(\succsim\) is mixture-continuous.
\prff

\prf[Proof of Corollary \ref{cor_behavioral_seu}]
The independence axiom implies weak separate independence. Hence Proposition
\ref{prop_hidden_complete_scalar} implies completeness and mixture-continuity.
Since we assume transitivity, \(\succsim\) is a complete, transitive,
independent and mixture-continuous relation on \(X=Y^S\). By the affine utility
representation theorem for mixture spaces, see \citet[Chapter 2, Theorem
1]{fi82}, there is an affine function \(V:X\to\Re\) representing
\(\succsim\). Since \(X=Y^{S}\) is a finite product of intervals, we can write
\[
        V(f)=c+\sum_{s\in S}a_sf_s .
\]
If \(a_s=0\) for all \(s\), then \(\succsim\) is fully indifferent, and any
\(\pi\in\Delta(S)\) together with any constant affine \(u\) gives the desired
representation. Suppose now that \(\succsim\) is non-trivial. Then some
\(a_s\neq0\). If \(s\) is null, then \(a_s=0\). If \(s\) and \(t\) are
non-null, state independence implies that \(a_s(\cdot)\) and \(a_t(\cdot)\)
induce the same order on \(Y\), so all non-zero coefficients have the same
sign. Let this common sign be \(\sigma\in\{-1,1\}\), and set $A=\sum_{s:a_s\neq0}|a_s|$. Define $\pi_s=|a_s|/A$ if $a_s\neq0$ and $\pi_s=0$ if $a_s=0$, and let \(u(r)=\sigma A r\). Then \(\pi\in\Delta(S)\), \(u\) is non-constant and affine, and $\sum_{s\in S}\pi_su(f_s)=\sum_{s:a_s\neq0}a_sf_s$. Hence the displayed subjective expected utility functional represents
\(\succsim\), and \(\pi_s=0\) for every null state.
\prff

We now turn to the results of Section \ref{sec_correspondence}, beginning with the open-graph theorem.

\prf[Proof of Theorem \ref{thm_separate_main}]
Open graph implies open upper and lower sections, and open upper sections are in particular separately open; this gives the forward implication. For the converse, assume $F$ has open lower sections and separately open upper sections.

\smallskip
\noindent\emph{Reduction of the open case to the product case.} Suppose first that $Y$ is open in $\Re^n$. Consider $F$ as a correspondence $\tilde F:X\tra\Re^n$ with $\tilde F(x)=F(x)$ for all $x\in X$. The lower sections are unchanged, hence open. For any $x\in X$ and any coordinate line $L$ of $\Re^n$, we have $\tilde F(x)\cap L=F(x)\cap(L\cap Y)$; this set is convex by separate convexity of $F$, and it is open in $L\cap Y$, which is itself open in $L$ since $Y$ is open, hence it is open in $L$. Thus $\tilde F$ has separately convex and separately open upper sections with range $\Re^n=\prod_{i=1}^n\Re$, a product set. If the product case of the theorem holds, then $\operatorname{gr}\tilde F$ is open in $X\times\Re^n$, and therefore $\operatorname{gr}F=\operatorname{gr}\tilde F\cap(X\times Y)$ is open in $X\times Y$. It therefore suffices to prove the theorem when $Y=\prod_{i=1}^nY_i$ is a product set; each $Y_i$ is then an interval, since $Y$ is convex.

\smallskip
Write $A:=\operatorname{gr}F$ and $Z:=X\times Y$. For $(x,y)\in A$ and $J\subseteq I$, let
$
S_J(x,y):=\{(x',y')\in Z\mid y'_i=y_i\ \text{for all}\ i\in I\setminus J\}
$
be the slice of $Z$ in which the $Y$-coordinates outside $J$ are frozen at $y$. Since $Y$ is a product set, $S_J(x,y)$ is homeomorphic to $X\times\prod_{j\in J}Y_j$, and we refer to $(x',(y'_j)_{j\in J})$ as the free part of a point of the slice. Note that every free part in $X\times\prod_{j\in J}Y_j$ corresponds to a point of $Z$, a fact special to the product case that we use repeatedly below. Put $A_J(x,y):=A\cap S_J(x,y)$. We prove, by induction on $|J|$,
\[
A_J(x,y)\ \text{is open in}\ S_J(x,y)\quad\text{for every}\ (x,y)\in A\ \text{and every}\ J\subseteq I. \tag{$\ast$}
\]
Taking $J=I$ gives $A_I(x,y)=A$ open in $S_I(x,y)=Z$, the desired conclusion.

\smallskip
\noindent\emph{Base case $J=\emptyset$.} Here $S_\emptyset(x,y)=X\times\{y\}$ and $A_\emptyset(x,y)=F^{-1}(y)\times\{y\}$, which is open in $X\times\{y\}$ because $F$ has open lower sections.

\smallskip
\noindent\emph{Inductive step.} Let $J\subsetneq I$, fix $i\in I\setminus J$, and assume $(\ast)$ for $J$. Fix $(x,y)\in A.$ We show that $A_{J\cup\{i\}}(x,y)$ is a neighbourhood of $(x,y)$ in $S_{J\cup\{i\}}(x,y)$. For $t\in Y_i$, let $y[t]\in Y$ be the point with $i$-th coordinate $t$ and all other coordinates equal to those of $y$.

Since $F$ has separately open upper sections, $F(x)$ meets the $i$-th coordinate section of $Y$ through $y$, a copy of the interval $Y_i$, in a set open in $Y_i$ and containing $y_i$. Hence there is an interval $[a,b]\subseteq Y_i$ that is a relative neighbourhood of $y_i$ in $Y_i$ (two-sided if $y_i$ is interior to $Y_i$, and one-sided if $y_i$ is an endpoint of $Y_i$) such that $y[t]\in F(x)$ for all $t\in[a,b]$; in particular, $(x,y[a]),\,(x,y[b])\in A$. Both $(x,y[a])$ and $(x,y[b])$ have their $Y$-coordinates outside $J$ frozen (coordinate $i$ at $a$, resp.\ $b$; the coordinates in $I\setminus(J\cup\{i\})$ at $y$). By the induction hypothesis, $A_J(x,y[a])$ and $A_J(x,y[b])$ are open in $S_J(x,y[a])$ and $S_J(x,y[b])$. In each slice the coordinate $i$ is frozen, so, writing $\langle x',(y'_j)_{j\in J},t\rangle$ for the point of $S_{J\cup\{i\}}(x,y)$ with free part $(x',(y'_j)_{j\in J})$ and $i$-th coordinate $t$, there are open neighbourhoods $N^a,N^b$ of the free part $(x,(y_j)_{j\in J})$ in $X\times\prod_{j\in J}Y_j$ with $\langle x',(y'_j)_j,a\rangle\in A$ for $(x',(y'_j)_j)\in N^a$, and $\langle x',(y'_j)_j,b\rangle\in A$ for $(x',(y'_j)_j)\in N^b$. Set $N:=N^a\cap N^b$. For $(x',(y'_j)_j)\in N$ and $t\in[a,b]$, the point $\langle x',(y'_j)_j,t\rangle$ lies in $Z$, since $t\in[a,b]\subseteq Y_i$ and the remaining coordinates lie in the corresponding factors of $Y$. Moreover, both $a$ and $b$ lie in the intersection of $F(x')$ with the $i$-th coordinate section of $Y$ through $\langle x',(y'_j)_j,a\rangle$; that intersection is convex by separate convexity of the upper sections, so it contains $[a,b]$, i.e.\ $\langle x',(y'_j)_j,t\rangle\in A$ for all $t\in[a,b]$. Hence
$\{\langle x',(y'_j)_j,t\rangle\mid (x',(y'_j)_j)\in N,\ t\in[a,b]\}\subseteq A.$
Since $S_{J\cup\{i\}}(x,y)$ is homeomorphic to the product $\big(X\times\prod_{j\in J}Y_j\big)\times Y_i$, and $[a,b]$ is a relative neighbourhood of $y_i$ in $Y_i$, the displayed set is a neighbourhood of $(x,y)$ in $S_{J\cup\{i\}}(x,y)$. Thus $A_{J\cup\{i\}}(x,y)$ is open at $(x,y)$. Since the same argument applies at every point of $A_{J\cup\{i\}}(x,y)$, and the corresponding slices coincide with $S_{J\cup\{i\}}(x,y)$, the whole trace $A_{J\cup\{i\}}(x,y)$ is open in $S_{J\cup\{i\}}(x,y)$. Thus $(\ast)$ holds for $J\cup\{i\}$. 
By induction $(\ast)$ holds for all $J\subseteq I$; with $J=I$, $\operatorname{gr}F$ is open in $Z$.

\smallskip
\noindent Two remarks on the argument. First, the product structure supplied by property {\bf \ref{PrA}} enters at exactly two points: it guarantees that every free part corresponds to a point of $Z$, so that  $\langle x',(y'_j)_j,t\rangle$ lies in $X\times Y$, and it guarantees that the coordinate sections of $Y$ are uniform across nearby points, so that $N\times[a,b]$ is a neighbourhood in the enlarged slice. Example \ref{ex_separate_S1} shows that without property {\bf \ref{PrA}} the argument fails at precisely these points. Second, the interval $[a,b]$ is a convex neighbourhood of $y_i$ in $Y_i$ generated by finitely many, indeed two, points of the section while Example \ref{ex_bilinear_pref} shows that the finite-dimensionality of the coordinate factors, which makes such finite convex generation possible, cannot be dispensed with either.\prff

\prf[Proof of Proposition \ref{thm_additional_directional_separate}] The implications (i) $\Rightarrow$ (ii) $\Rightarrow$ (iii) are immediate from the definitions. Conversely, assume $F$ has separately open upper sections, and separately convex upper sections in the coordinates other than $i$. Pick $x\in X$. We need to show that $F(x)$ is open. If $Y$ is a product set, then define a correspondence $H: Y_i\tra Y_{-i}$ such that $\operatorname{gr}H=F(x)$, where $i$ is the exempted coordinate. The lower sections of $H$ are the traces of $F(x)$ on the $i$-th coordinate sections of $Y$, which are open by separate openness and the upper sections of $H$ are separately convex and separately open in the remaining $n-1$ coordinates, by hypothesis. Then, by Theorem \ref{thm_separate_main} applied with domain $Y_i$ and range $Y_{-i}$, $H$ has an open graph, hence $F(x)$ is open in $Y$. Note that we do not use  the convexity of $F(x)$ along the coordinate $i$: coordinate $i$ serves as the domain of $H$ and only openness is needed there. If $Y$ is an open set in $\Re^n$, the reduction step in the proof of Theorem \ref{thm_separate_main} extends $F(x)$ to $\Re^n=\prod_{i=1}^n\Re$, a product set, preserving separate convexity in the coordinates other than $i$ and separate openness; the product case just treated then gives that $F(x)$ is open in $\Re^n$; since $F(x)\subseteq Y$, it is open in $Y$.
\prff

We now prepare for the proofs of Theorem \ref{thm_linear_main} and Proposition \ref{thm_additional_directional_linear}. Let $X$ be a subset of $\Re^n.$   The {\it closure of $X$} is denoted by cl$X$ and its {\it interior} by int$X$.   
Since any lower dimensional subset of $\Re^n$ has an empty interior, it is more convenient to work with the concept of relative interior.   
Recall that a subset $X$ of a (real) vector space is  affine  if for all $x,y\in X$ and $\lambda\in \Re,$ $\lambda x+ (1-\lambda)y\in X.$ It is clear that $A$ is affine if and only if $A-\{a\}$ is a subspace of $X$ for all $a\in A.$ The {\it affine hull of $X,$} aff$X,$ is  the smallest affine set containing $X$.  
The {\it relative interior} of a subset $X$ of $\Re^n$ is defined as $\text{\nf ri} X=\{x\in \mbox{\nf aff}X~|~\exists N_\varepsilon, \mbox{ an } \varepsilon \text{ neighborhood of } x, \text{ such that }  N_\varepsilon\cap \text{\nf aff}X\subseteq X\}$. That is, the relative interior of $X$ is the interior of $X$ with respect to the smallest affine subspace containing $X$. 
The following result is due to Rockafellar \citeyearpar[Theorem 6.1, p. 45]{ro70}.

\lm 
Let $X$ be a non-empty and convex subset of $\Re^n.$ Then $\mbox{\nf ri}X$ is non-empty, and for all $x\in \mbox{\nf ri}X, y\in \mbox{\nf cl}X$ and all $\lambda\in [0,1),$ $y\lambda x \in \mbox{\nf ri}X.$ 
\label{thm_rockafellar}
\lmm

\prf[Proof of Theorem \ref{thm_linear_main}]
The forward direction is obvious. For the backward direction, assume $F:X\tra Y$ has convex and linearly open upper sections and open lower sections. We first show that $F$ has open upper sections, that is, $F(x)$ is open in $Y$ for all $x\in X$; we then show that $\operatorname{gr}F$ is open.

\smallskip
\noindent{\it Step 1: open upper sections.} Fix $x_0\in X$ and write $A:=F(x_0)$. If $A$ is empty, it is open, so assume $A\neq\emptyset$. We first note two facts.

\smallskip
\noindent{\it Fact 1: $\mbox{\nf aff}A=\mbox{\nf aff}Y$.} Clearly $\mbox{\nf aff}A\subseteq\mbox{\nf aff}Y$. Suppose the inclusion is strict, and pick $y\in A$. If every $z\in Y$ satisfied $z-y\in\mbox{\nf aff}A-y$, then $\mbox{\nf aff}Y-y=\mbox{\nf span}(Y-y)\subseteq\mbox{\nf aff}A-y$, a contradiction; hence there exists $z\in Y$ with $z-y\notin\mbox{\nf aff}A-y$, and in particular $z\neq y$. Let $L$ be the straight line in $Y$ through $y$ and $z$. Then $L$ contains the non-degenerate segment joining $y$ and $z$, while $A\cap L\subseteq\mbox{\nf aff}A\cap L=\{y\}$, since $\mbox{\nf aff}A-y$ is a subspace not containing $z-y$. Thus $A\cap L=\{y\}$, which is not open in $L$, contradicting the linear openness of the upper sections. Hence $\mbox{\nf aff}A=\mbox{\nf aff}Y$.

\smallskip
\noindent{\it Fact 2: $A\cap\mbox{\nf ri}Y\subseteq\mbox{\nf ri}A$.} By Lemma \ref{thm_rockafellar}, $\mbox{\nf ri}A$ is non-empty; fix $w\in\mbox{\nf ri}A$, and pick $y\in A\cap\mbox{\nf ri}Y$. If $y=w$, there is nothing to prove. Otherwise, since $w\in A\subseteq\mbox{\nf aff}Y$ and $y\in\mbox{\nf ri}Y$, there exists $\bar\varepsilon>0$ such that $y+\varepsilon(y-w)\in Y$ for all $\varepsilon\in[0,\bar\varepsilon]$. Let $L$ be the straight line in $Y$ through $w$ and $y$. Then $L$ contains the segment joining $w$ and $y+\bar\varepsilon(y-w)$, and $A\cap L$ is open in $L$ and contains $y$; hence there exists $\varepsilon\in(0,\bar\varepsilon]$ with $z:=y+\varepsilon(y-w)\in A$. Since $y=z\lambda w$ for $\lambda=1/(1+\varepsilon)\in[0,1)$, with $w\in\mbox{\nf ri}A$ and $z\in A\subseteq\mbox{\nf cl}A$, Lemma \ref{thm_rockafellar} yields $y\in\mbox{\nf ri}A$.

\smallskip
By Fact 1, $\mbox{\nf ri}A$ is open in $\mbox{\nf aff}Y$, and hence open in $Y$; by Fact 2, it contains $A\cap\mbox{\nf ri}Y$. Thus every point of $A\cap\mbox{\nf ri}Y$ is an interior point of $A$ in $Y$. If $Y$ is open, then $\mbox{\nf ri}Y=Y$, so that $A=A\cap\mbox{\nf ri}Y\subseteq\mbox{\nf ri}A\subseteq A$, and $A=\mbox{\nf ri}A$ is open in $Y$: Step 1 is complete in this case. By property {\bf \ref{PrB}}, it remains to consider a polyhedron $Y=\{z\in\Re^n\mid \langle a_j,z\rangle\le b_j,\ j=1,\ldots,\ell\}$ and a point $y\in A$ on the relative boundary of $Y$.

Let $J:=\{j\mid \langle a_j,y\rangle=b_j\}$ be the set of active constraints at $y$. Since the inactive constraints are slack on a small ball, there exists $\delta>0$ with
\[
Y\cap B(y,\delta)=(y+T)\cap B(y,\delta),
\qquad
T=\{d\in\Re^n\mid \langle a_j,d\rangle\le 0\ \text{for all}\ j\in J\}.
\]
The cone $T$ is polyhedral, hence finitely generated, say $T=\mbox{\nf cone}\{d^1,\ldots,d^m\}$ with $d^k\neq0$ for all $k$; see \citet[Theorem 19.1]{ro70}. If $T=\{0\}$, then $Y\cap B(y,\delta)=\{y\}\subseteq A$, and we are done; so assume $m\geq1$. Each $d^k$ is admissible at $y$: for the active constraints, $\langle a_j,y+\varepsilon d^k\rangle=b_j+\varepsilon\langle a_j,d^k\rangle\le b_j$, and the inactive constraints remain slack for all sufficiently small $\varepsilon>0$. Hence the straight line in $Y$ through $y$ with direction $d^k$ contains a non-degenerate segment issuing from $y$ in the direction $d^k$, and, by linear openness, there exists $\varepsilon_k>0$ with $p^k:=y+\varepsilon_kd^k\in A$. Let $P:=\mbox{\nf co}\{y,p^1,\ldots,p^m\}$; by the convexity of $A$, $P\subseteq A$. Being a polytope, $P$ is a polyhedron, say $P=\{z\in\Re^n\mid \langle c_l,z\rangle\le\gamma_l,\ l=1,\ldots,q\}$, again by \citet[Theorem 19.1]{ro70}. Fix $l$. If $\langle c_l,y\rangle=\gamma_l$, then for each $k$, $p^k\in P$ gives $\langle c_l,y\rangle+\varepsilon_k\langle c_l,d^k\rangle\le\gamma_l=\langle c_l,y\rangle$, so that $\langle c_l,d^k\rangle\le0$; therefore $\langle c_l,d\rangle\le0$ for every $d\in T=\mbox{\nf cone}\{d^1,\ldots,d^m\}$. If $\langle c_l,y\rangle<\gamma_l$, then there exists $\delta_l>0$ with $\langle c_l,z\rangle<\gamma_l$ for all $z\in B(y,\delta_l)$. Let $\delta'>0$ be the minimum of $\delta$ and the $\delta_l$ over the inactive indices $l$. For any $w\in Y\cap B(y,\delta')$, the displayed identity gives $w=y+d$ with $d\in T$; then $\langle c_l,w\rangle=\langle c_l,y\rangle+\langle c_l,d\rangle\le\gamma_l$ for the active indices, and $\langle c_l,w\rangle<\gamma_l$ for the inactive ones. Hence $Y\cap B(y,\delta')\subseteq P\subseteq A$, and $y$ is an interior point of $A$ in $Y$. Therefore $F(x)$ is open in $Y$ for all $x\in X$.

\smallskip
\noindent{\it Step 2: open graph.} Pick $z_0=(x_0,y_0)\in \operatorname{gr}F$. Since $F(x_0)$ is open in $Y$, there is an open set $O\subseteq\Re^n$ with $y_0\in O$ and $Y\cap O\subseteq F(x_0)$. Let $Q\subseteq O$ be a closed cube centered at $y_0$. If $Y$ is open, we may take $Q\subseteq Y$, and set $V:=Q$, the convex hull of the finitely many vertices $p^1,\ldots,p^m$ of $Q$. If $Y$ is a polyhedron, then $Y\cap Q$ is a polytope, hence the convex hull of finitely many points $p^1,\ldots,p^m\in Y\cap Q$, by \citet[Theorem 19.1]{ro70}; set $V:=Y\cap Q$. In either case, $V=\mbox{\nf co}\{p^1,\ldots,p^m\}$ is a neighbourhood of $y_0$ in $Y$ with $V\subseteq F(x_0)$ and $p^1,\ldots,p^m\in F(x_0)$.

For each $k=1,\ldots,m$, we have $p^k\in F(x_0)$, hence $x_0\in F^{-1}(p^k)$. Since $F$ has open lower sections, for each $k$ there exists a neighbourhood $U^k$ of $x_0$ in $X$ such that $U^k\subseteq F^{-1}(p^k)$. Then $U=\bigcap_{k=1}^m U^k$ is a neighbourhood of $x_0$. For all $x\in U$ and all $k$, $p^k\in F(x)$; by convexity of $F(x)$, $V\subseteq F(x)$. Therefore $U\times V\subseteq \operatorname{gr}F$. Since $U\times V$ is a neighbourhood of $z_0$, $z_0$ is an interior point of $\operatorname{gr}F$. Hence $F$ has open graph.
\prff

\nt Since a subset of the product space $X\times Y$ can be defined as the graph of a correspondence from $X$ into $Y$, Theorems \ref{thm_separate_main} and \ref{thm_linear_main} provide characterizations of open sets when the sections of the set satisfy a suitable convexity assumption; see for example \cite{ha66amm}, \cite{uk23bams} and their references for characterizations of open sets in $\Re^n$, and also see \cite{fa66convex} for applications of sets with convex sections.

\prf[Proof of Proposition \ref{thm_additional_directional_linear}] Open upper sections immediately imply linearly open upper sections. Conversely, fix $x\in X$ and apply Step 1 of the proof of Theorem \ref{thm_linear_main} to the upper section $G(x)\subseteq Z$: the argument there uses only the convexity and the linear openness of the individual upper section, together with Lemma \ref{thm_rockafellar} and, on the relative boundary of a polyhedral range, the finite generation guaranteed by \citet[Theorem 19.1]{ro70}; it draws neither on the lower sections nor on the other upper sections of the correspondence. Hence $G(x)$ is open in $Z$ for every $x\in X$.
\prff

\medskip

\setlength{\bibsep}{4pt}
\setstretch{0.95}

\bibliographystyle{econometrica} 
\bibliography{References.bib}

\end{document}

%% file: figure_sep_ex.tex
\resizebox{0.45\textwidth}{!}{%
\begin{tikzpicture}
    \draw[->] (0,0) -- (6,0) node[right] {$x$};
    \draw[->] (0,0) -- (0,6) node[above] {$y$};

    \draw[domain=0.1:2, smooth, variable=\x, black]
        plot ({\x}, {pow(16/\x, 1/3)});
    \draw[domain=2:5, smooth, variable=\x, black]
        plot ({\x}, {16/pow(\x, 3)});

    \node at (5,0.2) [above] {$u_{1}(x,y)$};

    \draw[dashed] (0,0) -- (5,5) node[above right] {$x=y$};
\end{tikzpicture}%
}

%% file: figure4.tex
\begin{tikzpicture}
    \draw[->] (0,0) -- (6,0) node[right] {$x$};
    \draw[->] (0,0) -- (0,6) node[above] {$y$};



\draw[domain=0.1:2.56,smooth,variable=\x,black] plot ({\x},{(2 - 0.25*sqrt(\x))^2});

\draw[domain=2.56:3.8,smooth,variable=\x,black] plot ({\x},{16*(2 -sqrt(\x))^2});

\node at (4.5,0.2) [above] {$ u_{2}(x,y)$};

    \draw[dashed] (0,0) -- (5,5) node[above right] {$x=y$};
\end{tikzpicture}

%% file: References.bib
@article{bd04wp,
  title={Existence of Nash equilibria on convex sets},
  author={Banks, Jeffrey S and Duggan, John},
  journal={working paper}, 
  year={2004}
}

@article{to22geb, 
  title={Equilibrium non-existence in generalized games},
  author={T{\'o}bi{\'a}s, {\'A}ron},
  journal={Games and Economic Behavior},
  volume={135},
  pages={327--337},
  year={2022},
  publisher={Elsevier}
}

@article{ro65ecma,
  title={Existence and uniqueness of equilibrium points for concave n-person games},
  author={Rosen, J Ben},
  journal={Econometrica},
  pages={520--534},
  year={1965},
  publisher={JSTOR}
}

@article{fetal11jet,
  title={When is multidimensional screening a convex program?},
  author={Figalli, Alessio and Kim, Young-Heon and McCann, Robert J},
  journal={Journal of Economic Theory},
  volume={146},
  number={2},
  pages={454--478},
  year={2011},
  publisher={Elsevier}
}

@article{ki25wp,
  title={Concavity, partial concavity and quasiconcavity of functions: Characterizations using modularity and homogeneity},
  author={Kim, Sung Hyun},
  journal={working paper},
  year={2025} 
}

@article{yy25wp,
  title={Multidimensional monotonicity and economic applications},
  author={Yang, Frank and Yang, Kai Hao},
  journal={arXiv preprint arXiv:2502.18876},
  year={2025} 
}

@article{af83mor,
  title={Jointly constrained biconvex programming},
  author={Al-Khayyal, Faiz A and Falk, James E},
  journal={Mathematics of Operations Research},
  volume={8},
  number={2},
  pages={273--286},
  year={1983},
  publisher={INFORMS}
}

@article{rsy25wp,
  title={Extreme Equilibria: The Benefits of Correlation},
  author={Rudov, Kirill and Sandomirskiy, Fedor and Yariv, Leeat},
  journal={working paper},
year={2025}
}

@article{ah86ijm,
  title={Bi-convexity and bi-martingales},
  author={Aumann, Robert J and Hart, Sergiu},
  journal={Israel Journal of Mathematics},
  volume={54},
  pages={159--180},
  year={1986},
  publisher={Springer}
}

@article{getal07mmor,
  title={Biconvex sets and optimization with biconvex functions: a survey and extensions},
  author={Gorski, Jochen and Pfeuffer, Frank and Klamroth, Kathrin},
  journal={Mathematical Methods of Operations Research},
  volume={66},
  number={3},
  pages={373--407},
  year={2007},
  publisher={Springer}
}

@article{ba58adm,
  title={Minimalstellen von funktionen und extremalpunkte},
  author={Bauer, Heinz},
  journal={Archiv der Mathematik},
  volume={9},
  number={4},
  pages={389--393},
  year={1958},
  publisher={Springer}
}

@book{ch69bk,
    AUTHOR = {Choquet, Gustave},
     TITLE = {Lectures on analysis. {V}ol. {II}: {R}epresentation theory},
    SERIES = {Edited by J. Marsden, T. Lance and S. Gelbart},
 PUBLISHER = {W. A. Benjamin, Inc., New York-Amsterdam},
      YEAR = {1969},
     PAGES = {Vol. II: xx+315 pp.+xxi},
   MRCLASS = {46.00 (28.00)},
  MRNUMBER = {0250012 (40 \#3253)},
MRREVIEWER = {H. E. Lacey},
  BOEKCODE = {46A05},
}

@book{ho75bk,
  title={Geometric Functional Analysis and its Applications},
  author={Holmes, Richard B},
  volume={24 (Graduate Texts in Mathematics)},
  year={1975},
  publisher={Springer}
}

@article{setal21siam, 
  title={Optimization of quasi-convex function over product measure sets},
  author={Stenger, Jerome and Gamboa, Fabrice and Keller, Merlin},
  journal={SIAM Journal on Optimization},
  volume={31},
  number={1},
  pages={425--447},
  year={2021},
  publisher={SIAM}
}

@article{ketal21ecma,
  title={Extreme points and majorization: Economic applications},
  author={Kleiner, Andreas and Moldovanu, Benny and Strack, Philipp},
  journal={Econometrica},
  volume={89},
  number={4},
  pages={1557--1593},
  year={2021},
  publisher={Wiley Online Library}
}

@article{aetal23te,
  title={Optimal persuasion via bi-pooling},
  author={Arieli, Itai and Babichenko, Yakov and Smorodinsky, Rann and Yamashita, Takuro},
  journal={Theoretical Economics},
  volume={18},
  number={1},
  pages={15--36},
  year={2023},
  publisher={Wiley Online Library}
}

@article{ba23wp,
  title={Bauer's Maximum Principle for Quasiconvex Functions},
  author={Ball, Ian},
  journal={arXiv preprint arXiv:2305.04893},
  year={2023}
}

@article{gkr68pams,
  title={Convex functions on convex polytopes},
  author={Gale, David and Klee, Victor and Rockafellar, RT},
  journal={Proceedings of the American Mathematical Society},
  volume={19},
  number={4},
  pages={867--873},
  year={1968}
}

@article{er13pams,
  title={A converse of the Gale-Klee-Rockafellar theorem: continuity of convex functions at the boundary of their domains},
  author={Ernst, Emil},
  journal={Proceedings of the American Mathematical Society},
  volume={141},
  number={10},
  pages={3665--3672},
  year={2013}
}

@article{aa63,
  title={A definition of subjective probability},
  author={Anscombe, Francis J and Aumann, Robert J},
  journal={The Annals of Mathematical Statistics},
  volume={34},
  number={1},
  pages={199--205},
  year={1963}, 
  publisher={JSTOR}
}

@book{ab06,
  title={Infinite Dimensional Analysis: A Hitchhikers Guide},
  author={Aliprantis, Charalambos D and Border, Kim C},
  year={2006},
  publisher={Berlin: Springer-Verlag}
}

@book{ah71,
  title={General Competitive Analysis},
  author={Arrow, Kenneth J and Hahn, Frank H},
  year={1971},
  publisher={Amsterdam: North-Holland},
  address={}
}

@article{bl16,
  title={Nash was a first to axiomatize expected utility},
  author={Bleichrodt, Han and Li, Chen and Moscati, Ivan and Wakker, Peter P},
  journal={Theory and Decision},
  volume={81},
  number={3},
  pages={309--312},
  year={2016},
  publisher={Springer}
}

@book{bm95,
  title={Representations of Preference Orderings},
  author={Bridges, Douglas S and Mehta, Ghanshyam B},
  year={1995},
  publisher={Berlin: Springer-Verlag},
  address={}
}

@article{bpr76,
  title={Preferences which have open graphs},
  author={Bergstrom, Theodore C and Parks, Robert P and Rader, Trout},
  journal={Journal of Mathematical Economics},
  volume={3},
  number={3},
  pages={265--268},
  year={1976},
  publisher={Elsevier}
}

@book{ca21,
  title={Cours d'analyse de l'{\'E}cole Royale Polytechnique},
  author={Cauchy, Augustin--Louis},
  year={1821},
  publisher={Paris}
}

@article{ca11,
  title={Symposium on: Existence of Nash equilibria in discontinuous games},
  author={Carmona, Guilherme},
  journal={Economic Theory},
  volume={48},
  number={1},
  pages={1--4},
  year={2011},
  publisher={Springer}
}

@article{cm16,
  title={A continuous tale on continuous and separately continuous functions},
  author={Ciesielski, Krzysztof Chris and Miller, David},
  journal={Real Analysis Exchange},
  volume={41},
  number={1},
  pages={19--54},
  year={2016},
  publisher={JSTOR}
}

@article{kmu25et,
  title={Excess demand approach with non-convexity and discontinuity: 
A generalization of the Gale-Nikaido-Kuhn-Debreu lemma},
  author={Khan, M. Ali and McLean, Richard P. and Uyanik, Metin},
  journal={Economic Theory},
  volume={79},
  number={},
  pages={1167--1190},
  year={2025},
  publisher={}
}

@article{daz81,
  title={Nine kinds of quasiconcavity and concavity},
  author={Diewert, W Erwin and Avriel, Mordecai and Zang, Israel},
  journal={Journal of Economic Theory},
  volume={25},
  number={3},
  pages={397--420},
  year={1981},
  publisher={Elsevier}
}

@article{he73,
  title={On the convexity of the production possibility set under general production conditions},
  author={Herberg, Horst},
  journal={Zeitschrift f{\"u}r die gesamte Staatswissenschaft/Journal of Institutional and Theoretical Economics},
  pages={205--214},
  year={1973},
  publisher={JSTOR}
}

@article{de52,
  title={A social equilibrium existence theorem},
  author={Debreu, Gerard},
  journal={Proceedings of the National Academy of Sciences},
  volume={38},
  number={10},
  pages={886--893},
  year={1952},
  publisher={National Academy of Sciences}
}

@article{dm86,
  title={The existence of equilibrium in discontinuous economic games, I: Theory, II: Applications},
  author={Dasgupta, Partha and Maskin, Eric},
  journal={The Review of Economic Studies},
  volume={53},
  number={1},
  pages={1--41},
  year={1986},
  publisher={JSTOR}
}

@article{dmo04,
  title={Expected utility theory without the completeness axiom},
  author={Dubra, Juan and Maccheroni, Fabio and Ok, Efe A},
  journal={Journal of Economic Theory},
  volume={115},
  number={1},
  pages={118--133},
  year={2004},
  publisher={Elsevier}
}

@article{du11,
  title={Continuity and completeness under risk},
  author={Dubra, Juan},
  journal={Mathematical Social Sciences},
  volume={61},
  number={1},
  pages={80--81},
  year={2011},
  publisher={Elsevier}
}

@article{ei41,
  title={Ordered topological spaces},
  author={Eilenberg, Samuel},
  journal={American Journal of Mathematics},
  volume={63},
  number={1},
  pages={39--45},
  year={1941},
  publisher={JSTOR}
}

@article{fa59,
 ISSN = {00223808, 1537534X},
 URL = {http://www.jstor.org/stable/1825163},
 author = {Farrell, M. J.},
 journal = {Journal of Political Economy},
 number = {4},
 pages = {377--391},
 publisher = {University of Chicago Press},
 title = {The Convexity Assumption in the Theory of Competitive Markets},
 volume = {67},
 year = {1959}
}

@article{fa61a,
 ISSN = {00223808, 1537534X},
 URL = {http://www.jstor.org/stable/1828538},
 author = {Farrell, M. J.},
 journal = {Journal of Political Economy},
 number = {5},
 pages = {484--489},
 publisher = {University of Chicago Press},
 title = {On Convexity, Efficiency, and Markets: A Reply},
 volume = {69},
 year = {1961}
}

@article{fa66convex,
  title={Applications of a theorem concerning sets with convex sections},
  author={Fan, Ky},
  journal={Mathematische Annalen},
  volume={163},
  number={3},
  pages={189--203},
  year={1966},
  publisher={Springer}
}

@book{fi82,
  title={The Foundations of Expected Utility},
  author={Fishburn, Peter C},
  year={1982},
  publisher={Boston: D. Reidel Publishing Company},
  address={}
}

@article{fr78,
  title={Mixture axioms in linear and multilinear utility theories},
  author={Fishburn, Peter C and Roberts, Fred S},
  journal={Theory and Decision},
  volume={9},
  number={2},
  pages={161--171},
  year={1978},
  publisher={Springer}
}

@article{ge10,
  title={Consumer theory with bounded rational preferences},
  author={Gerasimou, Georgios},
  journal={Journal of Mathematical Economics},
  volume={46},
  number={5},
  pages={708--714},
  year={2010},
  publisher={Elsevier}
}

@article{ge13,
  title={On continuity of incomplete preferences},
  author={Gerasimou, Georgios},
  journal={Social Choice and Welfare},
  volume={41},
  number={1},
  pages={157--167},
  year={2013},
  publisher={Springer}
}

@book{tr84,
  title={Market demand: An analysis of large economies with non-convex preferences},
  author={Trockel, Walter},
  volume={223},
  year={1984},
  publisher={Springer Science \& Business Media}
}

@article{ge15,
  title={(Hemi)continuity of additive preference preorders},
  author={Gerasimou, Georgios},
  journal={Journal of Mathematical Economics},
  volume={58},
  pages={79--81},
  year={2015},
  publisher={Elsevier}
}

@article{gku19,
  title={Completeness and transitivity of preferences on mixture sets},
  author={Galaabaatar, T and Khan, M Ali and Uyanik, Metin},
  journal={Mathematical Social Sciences},
  volume={99},
  number={},
  pages={49--62},
  year={2019},
  publisher={}
}

@article{gku22games,
  title={The intermediate value theorem and decision-making in psychology and economics: An expositional consolidation},
  author={Ghosh, Aniruddha and Khan, M. Ali and Uyanik, Metin},
JOURNAL = {Games},
VOLUME = {13},
YEAR = {2022},
NUMBER = {4},
ARTICLE-NUMBER = {51} 
}

@article{gku23td,
  title={Continuity postulates and solvability axioms in economic theory and in mathematical psychology: a consolidation of the theory of individual choice},
  author={Ghosh, Aniruddha and Khan, M. Ali and Uyanik, Metin},
  journal={Theory and Decision},
  volume={94},
  number={2},
  pages={189--210},
  year={2023},
  publisher={Springer}
}

@article{gmms10,
  title={Objective and subjective rationality in a multiple prior model},
  author={Gilboa, Itzhak and Maccheroni, Fabio and Marinacci, Massimo and Schmeidler, David},
  journal={Econometrica},
  volume={78},
  number={2},
  pages={755--770},
  year={2010},
  publisher={Wiley Online Library}
}

@book{gp84,
  title={Calcolo Differentiale e Principii di Calcolo},
  author={Genocchi, Angelo and Peano, Giuseppe},
  year={1884},
  publisher={Torino: Fratelli Bocca}
}

@article{ha66amm,
  title={Necessary and sufficient condition for a convex set to be closed},
  author={Halkin, Hubert},
  journal={The American Mathematical Monthly},
  volume={73},
  number={6},
  pages={628--630},
  year={1966},
  publisher={JSTOR}
}

@article{hpz17,
  title={Non-parametric bounds for non-convex preferences},
  author={Halevy, Yoram and Persitz, Dotan and Zrill, Lanny},
  journal={Journal of Economic Behavior \& Organization},
  volume={137},
  pages={105--112},
  year={2017},
  publisher={Elsevier}
}

@article{hy16,
  title={Existence of Walrasian equilibria with discontinuous, non-ordered, interdependent and price-dependent preferences},
  author={He, Wei and Yannelis, Nicholas C},
  journal={Economic Theory},
  volume={61},
  number={3},
  pages={497--513},
  year={2016},
  publisher={Springer}
}

@article{ir12,
  title={The open graph theorem for correspondences: A new proof and some applications},
  author={Impicciatore, Galeazzo and Ruscitti, Francesco},
  journal={Theoretical Economics Letters},
  volume={2},
  number={3},
  pages={270--273},
  year={2012},
  publisher={Scientific Research Publishing}
}

@article{ka07,
  title={Archimedean and continuity},
  author={Karni, Edi},
  journal={Mathematical Social Sciences},
  volume={53},
  number={3},
  pages={332--334},
  year={2007},
  publisher={Elsevier}
}

@article{kd69amm,
  title={Joint continuity of monotonic functions},
  author={Kruse, Robert L. and Deely, John J.},
  journal={The American Mathematical Monthly},
  volume={76},
  number={1},
  pages={74--76},
  year={1969},
  publisher={Taylor \& Francis}
}

@incollection{ks02,
  title={Non-cooperative games with many players},
  author={Khan, M Ali and Sun, Yeneng},
  booktitle={Handbook of Game Theory},
  editor={Aumann, Robert J. and Hart, Sergiu},
  volume={3, Chapter 46},
  pages={1761--1808},
  year={2002},
  publisher={Amsterdam: Elsevier}
}

@book{ko57,
  title={Three Essays on the State of Economic Science},
  author={Koopmans, Tjalling C},
  year={1957},
  publisher={New York: McGraw-Hill}
}

@article{ko61,
  title={Convexity assumptions, allocative efficiency, and competitive equilibrium},
  author={Koopmans, Tjalling C},
  journal={Journal of Political Economy},
  volume={69},
  number={5},
  pages={478--479},
  year={1961},
  publisher={The University of Chicago Press}
}

@article{kr86,
  title={Nontransitive-nontotal consumer theory},
  author={Kim, Taesung and Richter, Marcel K},
  journal={Journal of Economic Theory},
  volume={38},
  number={2},
  pages={324--363},
  year={1986},
  publisher={Elsevier}
}

@article{ks15,
  title={Continuity, completeness, betweenness and cone-monotonicity},
  author={Karni, Edi and Safra, Zvi},
  journal={Mathematical Social Sciences},
  volume={74},
  pages={68--72},
  year={2015},
  publisher={Elsevier}
}

@article{ku19a, 
  title={Topological connectedness and behavioral assumptions on preferences: a two-way relationship},
  author={Khan, M Ali and Uyan{\i}k, Metin},
  journal={Economic Theory},
 number = {2},
 pages = {411--460},
 publisher = {Springer}, 
 urldate = {2026-07-10},
 volume = {71},
 year = {2021}
}

@article{ma50,
  title={Rational behavior, uncertain prospects, and measurable utility},
  author={Marschak, Jacob},
  journal={Econometrica},
  volume={18},
  number={2},
  pages={111--141},
  year={1950},
  publisher={JSTOR}
}

@book{hi74,
  title={Core and Equilibria of a Large Economy},
  author={Hildenbrand, Werner},
  volume={4},
  year={1974},
  publisher={Princeton: Princeton University Press}
}

@article{na50b,
  title={The bargaining problem},
  author={Nash, John F},
  journal={Econometrica},
  volume={18},
  number={2},
  pages={155--162},
  year={1950},
  publisher={JSTOR}
}

@article{nt95,
  title={Continuous linear representability of binary relations},
  author={Neuefeind, Wilhelm and Trockel, Walter},
  journal={Economic Theory},
  volume={6},
  number={2},
  pages={351--356},
  year={1995},
  publisher={Springer}
}

@article{pww24,
  title={Anticomonotonicity for preference axioms: The natural counterpart
to comonotonicity},
  author={Principi, Giuliu and Wakker, Peter P. and Wang, Ruodo},
  journal={unpublished},
  volume={},
  number={},
  pages={},
  year={2024},
  publisher={The Mathematical Society of the Republic of China}
}

@article{re20are,
  title={Nash equilibrium in discontinuous games},
  author={Reny, Philip J},
  journal={Annual Review of Economics},
  volume={12},
  pages={439--470},
  year={2020},
  publisher={Annual Reviews}
}

@article{re99,
  title={On the existence of pure and mixed strategy Nash equilibria in discontinuous games},
  author={Reny, Philip J},
  journal={Econometrica},
  volume={67},
  number={5},
  pages={1029--1056},
  year={1999},
  publisher={Wiley Online Library}
}

@article{ro55,
  title={On the continuity of functions of several variables},
  author={Rosenthal, Arthur},
  journal={Mathematische Zeitschrift},
  volume={63},
  number={1},
  pages={31--38},
  year={1955},
  publisher={Springer}
}

@book{ro70,
  title={Convex Analysis},
  author={Rockafellar, Ralph Tyrell},
  year={1970},
  publisher={New York: Princeton University Press},
  address={}
}

@article{sc69,
  title={Competitive equilibria in markets with a continuum of traders and incomplete preferences},
  author={Schmeidler, David},
  journal={Econometrica},
  volume={37},
  number={4},
  pages={578--585},
  year={1969},
  publisher={JSTOR}
}

@article{sc71,
  title={A condition for the completeness of partial preference relations},
  author={Schmeidler, David},
  journal={Econometrica},
  volume={39},
  number={2},
  pages={403--404},
  year={1971},
  publisher={Blackwell Publishing}
}

@article{sh74,
  title={The nontransitive consumer},
  author={Shafer, Wayne},
  journal={Econometrica},
  volume={42},
  number={5},
  pages={913--919},
  year={1974},
  publisher={Elsevier}
}

@article{so65,
  title={The relationship between transitive preference and the structure of the choice space},
  author={Sonnenschein, Hugo},
  journal={Econometrica},
 volume={33},
   number={3},
  pages={624--634},
  year={1965},
  publisher={JSTOR}
}

@article{uk22jme,
  title={ The continuity postulate in economic theory: a deconstruction and an integration},
  author={Uyanik, Metin and Khan, M. Ali},
  journal={Journal of Mathematical Economics},
  volume={101},
 pages = {Article 102704},
 year = {2022},
  publisher={},
  nameorder={random}
}

@article{uk23bams,
  title={On separate continuity and separate convexity: A synthetic treatment for functions and sets},
  author={Uyanik, Metin and Khan, M. Ali},
  journal={Bulletin of the Australian Mathematical Society},
  pages={1--11},
  year={2023}
}

@article{wa54a,
  title={Partially ordered topological spaces},
  author={Ward, Lewis E.},
  journal={Proceedings of the American Mathematical Society},
  volume={5},
  number={1},
  pages={144--161},
  year={1954},
  publisher={JSTOR}
}

@book{wa89,
  title={Additive Representations of Preferences: A New Foundation of Decision Analysis},
  author={Wakker, Peter},
  year={1989},
  publisher={Boston: Kluwer Academic Publishers}
}

@book{wj53,
  title={Demand Analysis},
  author={Wold, Herman and Jureen, Lars},
  year={1953},
  publisher={New York: John Wiley and Sons, Inc.}
}

@article{wo43,
  title={A synthesis of pure demand analysis, I--III},
  author={Wold, Herman},
  journal={Scandinavian Actuarial Journal},
  volume={26},
  pages={85--118, 220--263; 27, 69--120},
  year={1943--44}
}

@article{rr19,
  title={Convex preferences: A new definition},
  author={Richter, Michael and Rubinstein, Ariel},
  journal={Theoretical Economics},
  volume={14},
  number={4},
  pages={1169--1183},
  year={2019},
  publisher={Wiley Online Library}
}

@article{wz99,
  title={A unified derivation of classical subjective expected utility models through cardinal utility},
  author={Wakker, Peter P and Zank, Horst},
  journal={Journal of Mathematical Economics},
  volume={32},
  number={1},
  pages={1--19},
  year={1999},
  publisher={Elsevier}
}

@article{ya83b,
  title={On open preferences},
  author={Yamazaki, Akira},
  journal={Hitotsubashi Journal of Economics},
  volume={24},
  pages={149--152},
  year={1983},
  publisher={JSTOR}
}

@article{yo10qjpam,
  title={A note on monotone functions},
  author={Young, WH},
  journal={The Quarterly Journal of Pure and Applied Mathematics (Oxford Ser.)},
  volume={41},
  number={},
  pages={79–87.},
  year={1910},
  publisher={}
}

@article{yp83,
  title={Existence of maximal elements and equilibria in linear topological spaces},
  author={Yannelis, Nicholas C and Prabhakar, N D},
  journal={Journal of Mathematical Economics},
  volume={12},
  number={3},
  pages={233--245},
  year={1983},
  publisher={Elsevier}
}

@article{zh95jmaa,
  title={On the existence of equilibrium for abstract economies},
  author={Zhou, JX},
  journal={Journal of Mathematical Analysis and Applications},
  volume={193},
  number={3},
  pages={839--858},
  year={1995},
  publisher={Elsevier}
}
